# Terahertz Josephson plasma waves in layered superconductors: spectrum, generation, nonlinear, and quantum phenomena

S.E. Savel'ev[1,2], V.A. Yampol'skii[1,3], A.L. Rakhmanov[1,2,4], and Franco Nori[1,5]

[1] *Advanced Science Institute, the Institute of Physical and Chemical Research (RIKEN),*
*Wako-shi, Saitama, 351-0198, Japan*

[2] *Department of Physics, Loughborough University, Loughborough LE11 3TU, UK*

[3] *A.Ya. Usikov Institute for Radiophysics and Electronics,*
*National Academy of Sciences of Ukraine,*
*12 Acad. Proskura Str., 61085 Kharkov, Ukraine*

[4] *Institute for Theoretical and Applied Electrodynamics*
*Russian Academy of Sciences, 125412 Moscow, Russia*

[5] *Department of Physics, Center for Theoretical Physics,*
*Applied Physics Program, Center for the Study of Complex Systems,*
*The University of Michigan, Ann Arbor, MI 48109-1040, USA*

# Abstract

The recent growing interest in terahertz (THz) and sub-THz science and technology is due to its many important applications in physics, astronomy, chemistry, biology, and medicine, including THz imaging, spectroscopy, tomography, medical diagnosis, health monitoring, environmental control, as well as chemical and biological identification. We review the problem of linear and nonlinear THz and sub-THz Josephson plasma waves in layered superconductors and their excitations produced by moving Josephson vortices. We start by discussing the coupled sine-Gordon equations for the gauge-invariant phase difference of the order parameter in the junctions, taking into account the effect of breaking the charge neutrality, and deriving the spectrum of Josephson plasma waves. We also review surface and waveguide Josephson plasma waves. The spectrum of these waves is presented, and their excitation is discussed. We review the propagation of weakly nonlinear Josephson plasma waves below the plasma frequency, $\omega_J$, which is very unusual for plasma-like excitations. In close analogy to nonlinear optics, these waves exhibit numerous remarkable features, including a self-focusing effect, and the pumping of weaker waves by a stronger one. In addition, an unusual stop-light phenomenon, when $\partial\omega/\partial k \approx 0$, caused by both nonlinearity and dissipation, can be observed in the Josephson plasma waves. At frequencies above $\omega_J$, the current-phase nonlinearity can be used for transforming continuous sub-THz radiation into short, strongly amplified, pulses. We also present quantum effects in layered superconductors, specifically, the problem of quantum tunnelling of fluxons through stacks of Josephson junctions. Moreover, the nonlocal sine-Gordon equation for Josephson vortices is reviewed. We discuss the Cherenkov and transition radiations of the Josephson plasma waves produced by moving Josephson vortices, either in a single Josephson junction or in layered superconductors. Furthermore, the expression for the Cherenkov cone of the excited Josephson plasma waves is derived. We also discuss the problem of coherent radiation (superradiance) of the THz waves by exciting uniform Josephson oscillations. The effects reviewed here could be potentially useful for sub-THz and THz emitters, filters, and detectors.

PACS numbers: 74.80.DM, 74.50.+r, 74.60.Ec

**List of notations and abbreviations**

$J_c$ — maximum Josephson current density;

$\lambda_{ab}$ and $\lambda_c$ — in-plane and out-of-plane London penetration depths;

$\lambda_J$ — Josephson penetration depth;

$\omega_J$ — Josephson plasma frequency;

$D$ — period of the layered structure;

$s \ll D$ — thickness of a superconducting layer;

$\varepsilon$ – interlayer dielectric constant;

$\sigma_\perp$ — quasiparticle conductivity across the layers;

$\sigma_\parallel$ — quasiparticle conductivity along the layers;

$\omega_r$ — relaxation frequency;

$\varphi^{l+1,l}$ — gauge invariant phase difference between the $l$th and $(l+1)$th superconducting layers;

$\vec{E}$ and $\vec{H}$ — electric and magnetic fields;

$\vec{A}$ — vector potential;

$\omega$ — wave frequency;

$\Omega = \omega/\omega_J$ — dimensionless frequency;

$\theta$ — incident angle;

$c_{\text{sw}} = \lambda_J \omega_J$ — Swihart velocity;

$V_{\min}$ — minimum vortex velocity for emitting out-of-plane Cherenkov radiation;

$D^*$ — thickness of a weak junction;

$J_c^*$ — maximum Josephson current density through a weak junction;

$\Phi_0 = \pi c \hbar / e$ — magnetic flux quantum;

$c$ — speed of light;

$e$ — elementary charge;

$\mathcal{H}_0 = \Phi_0 / 2\pi D \lambda_c$ — characteristic scale of the magnetic fields in layered superconductors;

THz — terahertz;

JPW — Josephson plasma wave;

JPR — Josephson plasma resonance;

JV — Josephson vortex;

HTS — high-temperature superconductor;

LTS — low-temperature superconductor;

NWGM — nonlinear waveguide mode;

SJJ - stack of Josephson junctions;

SJPW — surface Josephson plasma wave;

CVC — current-voltage characteristics;

MQT — macroscopic quantum tunnelling.

$z$-axis is directed across the layers;

$y$-axis is directed along the magnetic field.

## I.  INTRODUCTION

The physical properties of layered superconductors have attracted a great deal of interest from many research groups. The strongly anisotropic high-$T_c$ (HTS) Bi$_2$Sr$_2$CaCu$_2$O$_{8+\delta}$ single crystals are characteristic members of this family. In these materials, very thin (0.2 nm) superconducting layers are separated by a thicker (1.5 nm) insulator which allows the propagation of electromagnetic waves with frequencies above the Josephson plasma frequency $\omega_J$ (about 0.1-0.4 THz).

Artificial stacks of Josephson junctions (SJJs), e.g., Nb/Al-AlO$_x$/Nb, represent another group of such materials. These artificial stacks can be prepared using a modified selective niobium anodization process and usually consist of thicker superconducting layers (the thickness $\sim$ 20 nm, is less or about the magnetic penetration depth) separated by thinner insulating tunnelling barriers ($\sim$ 2 nm). In such artificial systems, the coupling can be controlled by varying the thickness of the superconducting layers, thus resulting in a controllable change of the frequency (usually up to 0.7 THz) of the propagating electromagnetic waves.

Many experiments on the **c**-axis transport in layered HTS justify the use of a model in which the superconducting layers are coupled by the intrinsic Josephson effect through the layers [1–5]. The Josephson current flowing along the **c**-axis couples with the electromagnetic field. Due to this coupling, similarly to a single Josephson junction, electromagnetic waves (so-called Josephson plasma waves (JPWs) [5–16]) can propagate either in artificial multistacks or in layered superconductors.

A great challenge is to excite electromagnetic waves in Bi$_2$Sr$_2$CaCu$_2$O$_{8+\delta}$ samples in a controllable manner because of its sub-terahertz and terahertz frequency ranges [16, 17], which is still hardly reachable for both electronic and optical devices. The poorly controlled THz range of electromagnetic spectra sits between 300 GHz and 30 THz which corresponds to 1000-10 $\mu$m (wavelength), 1.25-125 meV (energy) or 14-1400 K (temperature). Thus, this THz gap covers temperatures of biological processes. Also, a substantial fraction of the luminosity of the Big Bang lies in the THz range [18]. During the last decade there have been many attempts to push THz science and technology forward because of many important potential applications in physics, astronomy, chemistry, biology and medicine, including THz imaging, spectroscopy, tomography, medical diagnosis, health monitoring, environmental control, chemical and biological identification (see, e.g., Refs. 18, 19).

There are several optical and electronic (microwave) THz devices competing for the THz market (see, e.g., Refs. 20–23). Optical devices employ several approaches to reduce their frequency. In the other side of the spectrum, the frequency of microwave devices (usually electronic devices based on semiconductors) has to be increased in order to reach the THz gap.

Recently, a wide variety of electronic THz sources are being investigated [24]. These include the following: resonant tunnelling diodes, THz plasma-wave photomixers, THz-quantum cascade lasers [25, 26], and Bloch oscillators. Numerous types of THz detectors [20–23] for time-domain systems have been studied so far including: bolometers; single-electron transistors; photoconductive antennas; and electro-optic sampling techniques for time-domain detection. THz waveguiding using conventional structures, such as plastic ribbons, metal tubes, and dielectric fibers has been demonstrated—however, these still have limited applications because of high losses. Moreover, despite of a variety of proposed and even constructed THz sources, detectors, and waveguides, there is still a lack of sufficiently controllable THz devices. Indeed, most of the devices mentioned above have problems for applications in electronics. This is because these are either rather large, or not easily assembled together, or non-tunable.

Superconducting devices employing the Josephson effect are now considered as potential candidates (see, e.g., Ref. 27) for making single-chip multifunctional THz devices. Indeed, the growing number of studies of Josephson devices is partly motivated by the current interest in the sub-THz and THz frequency range of electromagnetic waves. These electromagnetic waves interact nonlinearly with the Josephson medium itself and with magnetic flux quanta (Josephson vortices), which, in turn, can be manipulated by varying an in-plane magnetic field and/or an out-of-plane electric current [28]. Such a level of controllability can be used to propose a set of well-integrated classical and even quantum THz devices, including pulse and continuous wave generators, single photon THz generators, tunable filters, detectors, wave mixers, lenses, converters, and amplifiers.

We intend to focus in this review on basic mechanisms that are potentially useful for controlling THz radiation in either artificial superconducting/non-superconducting multi-layers or high-temperature layered superconductors.

Another reason for the scientific interest of the Josephson plasma waves is due to their resonant interaction with an external microwave electric field oriented along the **c**-axis, called the Josephson plasma resonance (JPR). This phenomenon plays an essential role in the microwave absorption and reflectivity of layered superconductors [29–39] near the resonance frequency $\omega_J$. Besides this, the JPR has been introduced as a powerful tool providing unique information on the Josephson coupling of superconducting layers in HTS and on the structure of vortex phases in the presence of an applied magnetic field (see, e.g., Refs. 11–13, 40–46.

## A. Cherenkov radiation from moving Josephson vortices

Since both Josephson plasma waves and Josephson vortices have a common nature and can be described by the same set of equations, called coupled sine-Gordon equations [37, 47–54], the Josephson vortex motion strongly couples to the plasma waves. As a result, Josephson plasma waves can be excited by the motion of Josephson vortices. It is common for electromagnetic processes that their emitted power increases with the speed of the moving object. Thus, the THz wave intensity produced by moving Josephson vortices should rise when increasing the vortex speed, $V$. The radiation should increase greatly if $V$ exceeds the value of the phase velocity of the waves, which is referred to as Cherenkov-effect [55]. In particular, Cherenkov radiation arises [17, 51, 56–67] in Josephson junctions if the vortex velocity $V$ exceeds the phase velocity $c_w(q)$ of the waves with a definite wave-vector $q$.

In the framework of the local theory [68], the maximum velocity of the vortex in a single Josephson junction cannot exceed the so-called Swihart velocity, $c_{\rm sw}$, which coincides with the minimal phase speed of the linear waves. This results in the absence of Cherenkov radiation in a single Josephson junction. However, even for a single junction, the local theory is valid only if the Josephson penetration length,

$$\lambda_J = \sqrt{\frac{c\Phi_0}{16\pi^2\lambda J_c}}, \qquad (1)$$

is much larger than the London penetration depth $\lambda$. Here $c$ is the speed of light, $\Phi_0 = \pi\hbar c/e$ is the flux quantum, and $J_c$ is the critical current density of the junction. To higher-order approximation with respect to $\lambda/\lambda_J$, a *nonlocal* sine-Gordon equation should be used for the description of the vortex dynamics [69] and the wave spectrum [56]. In this case, the maximum vortex velocity *exceeds* the phase velocity of the waves with high $q$ and Cherenkov radiation can be observed. The interaction between the junctions in SJJ gives rise to a similar effect [57].

## B. Out-of-plane Cherenkov radiation

The Cherenkov radiation in a single junction differs, in some essential features, from the classical Cherenkov effect. If the particle moves with a velocity $V$ higher than the speed $c_w$ of the emitted waves, the Cherenkov radiation propagates *inside* the Cherenkov cone with an angle $\theta$ determined by the relation $\cos\theta = c_w/V$. However, the Cherenkov radiation in a single junction has a wave vector directed *along* the layers in parallel to the vortex motion. Moreover, this radiation could be treated as a fine structure of the Josephson vortex moving with the same velocity $V$ as the vortex itself. Particularly interesting is to answer the question if the Cherenkov radiation can run *away* from the layers where it is generated. In other words, can the moving vortex generate *out-of-plane* radiation with its wave vector directed at a finite angle from its velocity? This question was answered in Refs. 70–72. The vortex motion in layered superconductors is described by the *nonlocal sine-Gordon equation*. The maximum vortex velocity $V_c$ is of the same order as the minimal velocity $V_{\min}$ necessary for the generation of out-of-plane waves. Therefore, the out-of-plane Cherenkov radiation can be certainly generated by a vortex moving in a junction weaker than others, since its maximum velocity is increased in this case. A subset of weaker intrinsic Josephson junctions in $Bi_2Sr_2CaCu_2O_{8+\delta}$-based samples can be made using either (i) the controllable intercalation technique [73–75], (ii) Chemical Vapor Deposition (see, e.g., Ref. 76), or (iii) via the admixture of $Bi_2Sr_2Cu_2O_{6+\delta}$ and $Bi_2Sr_2Ca_2Cu_3O_{10+\delta}$ [77]. Also, such a system can be created using artificial stacks of layers of low temperature superconductors. It is necessary to emphasize that the out-of-plane Cherenkov radiation can be more preferable for applications than propagating along the **ab** plane. The ratio of the tangential electric field and magnetic field appears to be of the same order in the most intense part of the spectrum of emitted out-of-plane waves. This a very peculiar property, unusual for conducting media, indicating that there is no impedance mismatch for the out-of-plane radiation.

### C. Nonlinear Josephson plasma waves

The set of coupled sine-Gordon equations is essentially nonlinear due to the Josephson relation between the current density $J$ across the layers and the gauge-invariant interlayer phase difference $\varphi$, i.e., $J = J_c \sin \varphi$. In the strongly nonlinear regime ($\varphi \sim \pi$), the sine-Gordon equation possesses soliton and breather solutions [78, 79]. However, the nonlinearity becomes crucial even for much smaller wave amplitudes, i.e., $|\varphi| \ll 1$, due to a gap in the spectrum of JPWs. In Refs. 80–83 we discussed such phenomena. Some of these, (e.g., JPWs self-focusing effects, the pumping of weaker waves by a stronger one, nonlinear plasma resonances, nonlinear surface waves and waveguide propagation) have analogues in traditional nonlinear optics. In addition, the unusual stop-light phenomenon, caused by both nonlinearity and dissipation, was predicted in Ref. 80.

### D. Overwiev of this article

This review aims to give an update physical picture of Josephson plasma waves, discuss their role in the electrodynamic properties of layered superconductors, and describe the Cherenkov mechanism of the wave excitation. About half of the review emphasizes work done by our group. In Sec. II we derive the coupled sine-Gordon equations, obtain the dispersion relation for the *linear* Josephson plasma waves, discuss the role of the dynamical breaking of charge neutrality in the JPW spectrum, study the effect of an external magnetic field on the linear JPWs, and discuss surface Josephson plasma waves. We describe the nonlinear Josephson plasma waves in Sec. III. Here we show the possible propagation of nonlinear waves below the plasma frequency, describe the stop-light phenomenon that occurs due to both nonlinearity and wave dissipation, study nonlinear Josephson plasma resonances, and discuss nonlinear waveguide modes in plates of layered superconductors. Section IV is devoted to the Cherenkov radiation of Josephson plasma waves in a single Josephson junction as well as in layered superconductors. In this section we also show that the Josephson vortex can emit out-of-plane electromagnetic waves even when $V < V_c$, if the critical current $J_c^*$ in the weak junction is nonhomogeneous. The problem of coherent radiation (superradiance) of the THz waves in layered superconductors is discussed in Section V. Quantum effects in layered superconductors, specifically, the problem of quantum tunnelling of fluxons through stacks of Josephson junctions, is reviewed in Section VI.

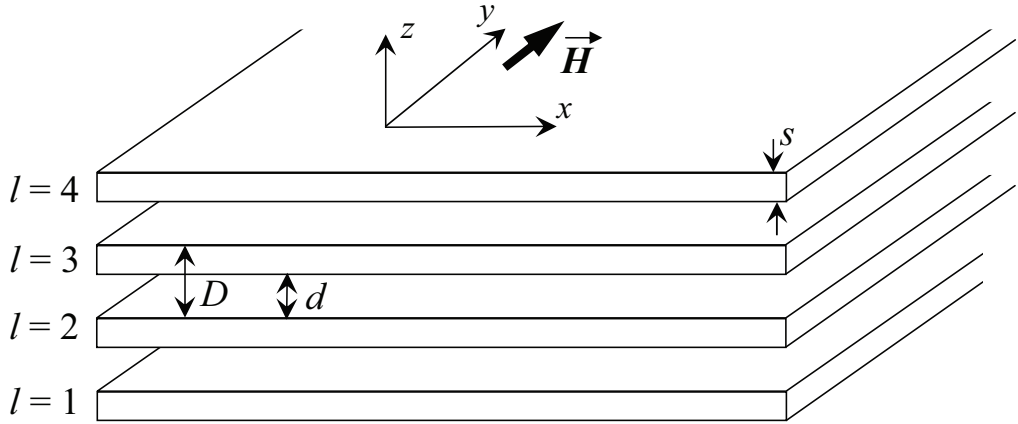

FIG. 1: Geometry of the problem and coordinate system.

## II. LINEAR JOSEPHSON PLASMA WAVES

### A. Coupled sine-Gordon equations. Dispersion relation for the Josephson plasma waves

To the best of our knowledge, Sakai et al. [47] were the first who derived the set of coupled sine-Gordon equations for the phenomenological description of the electrodynamics of layered superconductors. Later on, many authors (see, e.g., Refs. 37, 48–51, 53, 54 rederived the coupled sine-Gordon equations using other approaches. Reference 52 presented a microscopic theory of the superconducting phase and charge dynamics in intrinsic Josephson-junction systems based on a BCS functional integral formalism. Under some conditions, their theory also rederives the coupled sine-Gordon equations.

Consider now an infinite layered superconductor in the geometry shown in Fig. 1. Superconducting and insulating layers have thicknesses $s$ and $d$, respectively. Following Refs. 50, 53, we assume that the superconducting layers are extremely thin, $s \ll d$, so that, inside them, the spatial variations of the electric field in the direction perpendicular to the layers may be neglected. The coordinate system is chosen in such a way that the crystallographic **ab**-plane coincides with the $xy$-plane and the **c**-axis is along the $z$-axis. Superconducting layers are numbered by the index $l$. The electric, $\vec{E}$, and magnetic, $\vec{H}$, fields have components,

$$\vec{E} = \{E_x, 0, E_z\}, \qquad \vec{H} = \{0, H, 0\}. \tag{2}$$

The total tunnelling current between the $(l+1)$th and $l$th superconducting layers is the sum of the Cooper pair current and the quasiparticle current caused by the electric field $E_z^{l+1,l}$. Its density obeys the usual Josephson relation [68]

$$J^{l+1,l} = J_c \sin(\varphi^{l+1,l}) + \sigma_\perp E_z^{l+1,l}, \tag{3}$$

where $\varphi^{l+1,l}$ is the gauge-invariant interlayer phase difference of the superconducting order parameter

$$\varphi^{l+1,l} = \chi_{(l+1)} - \chi_l - \frac{2\pi}{\Phi_0} \int_l^{l+1} A_z^{l+1,l} dz, \tag{4}$$

$\sigma_\perp$ is the quasiparticle conductivity in the direction orthogonal to the layers, $\chi_l$ is the phase of the order parameter in the $l$th superconducting layer, $A_z^{l+1,l}$ is the $z$-component of the vector potential. The values $H^{l+1,l}$ and $E_z^{l+1,l}$ of the electric and magnetic fields within the junction between the $(l+1)$th and $l$th layers are related to each other by the Maxwell equation,

$$\frac{\partial H^{l+1,l}}{\partial x} = \frac{4\pi}{c}[J_c \sin(\varphi^{l+1,l}) + \sigma_\perp E_z^{l+1,l}] + \frac{\varepsilon}{c}\frac{\partial E_z^{l+1,l}}{\partial t}, \tag{5}$$

where $\varepsilon$ is the dielectric constant of the insulating layers.

We use the Maxwell equation and the quantum mechanical expression for the $x$-component of the current in the $l$th superconducting layer and find, for the discrete case considered here, the relation between the magnetic fields in neighboring junctions in the form,

$$-\frac{H^{l+1,l} - H^{l,l-1}}{s} = \frac{\Phi_0}{2\pi\lambda^2}\left(\frac{\partial \chi_l}{\partial x} - \frac{2\pi}{\Phi_0}A_{xl}\right)$$
$$+ \frac{4\pi}{c}\sigma_\| E_{xl}, \tag{6}$$

where $\lambda$ is the London penetration depth of the *bulk* superconductor, $E_{xl}$ and $A_{xl}$ are the $x$-components of the electric field and vector potential in the $l$th superconducting layer, $\sigma_\|$ is the quasiparticle conductivity along the layers. The expressions for the electric and magnetic fields via the vector potential can be written as,

$$E_{xl} = -\frac{1}{c}\frac{\partial A_{xl}}{\partial t} - \frac{\partial A_{0l}}{\partial x}, \tag{7}$$

$$E_z^{l+1,l} = -\frac{1}{c}\frac{\partial A_z^{l+1,l}}{\partial t} - \frac{A_{0(l+1)} - A_{0l}}{D}, \tag{8}$$

$$H^{l+1,l} = -\frac{\partial A_z^{l+1,l}}{\partial x} - \frac{A_{x(l+1)} - A_{xl}}{D}. \tag{9}$$

Here $A_{0l}$ is the scalar potential in the $l$th layer, $D = s + d \approx d$ is the period of layered structure.

Using Eqs. (4), (8), and assuming that $\int_l^{l+1} A_z^{l+1,l} dz = A_z^{l+1,l} D$, one can easily obtain the relationship between the electric field across the layers and the phase difference $\varphi^{l+1,l}$,

$$E_z^{l+1,l} = \frac{\Phi_0}{2\pi cD}\frac{\partial \varphi^{l+1,l}}{\partial t} - \frac{\psi_{l+1} - \psi_l}{D}, \tag{10}$$

where

$$\psi_l = \frac{\Phi_0}{2\pi c}\frac{\partial \chi_l}{\partial t} + A_{0l} \tag{11}$$

is the gauge-invariant scalar potential. This potential can be obtained from the Poisson equation. The contribution of the gradient of $\psi$ to the electric field is caused by breaking the charge neutrality in the layered superconductor. Because of the smallness of the Debye length in any superconductor, with respect to the London penetration depth, this effect can be neglected in many cases. For example, the breaking of charge neutrality is negligible for frequencies far from $\omega_J$. However, breaking the charge neutrality can play an important role in the dispersion properties of the Josephson plasma waves when $\omega$ is very close to $\omega_J$ (see, e.g., Ref. 37), and we will consider this specific situation in the next section.

Now, following Artemenko and Remizov [50, 53], we omit the gauge-invariant scalar potential in Eq. (10). Thus, excluding the vector potential, as well as the electric and magnetic fields from the set of equations (5)–(10), we derive the coupled sine-Gordon equations for the phase differences $\varphi^{l+1,l}$,

$$\left(1 - \frac{\lambda_{ab}^2}{D^2}\partial_l^2\right)\left[\frac{\partial^2 \varphi^{l+1,l}}{\partial t^2} + \omega_r\frac{\partial \varphi^{l+1,l}}{\partial t} + \omega_J^2 \sin\left(\varphi^{l+1,l}\right)\right]$$

$$- \frac{c^2}{\varepsilon}\frac{\partial^2 \varphi^{l+1,l}}{\partial x^2} = 0. \tag{12}$$

Here $\lambda_{ab} = \lambda(D/s)^{1/2}$ is the London penetration depth into the $z$-direction of the *layered* HTS, the second-order discrete differential operator $\partial_l^2$ is defined as $\partial_l^2 f_l = f_{l+1} + f_{l-1} - 2f_l$,

$$\omega_J = \sqrt{\frac{8\pi e D J_c}{\hbar \varepsilon}} \tag{13}$$

is the Josephson plasma frequency, $\omega_r = 4\pi\sigma_\perp/\varepsilon$ is the relaxation frequency proportional to the **c**-axis quasi-particle conductivity. As was shown in Ref. 84, the intralayer quasiparticle conductivity $\sigma_\parallel$ appearing in Eq. (6) should be accounted for when $\omega$ is far enough from the plasma frequency. The contribution of the in-plane conductivity can be easily incorporated in our analysis. However, for the frequency range close to $\omega_J$, the term with $\sigma_\parallel$ is strongly suppressed and can be omitted if

$$\frac{\varepsilon \omega_J^2 \lambda_{ab}^2}{c^2} \frac{\sigma_\parallel}{\sigma_\perp} \left| 1 - \frac{\omega^2}{\omega_J^2} \right| \ll 1. \tag{14}$$

The electric and magnetic field in a layered superconductor can be obtained from the distribution of the gauge-invariant phase difference using equations,

$$E_z^{l+1,l} = \mathcal{H}_0 \frac{1}{\omega_J \sqrt{\varepsilon}} \frac{\partial \varphi^{l+1,l}}{\partial t}, \quad \mathcal{H}_0 = \frac{\Phi_0}{2\pi D \lambda_c}, \tag{15}$$

$$\frac{\partial H^{l+1,l}}{\partial x} = -\frac{\mathcal{H}_0}{\lambda_c} \left[ \frac{1}{\omega_J^2} \frac{\partial^2 \varphi^{l+1,l}}{\partial t^2} + \frac{\omega_r}{\omega_J^2} \frac{\partial \varphi^{l+1,l}}{\partial t} \right.$$
$$\left. + \sin\left(\varphi^{l+1,l}\right) \right], \tag{16}$$

$$E_{xl} = -\frac{\lambda_{ab}^2}{cD} \frac{\partial}{\partial t}(H^{l+1,l} - H^{l,l-1}), \tag{17}$$

where we introduce the London penetration depth of the dc magnetic field in the **c**-direction,

$$\lambda_c = \frac{c}{\sqrt{\varepsilon} \, \omega_J}. \tag{18}$$

The coupled sine-Gordon equations can be used to describe both the Josephson vortices in layered superconductors and the Josephson plasma waves. In the case of small-amplitude waves, Eq. (12) can be linearized, i.e., $\sin\left(\varphi^{l+1,l}\right)$ can be replaced by $\varphi^{l+1,l}$. Then, Eq. (12) has wave solutions,

$$\varphi^{l+1,l} \propto \exp\left[i(qx - \omega t + k(q,\omega)lD)\right]. \tag{19}$$

Substituting Eq. (19) into the linearized Eq. (12), we obtain the dispersion law for the Josephson plasma waves,

$$\sin^2\left(\frac{kD}{2}\right) = \frac{D^2}{4\lambda_{ab}^2} \left[ \frac{c^2 q^2}{\varepsilon(\omega^2 - \omega_J^2 + i\omega\omega_r)} - 1 \right]. \tag{20}$$

Equation (20) shows that the Josephson plasma waves can propagate in the layered superconductor if

$$\omega > \omega_J. \tag{21}$$

Thus, a gap exists in the frequency spectrum of the electromagnetic waves in layered superconductors. This means, that the small-amplitude incident wave is completely reflected from the surface of the superconductor if the wave frequency is lower than $\omega_J$. Note also that the dispersion equation (20), obtained under the assumption of charge neutrality, predicts the minimum possible value of the longitudinal wave vector,

$$q_{\min} = \left[\frac{\varepsilon(\omega^2 - \omega_J^2)}{c}\right]^{1/2}. \tag{22}$$

The wave propagation deep into the superconductor is accompanied by a decay controlled by the term $\omega_r$ in Eq. (20), i.e., by the value of the quasiparticle conductivity $\sigma_\perp$. At low temperatures, this conductivity is small for HTS . Thus, for many problems of interest, the decay of plasma waves can be neglected.

### B.  Breaking of charge neutrality

One can see from Eq. (20) that the value of $1/k(q,\omega)$ becomes comparable with the spacing $D$ in the layered structure of a superconductor if the frequency $\omega$ is close to the Josephson plasma frequency $\omega_J$. In this case, the effect of breaking the charge neutrality of the superconducting layers can play an important role in forming the spectrum of the Josephson plasma waves [34, 37, 49, 51, 85–87] and in transport properties [88, 89]. In particular, as we review below, a new branch in the wave spectrum appears due to this effect.

#### 1.  Longitudinal plane wave

First, following Ref. 49, we derive the equation for the phase difference $\varphi^{l+1,l}$ assuming that the wave is uniform along the layers, i.e., we consider a plane wave (19) with $q = 0$. It follows from the charge conservation law that

$$J_c \sin\left(\varphi^{l+1,l}\right) + \sigma_\perp E_z^{l+1,l}$$

$$= J_c \sin\left(\varphi^{l,l-1}\right) + \sigma_\perp E_z^{l,l-1} - s\frac{\partial \rho_l}{\partial t} \tag{23}$$

where $\rho_l$ is the charge density in the $l$th superconducting layer. From the Maxwell equation

$$\text{div}(\varepsilon \vec{E}) = 4\pi\rho, \tag{24}$$

we obtain the following relation:

$$E_z^{l+1,l} - E_z^{l,l-1} = \frac{4\pi s}{\varepsilon}\rho_l. \tag{25}$$

From Eqs. (23) and (25) we derive the conservation law for the case when waves propagate only along the **c**-axis,

$$J_c \sin\left(\varphi^{l+1,l}\right) + \sigma_\perp E_z^{l+1,l} + \frac{\varepsilon}{4\pi}\frac{\partial E_z^{l+1,l}}{\partial t}$$
$$= J_c \sin\left(\varphi^{l,l-1}\right) + \sigma_\perp E_z^{l,l-1} + \frac{\varepsilon}{4\pi}\frac{\partial E_z^{l,l-1}}{\partial t}. \tag{26}$$

In other words, the total **c**-axis current, including the displacement current, should be the same in each junction. In particular, in the absence of the external bias current, Eq. (26) gives

$$J_c \sin\left(\varphi^{l+1,l}\right) + \sigma_\perp E_z^{l+1,l} + \frac{\varepsilon}{4\pi}\frac{\partial E_z^{l+1,l}}{\partial t} = 0. \tag{27}$$

If we assume the usual local Josephson relation between the phase difference and voltage,

$$\frac{\partial \varphi^{l+1,l}}{\partial t} = \frac{2e}{\hbar}V^{l+1,l} = \frac{2\pi c}{\Phi_0}DE_z^{l+1,l}. \tag{28}$$

we obtain from Eqs. (27) and (26) a set of *independent* equations for $\varphi^{l+1,l}$ in each junction, i.e., no interference effect takes place among the junctions. However, as was argued in Refs. 90, 91, Eq. (26) becomes incorrect for systems where the charge neutrality breaking effect is important. Taking the time derivative of Eq. (4) and using Eq. (8), we obtain, instead of Eq. (28), the following relation:

$$\frac{\Phi_0}{2\pi c}\frac{\partial \varphi^{l+1,l}}{\partial t} = V^{l+1,l} + \left(A_{0(l+1)} + \frac{\Phi_0}{2\pi c}\frac{\partial \chi_{(l+1)}}{\partial t}\right)$$
$$- \left(A_{0l} + \frac{\Phi_0}{2\pi c}\frac{\partial \chi_l}{\partial t}\right). \tag{29}$$

We assume a rather natural relation between the charge density and the gauge-invariant scalar potential [91, 92]:

$$\rho_l = -\frac{1}{4\pi R_D}\left(A_{0l} + \frac{\Phi_0}{2\pi c}\frac{\partial \chi_l}{\partial t}\right), \tag{30}$$

where $R_D$ is the Debye length for a charge in a superconductor, which is usually much shorter than the London penetration depth. Substituting Eq. (30) into Eq. (29) and using Eq. (25), we obtain the relation between the gauge-invariant phase difference and the voltage,

$$\frac{\hbar}{2e}\frac{\partial \varphi^{l+1,l}}{\partial t} = \frac{\varepsilon R_D^2}{sD}\left[-V^{l,(l-1)} + \left(2 + \frac{sD}{\varepsilon R_D^2}\right)V^{(l+1),l}\right.$$
$$\left. - V^{(l+2),(l+1)}\right]. \tag{31}$$

Note that Eq. (31) is reduced to Eq. (28) in the limiting case $R_D^2 \ll sD/\varepsilon$. Otherwise, we cannot neglect the correction for the Josephson relation. The equation for the gauge-invariant phase difference can be derived from Eq. (27) and Eq. (31),

$$\frac{\partial^2 \varphi^{l+1,l}}{\partial t^2} = \omega_J^2 \left[\alpha \partial_l^2 \sin\left(\varphi^{l+1,l}\right) - \sin\left(\varphi^{l+1,l}\right)\right.$$
$$\left. -\frac{\omega_r}{\omega_J^2}\frac{\partial \varphi^{l+1,l}}{\partial t}\right] \tag{32}$$

with

$$\alpha = \frac{\varepsilon R_D^2}{sD}. \tag{33}$$

For small $\varphi^{l+1,l}$, this equation has the longitudinal-wave solution of the form Eq. (19) with $q = 0$ and the dispersion law,

$$\sin^2\left(\frac{kD}{2}\right) = \frac{1}{4\alpha\omega_J^2}(\omega^2 - \omega_J^2 + i\omega\omega_r). \tag{34}$$

This wave exists in the frequency interval,

$$\omega_J^2 < \omega^2 < \omega_J^2(1 + 4\alpha). \tag{35}$$

In contrast to the dispersion equation (20), the wave considered here can propagate at $q = 0$.

2. *Dispersion relation for Josephson plasma waves propagating in an arbitrary direction*

When the wave vector has an arbitrary inclination to the layers, the dispersion relation for the Josephson plasma waves can be derived in a manner similar to Eq. (34). When neglecting the relaxation term, the result can be written as [37]

$$\frac{\omega^2(q,k)}{\omega_J^2} = 1 + \frac{\lambda_c^2 q^2}{1 + (4\lambda_{ab}^2/D^2)\sin^2(kD/2)}$$
$$+ 4\alpha \sin^2(kD/2). \tag{36}$$

Obviously, Eq. (36) coincides with Eq. (20) when $\alpha = 0$ and $\omega_r = 0$.

Following Ref. 37, we consider the excitation of Josephson plasma waves by an external electromagnetic wave incident onto a layered superconductor at some angle $\theta$ with respect to the **ab**-plane. In such a situation, the frequency $\omega$ and the longitudinal component $q$ of the wave vector are related by

$$cq = \omega \sin\theta. \tag{37}$$

Substituting Eq. (37) into Eq. (36) and solving the obtained equation with respect to $\sin^2(kD/2)$, one obtains two branches of the dispersion law,

$$\sin^2(k^{\pm}D/2) = \frac{1}{8\alpha}\left[\Delta - \alpha\frac{D^2}{\lambda_{ab}^2}\right.$$
$$\left.\pm\sqrt{(\Delta+\alpha\frac{D^2}{\lambda_{ab}^2})^2 - 4\alpha\frac{D^2}{\lambda_{ab}^2}\frac{\sin^2\theta}{\varepsilon}(1+\Delta)}\right]$$
$$\approx \frac{1}{8\alpha}\left[\Delta \pm \sqrt{\Delta^2 - 4\alpha\frac{D^2}{\lambda_{ab}^2}\frac{\sin^2\theta}{\varepsilon}}\right], \tag{38}$$

$$\Delta = \frac{\omega^2}{\omega_J^2} - 1. \tag{39}$$

The branch "+" is characterized by a normal dispersion ($v_g = \partial\omega/\partial k > 0$). It exists at any incident angle $\theta$ in the frequency range,

$$\Delta_{\min} < \Delta < \Delta_{\max}^+, \tag{40}$$

$$\Delta_{\min} = 2\sqrt{\frac{\alpha}{\varepsilon}}\frac{D}{\lambda_{ab}}\sin\theta, \qquad \Delta_{\max}^+ = 4\alpha. \tag{41}$$

It is of interest that this branch disappears when $\alpha \to 0$, i.e., the effect of breaking charge neutrality is responsible for this branch.

In the limit of charge neutrality, $\alpha = 0$, the branch "−" coincides with the mode predicted by the dispersion law in Eq. (20). This branch [37] is characterized by an anomalous dispersion ($v_g < 0$). It exists for

$$\frac{\sin\theta}{\varepsilon - \sin^2\theta} > 2\sqrt{\frac{\alpha}{\varepsilon}}\frac{D}{\lambda_{ab}}, \tag{42}$$

in the following frequency range:

$$\Delta_{\min} < \Delta < \Delta_{\max}^- = \frac{\sin^2\theta}{\varepsilon - \sin^2\theta}. \tag{43}$$

This analysis [37] shows that layered superconductors represent an example of conducting media where incident light with a given frequency excites several eigen-modes with different wave vectors **k**. This poses the fundamental problem that the Maxwell boundary conditions, i.e., the continuity of the electric and magnetic field components parallel to the surface, are insufficient to calculate the relative amplitudes of these modes, and one should use the so-called additional boundary conditions. The additional boundary conditions for layered superconductors can be obtained considering the interlayer charge oscillations due to the tunnelling of Cooper pairs and quasiparticles, as was done in Ref. 37.

When the wave numbers differ significantly, at $k^+ \gg k^-$, the incident wave excites mainly the branch "−", and the usual Fresnel approach is valid. However, this approach becomes invalid near frequencies where the group velocity of the wave packets inside the crystal vanishes. Near this particular frequency, $\omega \approx \omega_J(1+2\alpha)$, the reflectivity depends on the atomic structure of the crystal. Reference 37 also noted that the spatial dispersion of the Josephson plasma waves provides a method to stop light pulses with $\omega \approx \omega_J$.

According to estimates from Ref. 37, $\alpha \sim 0.05$–$0.1$ for Bi-2212 or Tl-2212 crystals. However, the value of $\alpha$ can be reliably extracted from experiments. According to experimental results in Ref. 93, this constant is much smaller than the one predicted in Ref. 37. Thus, the capacitive coupling parameter $\alpha$ is negligible and does not significantly affect the distribution of the gauge-invariant phase difference and the electromagnetic field inside a superconductor. Therefore, below we do not take into account the effect of breaking charge neutrality.

### C. Surface Josephson plasma waves

Equation (20) shows that the value of the transverse wave number $k(q,\omega)$ of the Josephson plasma waves becomes imaginary when $\omega < \omega_J$. This corresponds to the damping of the wave amplitude inside the superconductor. This is an inherent feature of surface waves [94–98]. In general, surface waves play a very important role in many fundamental resonance phenomena, such as the Wood anomalies in the reflectivity [95–97, 99] and transmissivity [98, 100–106] of periodically-corrugated metal and semiconductor samples. Therefore, it is important to describe how to excite surface waves in layered superconductors and to study the resonances associated with these surface waves.

In this subsection, closely following Refs. 71, 107–109, we review the surface Josephson plasma waves (SJPWs) propagating along the interface separating the vacuum and a layered superconductor (see Fig. 2). In particular, below we obtain the dispersion relation for SJPWs, and show that SJPWs can be excited via the so-called "attenuated total reflection method" (Otto-configuration [95–97, 110, 111]) in a frequency range below $\omega_J$, i.e., by an *evanescent* wave in the vacuum gap between the superconductor and a dielectric prism. Due to the resonant excitation of the SJPW, the reflectivity of the incident wave depends sharply on its frequency and incident angle. This resonance effect can be useful for *filtering and detecting* THz and sub-THz radiation using layered superconductors. Reference 109 finds the optimal conditions for the best matching of the incident wave and the SJPWs, as well as for the total suppression of the specular reflection.

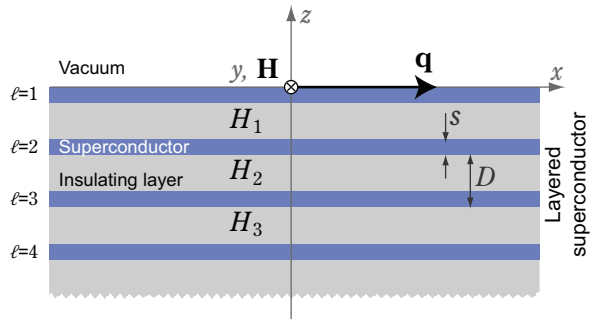

FIG. 2: (Color online) Geometry for studying surface waves (From Ref. 109).

1. *Dispersion relation for surface Josephson plasma waves*

Now we consider a plane-shaped interface (the $xy$-plane) separating the vacuum ($z > 0$ in Fig. 2) and a layered superconductor ($z \leq 0$). We study a linear surface transverse-magnetic (TM) monochromatic electromagnetic wave propagating along the $x$-axis with the electric, $\mathbf{E} = \{E_x, 0, E_z\}$, and magnetic, $\mathbf{H} = \{0, H, 0\}$, fields proportional to $\exp[i(qx - \omega t)]$ and *decaying* into both the vacuum and layered superconductor away from the interface $z = 0$. When $q > \omega/c$, the Maxwell equations yield an exponential decay of the wave amplitude into the vacuum,

$$H^{\text{vac}}, E_x^{\text{vac}}, E_z^{\text{vac}} \propto \exp(iqx - i\omega t - k_v z), \quad z > 0 \tag{44}$$

with the decay constant

$$k_v = \sqrt{q^2 - \frac{\omega^2}{c^2}} > 0.$$

Moreover, the Maxwell equations provide the ratio of amplitudes for the tangential electric and magnetic fields at the interface $z = +0$ (i.e., right above the sample surface):

$$\frac{E_x^{\text{vac}}}{H^{\text{vac}}} = \frac{ic}{\omega} k_v = \frac{ic}{\omega} \sqrt{q^2 - \frac{\omega^2}{c^2}}. \tag{45}$$

The electromagnetic field inside the layered superconductor, $z < 0$, is defined by the distribution of the gauge-invariant phase difference $\varphi_l(x,t)$ of the order parameter between the $l$th and $(l+1)$th layers.

The linearized version of the coupled sine-Gordon equations (12), together with Eqs. (15)–(17), have a solution of the form

$$\varphi_l,\ H_l^s,\ E_{x,l}^s,\ E_{z,l}^s \propto \exp(iqx - i\omega t - k_s l D) \tag{46}$$

inside a layered superconductor and give the relation between the decay constant $k_s$ ($\operatorname{Re}(k_s) > 0$), wave-number $q$, and dimensionless frequency

$$\Omega = \frac{\omega}{\omega_J},$$

$$\sinh^2\left(\frac{k_s D}{2}\right) = \frac{D^2}{4\lambda_{ab}^2}\left(1 + \frac{q^2 \lambda_c^2}{1 - \Omega^2 - ir\Omega}\right). \tag{47}$$

The dispersion relation, $q(\omega)$, for the surface Josephson plasma wave can be obtained by matching the in-plane fields $H$ and $E_x$ at the vacuum-superconductor interface. Thus, in order to find the spectrum of the surface JPW, we should derive the ratio $E_x^s/H^s$ at $z = 0$ and use the impedance matching,

$$\frac{E_x^{\text{vac}}}{H^{\text{vac}}} = \frac{E_x^s}{H^s}.$$

The difference between the magnetic field $H^s(z = 0)$ at the sample surface and its value $H_1^s$ between the first and second superconducting layers is defined by the $x$-component of the supercurrent density $J_x(l = 1)$. In the London approximation, the value of $J_x$ is proportional to the $x$-component of the vector-potential, $A_x(l = 1)$, and, therefore, to the electric field $E_{x,1}^s$. As a result, we obtain the relation,

$$\frac{H^s(z=0) - H_1^s}{D} \approx \frac{A_x(l=1)}{\lambda_{ab}^2} \approx \frac{-ic}{\lambda_{ab}^2 \omega} E_{x,1}^s. \tag{48}$$

Moreover, for $l = 1$, Eq. (46) implies that

$$H^s(z=0) - H^s_1 = H^s(z=0)[1 - \exp(-k_s D)]. \tag{49}$$

Using Eqs. (48) and (49), we obtain the ratio between the electric and magnetic fields at $z = -0$ (i.e., right below the sample surface),

$$\frac{E^s_x(z=0)}{H^s(z=0)} = i\frac{\Omega \lambda_{ab}^2 \omega_J}{cD}[1 - \exp(-k_s D)]$$

$$= 2ib\,\Omega Z\left(\sqrt{1 + \frac{1}{Z}} - 1\right), \tag{50}$$

with

$$b = \frac{\lambda_{ab}^2 \omega_J}{cD}, \quad Z = \Gamma^2\left[1 + \frac{\kappa^2}{\varepsilon(1 - \Omega^2 - ir\Omega)}\right],$$

$$\Gamma = \frac{D}{2\lambda_{ab}}, \quad \kappa = \frac{cq}{\omega_J}. \tag{51}$$

Matching the impedances in Eqs. (50) and (45), we obtain the *dispersion relation* for the surface Josephson plasma waves,

$$\sqrt{\kappa^2 - \Omega^2} = 2b\,\Omega^2 Z\left(\sqrt{1 + \frac{1}{Z}} - 1\right). \tag{52}$$

For $Bi_2Sr_2CaCu_2O_{8+\delta}$ superconductors, one can use the following values of the parameters: $b \approx 0.7$, $\omega_J/2\pi = 1$ THz, $D = 20$ Å, and $\lambda_{ab} = 2000$ Å.

For simplicity, below we consider surface waves in the continuum limit ($lD \to -z$), when the damping length, $k_s^{-1}$, in the superconductor is large compared with the interlayer spacing $D$:

$$k_s D \ll 1. \tag{53}$$

Under such a condition, the value of $Z$ in Eq. (52) is small. So, in the *continuum limit*, the dispersion relation takes the form,

$$\kappa^2 = \Omega^2 + 4b^2\Omega^4\Gamma^2\left(1 + \frac{\kappa^2}{\varepsilon(1 - \Omega^2 - ir\Omega)}\right). \tag{54}$$

2. *Excitation of surface Josephson plasma waves: resonant electromagnetic absorption*

Here we describe how to excite a surface Josephson plasma wave by a wave incident from a dielectric prism onto a superconductor separated from the prism by a thin vacuum gap (see

Fig. 3). In the absence of the superconductor, the incident wave completely reflects from the bottom of the prism, if the incident angle $\theta$ exceeds the limit angle $\theta_t$ for total internal reflection. However, *the evanescent wave penetrates under the prism* a distance about a wavelength. The wave vector of the evanescent mode is along the bottom surface of the prism and its value is higher than $\omega/c$. This feature is the same as for surface waves. So, it is natural to expect the spatial-and-temporal matching (coincidence of both, the frequencies and wave vectors) of evanescent modes and surface Josephson plasma waves for a certain incident angle. When the resonant excitation of SJPWs by the incident wave occurs, this results in a *strong suppression of the reflected wave*. This is the well known "attenuated-total-reflection method" for *generating surface waves*. Below we present a detailed description of this method for SJPWs propagating along the superconducting layers. The geometry is shown in Fig. 3.

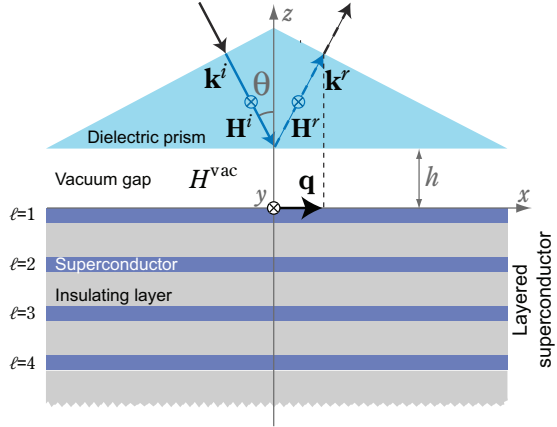

FIG. 3: (Color online) Geometry considered in Ref. 109 (Otto configuration): a dielectric prism is separated from a layered superconductor by a vacuum gap of thickness $h$. An electromagnetic wave with incident angle $\theta > \theta_t$ can excite SJPWs that satisfy the following resonant condition: $\omega \sin\theta/c = q$. Here $\mathbf{k}^i$ and $\mathbf{k}^r$ are the wave vectors of the incident and reflected waves associated with the magnetic field amplitudes $\mathbf{H}^i$ and $\mathbf{H}^r$. The resonant excitation of SJPWs by the incident wave produces a strong suppression of the reflected wave. This method for producing surface waves is known as the "attenuated-total-reflection" method.

We now consider an electromagnetic wave with the electric, $\mathbf{E}^d = \{E_x^d, 0, E_z^d\}$, and magnetic, $\mathbf{H}^d = \{0, H^d, 0\}$, fields incident from the dielectric prism. The prism has permittivity $\epsilon$ and is separated from the layered superconductor by a vacuum interlayer of thickness $h$.

The wave frequency $\omega$ is assumed to be below the Josephson plasma frequency $\omega_J$.

The magnetic field $H^d$ in the dielectric prism can be represented as a sum of incident and reflected waves with amplitudes $H^i$ and $H^r$, respectively,

$$H^d = H^i \exp[iqx - ik_d(z - h)]$$

$$+ H^r \exp(iqx + ik_d(z - h)), \quad z > h. \tag{55}$$

Here and below in this subsection we omit the time-dependent multiplier, $\exp(-i\omega t)$. The plane $z = 0$ corresponds to the vacuum-superconductor boundary. The tangential $q$ and normal $k_d$ components of the wave vector, for waves in the prism, are defined by

$$q = k\sqrt{\epsilon}\sin\theta, \quad k_d = \sqrt{k^2\epsilon - q^2} = k\sqrt{\epsilon}\cos\theta, \tag{56}$$

with $k = \omega/c$. The condition for total internal reflection of the wave in the dielectric prism is assumed to be fulfilled, i.e.,

$$\sin^2\theta > \frac{1}{\epsilon}. \tag{57}$$

The magnetic field

$$H^{\text{vac}} = H^i \left[ h^+ \exp(iqx + k_v z) \right.$$

$$\left. + h^- \exp(iqx - k_v z) \right], \tag{58}$$

of the evanescent mode in the vacuum gap is generated by the wave from the dielectric prism. Here $h^+$ ($h^-$) are the dimensionless amplitudes of the evanescent waves that exponentially increase/decrease with the spatial increment rate

$$k_v = \sqrt{q^2 - k^2} = k\sqrt{\epsilon \sin^2\theta - 1}. \tag{59}$$

Using Maxwell's equations, one can express the $x$-components, $E_x^d$ and $E_x^{\text{vac}}$, of the electric field in the dielectric prism and in the vacuum gap via the magnetic field amplitudes,

$$E_x^d = \frac{k_d}{k\epsilon} H^i \left[ h^r \exp(iqx + ik_d(z - h)) \right.$$

$$\left. - \exp(iqx - ik_d(z - h)) \right], \quad h^r = H^r/H^i,$$

$$E_x^{\text{vac}} = -i\frac{k_v}{k} H^i \left[ h^+ \exp(iqx + k_v z) \right.$$

$$\left. - h^- \exp(iqx - k_v z) \right]. \tag{60}$$

In the layered superconductor, the electromagnetic field is described by Eqs. (46), (47).

Using the conditions of continuity of the tangential components of the electric and magnetic fields at the dielectric-vacuum and vacuum-layered superconductor interfaces, one obtains a set of four linear algebraic equations for four unknown wave amplitudes, $h^r$, $h^+$, $h^-$, and $H^s$. Solving this set gives the *reflection coefficient*

$$R \equiv h^r = \frac{R_F\,(k_v/k - a) + (k_v/k + a)\,C(h,\theta)}{(k_v/k - a) + (k_v/k + a)\,R_F\,C(h,\theta)}, \tag{61}$$

for the wave reflected from the bottom of the prism. Here

$$R_F = \frac{k_d - ik_v\epsilon}{k_d + ik_v\epsilon} \equiv \exp(-i\psi) \tag{62}$$

is the *Fresnel reflection coefficient*,

$$C(h,\theta) = \exp(-2k_v h) \tag{63}$$

is the parameter that provides the *coupling* between waves in the dielectric prism and the layered superconductor. Also,

$$a \equiv a(\Omega,\theta) = 2b\,\Omega Z\left(\sqrt{1+\frac{1}{Z}} - 1\right) \tag{64}$$

is the *effective surface impedance* of the superconductor (see Eq. (50)). Below we assume the coupling parameter $C$ to be small. However, even when $C \ll 1$, the coupling of the waves in the dielectric prism and superconductor plays a very important role in the excitation of SJPWs and anomalies in the reflection properties (Wood's anomalies). First, the dispersion relation of the surface Josephson plasma waves is modified, involving the radiation leakage through the dielectric prism. The new spectrum of the SJPWs is defined by the denominator in Eq. (61). Actually, the region where the coupling $C \ll 1$ (when the radiation leakage of the excited SJPW through the prism does not dominate) corresponds to the strongest excitation of the surface waves by the incident waves. Furthermore, the coupling results in breaking the total internal reflection of the electromagnetic waves from the dielectric-vacuum interface. Due to this coupling, the reflection coefficient $R$ in Eq. (61) differs from the Fresnel one $R_F$, its modulus becoming less than unity. Moreover, as we show below, the reflection of waves with any frequency $\omega < \omega_J$ can be completely suppressed, for the specific incident angle $\theta$ and depth $h$ of the vacuum gap. This provides a way to control, detect, and filter the THz radiation.

To study the phenomenon of attenuated total reflection, we consider the case which is most suitable for observations, when the following inequalities are satisfied:

$$\frac{b^2\Gamma^2\epsilon \sin^2\theta}{\varepsilon} \ll \left(1 - \Omega^2\right) \ll \frac{\epsilon \sin^2\theta}{\varepsilon}. \tag{65}$$

Here, the left inequality corresponds to the continuum limit for the field distribution in the $z$ direction, whereas the right one allows neglecting unity in the square brackets in Eq. (51). Also, we assume the dissipation parameter $r$ in Eq. (51) to be small compared with $(1-\Omega^2)$,

$$r \ll (1 - \Omega^2). \tag{66}$$

For this frequency region, the complex parameter $a(\Omega, \theta)$, Eq. (64), can be presented as

$$a(\Omega, \theta) \equiv a' + ia''$$
$$= 2b\Gamma \sqrt{\frac{\epsilon \sin\theta}{\varepsilon(1-\Omega^2)}} \left(1 + \frac{ir}{2(1-\Omega^2)}\right). \tag{67}$$

When the inequalities in Eqs. (65), (66) are satisfied, the expression for the reflectivity coefficient $R$ can be significantly simplified. First, the phase $\psi$ of the Fresnel reflectivity coefficient $R_F$, Eq. (62), is small. In the vicinity of the SJPW spectrum, at $k_v/k \simeq a'$,

$$\psi \simeq \frac{4b\Gamma\epsilon}{\sqrt{\varepsilon(\epsilon-1)(1-\Omega^2)}} \ll 1. \tag{68}$$

Second, the main changes of the reflectivity coefficient $R$ in Eq. (61) occur in the region of incident angles $\theta$ close to the limit-angle $\theta_t$ for total-internal-reflection,

$$\vartheta \equiv (\theta - \theta_t) \ll 1, \qquad \sin^2\theta_t = \frac{1}{\epsilon}. \tag{69}$$

Third, the parameter $a'$ in Eq. (67) is almost independent of the angle $\vartheta$ in the essential region where $\vartheta$ changes, whereas it depends very strongly on the frequency detuning $(1-\Omega)$. Using the properties mentioned above, the reflection coefficient $R$ can be rewritten in the form,

$$R = \frac{X(\Omega, \vartheta) - iB(\Omega)(C_{\text{opt}}(\Omega) - C(h, \vartheta))}{X(\Omega, \vartheta) - iB(\Omega)(C_{\text{opt}}(\Omega) + C(h, \vartheta))}, \tag{70}$$

with

$$X(\Omega, \vartheta) \simeq \sqrt{2}(\epsilon - 1)^{1/4}\sqrt{\vartheta} - \frac{2b\Gamma}{\sqrt{\varepsilon(1-\Omega^2)}}, \tag{71}$$

$$B(\Omega) \simeq \frac{16b^2\Gamma^2\epsilon}{\varepsilon\sqrt{\epsilon-1}(1-\Omega^2)}, \tag{72}$$

$$C_{\text{opt}}(\Omega) \simeq \frac{r\sqrt{\epsilon-1}\sqrt{\varepsilon}}{16b\Gamma\epsilon\sqrt{1-\Omega^2}}. \tag{73}$$

Equations (70), (71) show that the modulus of the reflectivity $R(\theta)$ has a *sharp resonance minimum* at

$$\vartheta = \vartheta_{\text{res}} \simeq \frac{2b^2\Gamma^2}{\varepsilon\sqrt{\epsilon-1}(1-\Omega^2)}. \tag{74}$$

The minimum value of $R$ is

$$|R|_{\min} \simeq \frac{|C_{\text{opt}}(\Omega) - C(h,\vartheta_{\text{res}})|}{C_{\text{opt}}(\Omega) + C(h,\vartheta_{\text{res}})}. \tag{75}$$

It is clearly seen that this value depends strongly on the frequency detuning $(1-\Omega)$, dissipation parameter $r$, and the coupling between the waves in the dielectric prism and the layered superconductor, i.e., on the thickness $h$ of the vacuum gap. This offers several important applications of the predicted anomaly of the reflectivity in the THz range. For instance, if the coupling parameter $C(h,\vartheta_{\text{res}})$ is equal to the optimal value $C_{\text{opt}}$, i.e., the thickness $h$ takes on the optimal value,

$$h_{\text{opt}} = \frac{c}{\omega}\frac{\sqrt{\varepsilon(1-\Omega^2)}}{4b\Gamma}\ln\left(\frac{16b\Gamma\epsilon\sqrt{1-\Omega^2}}{r\sqrt{\varepsilon(\epsilon-1)}}\right), \tag{76}$$

the reflection coefficient $R$ at $\vartheta = \vartheta_{\text{res}}$ vanishes. This means that a complete suppression of the reflectivity can be achieved by an appropriate choice of the parameters, due to the resonant excitation of the surface Josephson plasma waves.

We emphasize that Eqs. (74), (76) describe the conditions for the *best matching* of the incident wave and SJPWs. Under such conditions, the amplitude $H^s_{\max}$ of the excited surface wave is much higher than the amplitude $H^i$ of the incident wave:

$$\frac{|H^s_{\max}|}{H^i} \sim \left(\frac{1-\Omega^2}{r}\right)^{1/2}\left(\frac{(1-\Omega^2)\varepsilon}{b^2\Gamma^2\epsilon}\right)^{1/4} \gg 1. \tag{77}$$

Thus, we can achieve a high concentration of energy in the THz SJPW. This proposed experimental setup could provide an unusual THz *resonator* or *cavity*.

For these optimal conditions, the total energy coming to the layered superconductor from the dielectric prism is transformed into Joule heat due to the quasiparticle resistance. Thus, if the conditions for the total suppression of the reflectivity are satisfied, the energy flux (i.e., the $z$-component of the Pointing vector of the incident wave) is completely absorbed.

The dependence of the *absorbtivity coefficient* $A$ on the wave frequency and the incident angle is described by a resonance curve,

$$A(\Omega, \vartheta) = 1 - |R(\Omega, \vartheta)|^2$$
$$\simeq \frac{4B^2(\Omega)C(h,\vartheta)C_{\text{opt}}(\Omega)}{X^2(\Omega,\vartheta) + B^2(\Omega)(C_{\text{opt}}(\Omega) + C(h,\vartheta))^2}. \tag{78}$$

The half-width $\delta\vartheta$ of the resonance line is much less than $\vartheta_{\text{res}}$,

$$\frac{\delta\vartheta}{\vartheta_{\text{res}}} \simeq \frac{16b\Gamma\epsilon(C_{\text{opt}}(\Omega) + C(h,\vartheta_{\text{res}}))}{\sqrt{\varepsilon(\epsilon-1)(1-\Omega^2)}} \ll 1. \tag{79}$$

If the total suppression of the reflectivity occurs, Eq. (79) can be simplified,

$$\frac{\delta\vartheta}{\vartheta_{\text{res}}} \simeq \frac{4b^2\Gamma^2 r}{\varepsilon\sqrt{\epsilon-1}} \ll 1. \tag{80}$$

Inequalities (65) are not necessary for the observation of the total suppression of the reflectivity and the resonant increase of the electromagnetic absorption. Departing from the strong inequalities (65), towards the region of parameters where $B \sim 1$, we perform numerical calculations. Figure 4 demonstrates the resonant suppression of the reflectivity for the parameter $B(\Omega) \approx 1.9$. Nevertheless, the asymptotic formulae Eqs. (70) – (73) describe rather well the resonant behavior of the reflectivity $R$.

Figure 5 shows the sharp decrease of the reflectivity in the $(\theta, (1-\Omega))$ plane, due to the resonant excitation of the surface Josephson plasma waves. Obviously, the suppression of the reflectivity can be observed by changing the frequency at a given incident angle, as is demonstrated in Fig. 6.

We also numerically calculate the total magnetic field distribution, Fig. 7. The interference pattern is seen in the non-resonant case, when the amplitudes of the incident and reflected waves practically coincide. For the resonant conditions, the reflected wave is suppressed and there is no interference of waves in the far-field zone (prism region). Otherwise, the near-field "torch" structure of the SJPW is clearly seen in the vacuum gap.

Note that the suppression of the reflectivity $|R|^2$ is accompanied by the resonant increase of the electromagnetic absorption in the layered superconductor. This process can result in a transition of the superconductor into the resistive or even into the normal state. Thus, a new kind of resonance phenomena can be observed due to the excitation of the SJPW. Moreover, this strongly-selective interaction of SJPWs, with the incident wave having a certain frequency and direction of propagation, can be used for designing THz detectors and filters.

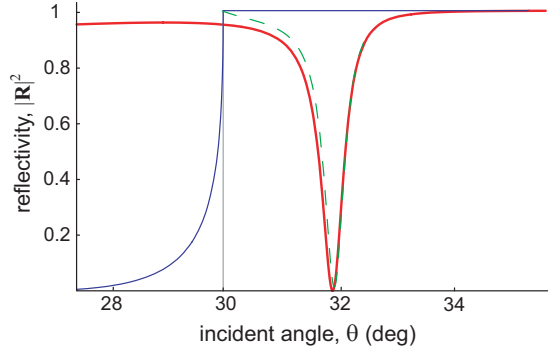

FIG. 4: (Color online) The dependence of the reflectivity coefficient $|R|^2$ on the incident angle $\theta$, calculated in Ref. 109 for the parameters $b = 0.7$, $\Gamma = 0.005$, $r = 10^{-6}$, $1 - \Omega^2 = 1.2 \cdot 10^{-5}$, $\varepsilon = 16$, and $\epsilon = 4$. The thickness of the vacuum gap is one wavelength, $hk = 2\pi$. The solid red curve presents the results of numerical calculations using Eqs. (61)–(63). The dashed green curve (that almost overlaps the red curve) describes the analytically obtained asymptotic dependence Eqs. (70)–(73). The vertical line at $\vartheta = 30^o$ corresponds to the limiting angle of the total internal reflection. The blue thin solid curve presents the Fresnel reflectivity coefficient.

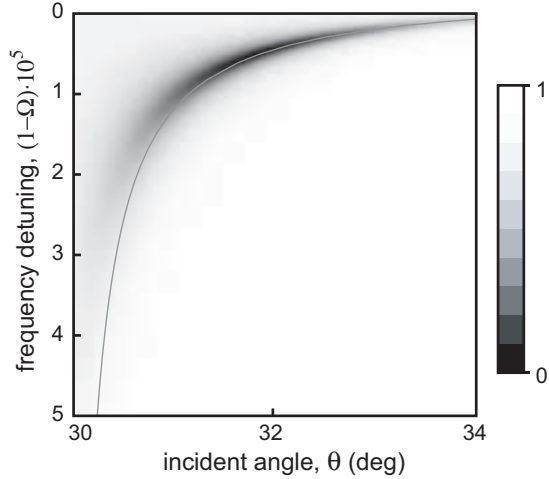

FIG. 5: The reflectivity coefficient in the plane $(\theta, (1 - \Omega))$ shown in gray levels for the same values of the parameters as in Fig. 4. The dispersion relation of the dielectric–vacuum–layered superconductor is presented by the solid curve (From Ref. 109).

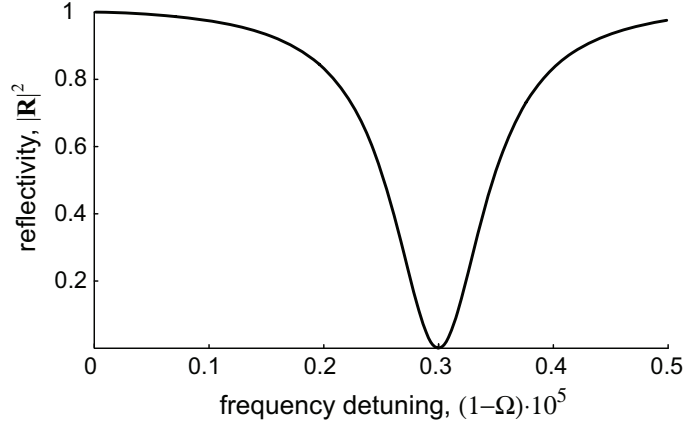

FIG. 6: The frequency dependence of the reflectivity coefficient $|R|^2$ obtained in Ref. 109 for $\theta = 31.867^o$ (From Ref. 109). Other parameters are the same as in Fig. 4.

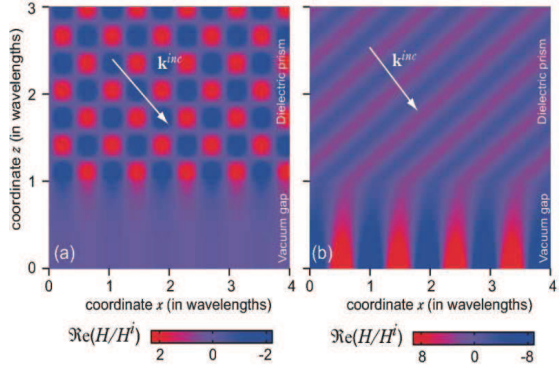

FIG. 7: (Color online) The magnetic field distribution for the non-resonant case, $\theta \neq \theta_{\text{res}}$, (a), and for the resonant condition, $\theta = \theta_{\text{res}} = 31.867^o$, (b) (From Ref. 109). Other parameters are the same as in Fig. 4.

We would like to note that the nonlinear regime of the electromagnetic wave propagation can be easily achieved during the resonant excitation of SJPW. Indeed, under the resonance conditions, the electromagnetic field in the superconductor is significantly increased with respect to the amplitude of the incident wave. Therefore, the value of the gauge-invariant phase of the order parameter increases also. A simple evaluation made by means of Eqs. (16)

and (77) gives, for $\varphi$ in the resonance region,

$$\varphi_{\max} \sim \frac{H^i}{\mathcal{H}_0}\left[(1-\Omega^2)\varepsilon r^2 b^2 \Gamma^2 \epsilon\right]^{1/4} \sim \frac{H^i}{\mathcal{H}_0} \cdot 10^5, \tag{81}$$

at $b = 0.7$, $\Gamma = 0.005$, $r = 10^{-6}$, $1 - \Omega^2 = 1.2 \cdot 10^{-5}$, $\varepsilon = 16$, and $\epsilon = 4$. Under such conditions, the nonlinear regime can be observed when $H^i \sim 10^{-3}$ Oe.

### D. Josephson plasma resonance

The study of Josephson plasma waves was mainly stimulated by the discovery of Josephson plasma resonances in layered HTS, low-$T_c$ (LTS), and artificial structures [11, 29–33, 112]. It was found that the reflectivity and absorption of electromagnetic waves in these systems exhibit a resonance-type behavior at some frequency (or at two-three frequencies) in the THz range. The resonance was observed only for the case when the electric field of the electromagnetic wave had a component parallel to the **c**-axis of the sample. Typical manifestations of such behavior are shown in Fig. 8 for HTS Bi$_2$Sr$_2$CaCu$_2$O$_8$ 11 and in Fig. 9 for organic $\kappa$-(ET)$_2$Cu(NCS)$_2$ single crystal [39]. As seen from the measurements shown in Figs. 8 and 9, the resonant frequency strongly depends on the temperature and the magnetic field. In particular, the effect disappears completely when $T > T_c$.

The observed (e.g., in Refs. 11, 38, 39) resonance behavior can be explained within the framework of the theory of the Josephson plasma waves presented in the previous sections. As an illustration, we consider here the simplest example. We consider an electromagnetic wave with the magnetic field,

$$\vec{H}_i = H_i \mathbf{e}_y \exp\left[i(q\sin\theta x + q\cos\theta z - \omega t)\right], \tag{82}$$

incident at an angle $\theta$ from the vacuum onto the surface $xy$ of a semi-infinite sample of a layered superconductor. Here $\mathbf{e}_y$ is the unit vector along the $y$-axis, and the usual vacuum dispersion relation, $\omega = cq$, is assumed.

For $\omega > \omega_J$, the wave (82) induces a transmitted electromagnetic wave that propagates in the sample,

$$\vec{H}_t = H_t \mathbf{e}_y \exp\{-i\omega t + iq\sin\theta x + ik(q\sin\theta, \omega)z\},$$

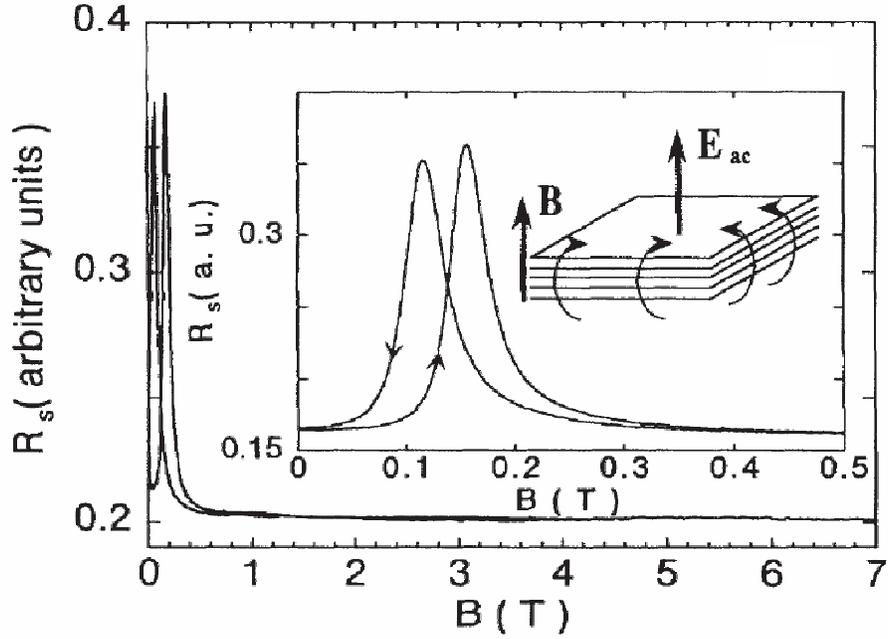

FIG. 8: Josephson plasma resonances in $Bi_2Sr_2CaCu_2O_8$ (From Ref. 11). The surface resistance, $R_s$, was measured at $f = 45$ GHz as a function of the magnetic field $B$ parallel to the **c**-axis at different temperatures; $R_s$ was measured for the wave with the electric field parallel to the **c**-axis.

with the dispersion law (20), and a reflected wave

$$\vec{H}_r = H_r \mathbf{e}_y \exp\left[i(q \sin\theta x - q \cos\theta z - \omega t)\right].$$

In the vacuum, the amplitudes of the $x$-component of the electric field, $E_x$, and magnetic field, $H$, at the sample surface ($z = 0$) are related to each other by

$$E_x = (H_r - H_i) \cos\theta. \tag{83}$$

In the superconductor, the relation between these amplitudes can be written as

$$E_x^s = -H_t \frac{\Omega \lambda_{ab}^2}{\lambda_c \varepsilon^{1/2}} k(q \sin\theta, \omega). \tag{84}$$

The continuity of the magnetic field and the tangential component $E_x$ of the electric field at the sample surface, combined with Eq. (20) taken in continuum limit, gives the expression for the reflection coefficient,

$$R = \frac{|H_r|^2}{|H_i|^2} = \left|\frac{1 - \Psi(\theta, \Omega)}{1 + \Psi(\theta, \Omega)}\right|^2. \tag{85}$$

with

$$\Psi(\theta, \Omega) = \frac{\lambda_{ab}}{\lambda_c} \frac{\sin\theta}{\varepsilon \cos\theta} \frac{\Omega^2}{\sqrt{\Omega^2 + ir\Omega - 1}}. \tag{86}$$

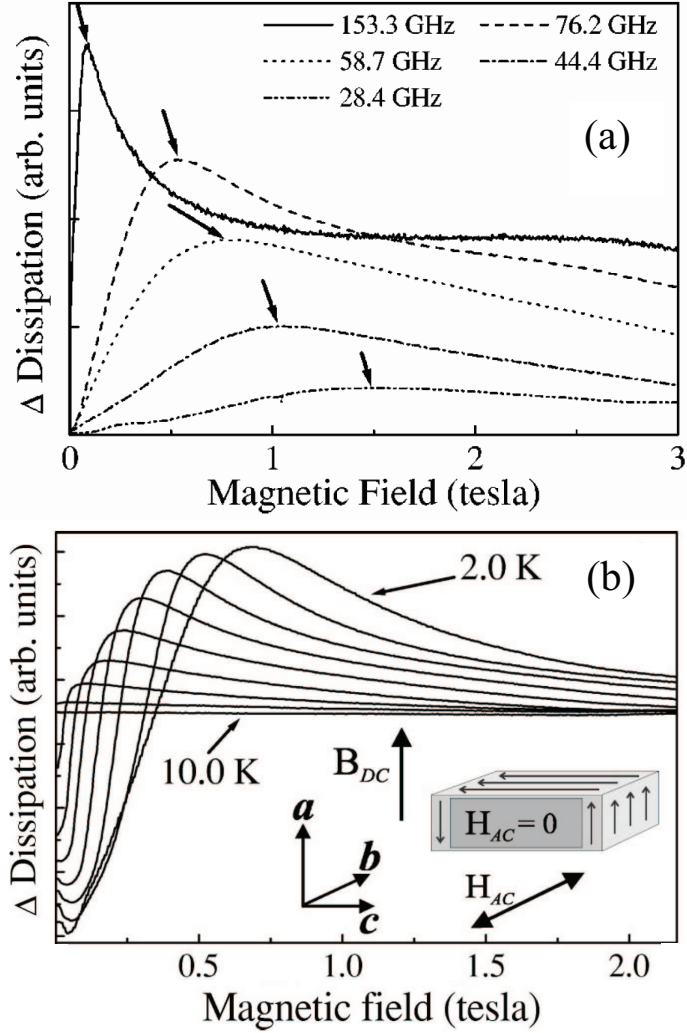

FIG. 9: Josephson plasma resonances in $\kappa$-$(ET)_2Cu(NCS)_2$ (From Ref. 39). (a) Dissipation versus magnetic field for different frequencies, at $T = 2$ K. When increasing the frequency, the resonance is observed at lower magnetic fields. (b) Dissipation versus magnetic field at $f = 76$ GHz, for different values of the currents flowing both parallel and perpendicular to the superconducting layers. The temperature varies from 2 K to 10 K. A resonant structure is clearly seen for this polarization of the microwave radiation.

Because of the large anisotropy, $\lambda_c/\lambda_{ab} \gg 1$, the reflection coefficient is close to one for frequencies far from the resonance. Also, the reflectivity can be depressed significantly near the plasma resonance, when $|\Omega - 1|/\Omega \ll 1$. Moreover, it can be suppressed completely slightly below the plasma frequency, when neglecting the dissipation term $r$ in Eq. (86).

As with other resonance methods, the Josephson plasma resonance is a sensitive and convenient tool for the study of the material properties [11–13, 40–46]. Evidently, the JPR gives *direct information about the Josephson coupling* of the superconducting layers. The resonance frequency $\omega_J$ depends on the transverse component of the critical current density, Eq. (13). This value is a function of the temperature and magnetic induction in the sample [68]. The details of the magnetic induction distribution in the superconductor, in turn, depend on the state of the flux line structure. Thus, JPRs can be used for the characterization of vortex matter state in layered superconductors.

### E. Josephson plasma waves in the presence of an external dc magnetic field. THz photonic crystal

When an external dc magnetic field $H_{ab}$ is applied parallel to the **ab** plane of a layered superconductor, the Josephson vortices (JVs) penetrate the sample and form a triangular lattice. In contrast to Abrikosov or pancake vortices, the interaction between JVs and crystal defects is weak, and the JV lattice is near perfect at low enough temperatures. Note also that the JV lattice can be easily pinned by pancake vortices generated by a low out-of-plane magnetic field (see, e.g., Refs. 113, 114), which allows a way for tuning JV arrays. Here we focus on the scattering and filtering of the Josephson plasma waves by a lattice of vortices, which is fixed or slowly moves inside a layered superconductor [115]. The JVs are objects of electromagnetic nature and should interact strongly with electromagnetic waves. Thus, we can expect strong magneto-optic effects in layered superconductors. In particular, the so-called photonic crystal state [116–119] can appear as a result of the periodicity of the JV lattice. A general comparison between the usual optical photonic crystals and JV photonic crystals is presented in Table 1. Note that photonic crystals formed due to the interaction of electromagnetic waves with vortices were also studied for a single Josephson junction [120] and for type-II superconductors [121].

Following Refs. 115, 122, we consider a layered superconductor placed in a dc magnetic field $H_{ab}$ parallel to the $y$-axis. Thus, the JVs parallel to the $y$-axis penetrate the superconductor. They form a triangular lattice with the distance $d_x$ between vortices within a layer (i.e., along the $x$-direction) and $d_z$ in the $z$-direction (see Fig. 10). Due to the high anisotropy of the superconducting medium, the distance $d_x$ is much larger than $d_z$, and $d_x/d_z = \gamma \gg 1$. Here $\gamma$ is the anisotropy coefficient of the sample. The usual value of $\gamma$ for Bi2212 single crystals is about 300–600. As a result, the JV lattice consists of dense vortex rows along the $z$-axis.

TABLE I: Comparison between standard photonic crystals and tunable THz photonic crystals using Josephson vortices [122].

| | Standard photonic crystals | Josephson-vortex (JV) photonic crystals |
|---|---|---|
| **Materials** | Various (e.g., semiconductors, polymers, insulators) | Layered superconductors (SCs) |
| **Frequency range** | Typically optical; Not in THz | Sub-THz and THz |
| **Scatterers** | Typically holes in materials | Josephson vortices in SCs |
| **Scatterers made by** | An often very complicated and cumbersome fabrication | Applying $H_{ab}$ |
| **Near-perfect periodicity** | Difficult to realize | Automatically provided |
| **Easily tunable?** | *No* | *Tunable* via applied magnetic field or current |
| **Moveable scatterers?** | No | Yes, producing a Doppler effect |
| **Gap size** | Can be large | Typically small |
| **Operating temperature** | Typically room temperature | $T < 90$ K |
| **Intrinsic nonlinearity** | Usually not | Yes, due to nonlinear current-phase relation |
| **Higher harmonic generation** | Usually not | Yes, due to nonlinearity |
| **Wave localization** | Requires introducing defects | Can be produced by nonlinearities |
| **Magneto-optical effect?** | No | Yes |

Consider JPWs propagating along the superconducting layers with the magnetic field

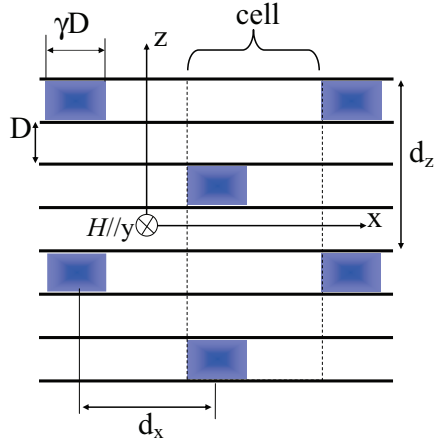

FIG. 10: (Color online) Schematics of JV lattice with notations used in the text (From Ref. 115).

along the direction of JVs,

$$\mathbf{H}(x, z, t) = \mathbf{e}_y H(x) \exp(ikz - i\omega t). \tag{87}$$

We assume that the amplitude $H$ of the wave is small compared to the external in-plane field $H_{ab}$ (responsible for the generation of the JV lattice). Therefore, the solution of Eq. (12) can be obtained perturbatively as

$$\varphi^{n+1,n} = \varphi_0^{n+1,n} + \psi_n(x, z, t), \tag{88}$$

where $\varphi_0^{n+1,n}$ corresponds to the steady JV lattice and $|\psi_n(x, z, t)| \ll |\varphi_0^{n+1,n}|$. For moderate magnetic fields, the steady-state solution can be approximated as a sum,

$$\varphi_0^{n+1,n} = \sum_m \varphi_0(x - x_{mn}),$$

of solitons [69],

$$\varphi_0 = \pi + 2\tan^{-1}(x/l_0),$$

where $2l_0 = \gamma D$. Here $x_{nm}$ is the position of the $m$th JV in the $n$th layer. The sum of the unperturbed solution $\varphi_0^{n+1,n}$ and the perturbation $\psi_n(x, z, t)$ can be viewed either as an electromagnetic wave propagating on the background of the fixed Josephson vortex lattice or as a sum of small oscillations of this lattice.

Substituting

$$\psi_n(x, z, t) = \psi(x) \exp(ikz - i\omega t) \tag{89}$$

into Eq. (12), averaging over $z$ for $|k| < \pi/D$, and neglecting the dissipation, we derive an equation for the wave amplitude $\psi(x)$ in the linear approximation,

$$\psi''(\eta) - \kappa_0^2(k) \left[\tilde{\omega}_J^2(\eta) - \Omega^2\right] \psi(\eta) = 0, \tag{90}$$

where the following dimensionless variables are introduced: $\eta = x/\gamma D$, $\Omega = \omega/\omega_J$,

$$h_{ab} = \frac{\gamma D^2 H_{ab}}{2\Phi_0}, \quad \kappa_0^2(k) = \left(\frac{D}{\lambda_{ab}}\right)^2 \left(1 + k^2 \lambda_{ab}^2\right). \tag{91}$$

When deriving Eqs. (90, 91), we use the relation $\lambda_c/\lambda_{ab} = \gamma$. The function $\tilde{\omega}_J^2(\eta)$ is defined as

$$\tilde{\omega}_J^2(\eta) = \left\langle \sum_m \cos\left(\varphi_0^{n+1,n}\right) \right\rangle_n, \tag{92}$$

where $\langle \ldots \rangle_n$ denotes averaging over the layers. $\tilde{\omega}_J^2(\eta)$ has a period $d_x$ along the $x$ direction. The modulation of the Josephson plasma frequency $\tilde{\omega}_J(\eta)$ results from the suppression of the Josephson current near the JV cores.

Equation (90) is an ordinary linear differential equation that has a form of the Schrödinger equation with a periodic "potential" $\tilde{\omega}_J^2(\eta)$. It can be solved numerically or approximately by the WKB method. For qualitative analysis and estimations, one can approximate the dependence $\tilde{\omega}_J^2(\eta)$ by an appropriate stepwise function:

$$\tilde{\omega}_J^2(\eta) = 1 - \frac{3\sqrt{h_{ab}}}{2} \sum_m F\left(\eta - \frac{m}{\sqrt{h_{ab}}}\right), \tag{93}$$

where $F(\eta) = 1$ if $|\eta| < 1$ and $F = 0$ if $|\eta| > 1$. When deriving Eq. (93), we use the relation $2\Phi_0/(d_x d_z) = H_{ab}$ and assume, as usual, that the core of each JV has a size $\gamma D$ along the $x$-direction and $D$ (i.e., one layer) along the $z$-direction. Outside the cores, $\cos(\varphi_0^{n+1,n}) = 1$, while inside the cores (shadowed regions in Fig. 10) $\cos(\varphi_0^{n+1,n}) = -1/2$. A detailed derivation of Eq. (93) was done in Ref. 122 and will not be repeated here.

The second order differential equation (90) requires the continuity of the functions $\psi(\eta)$ and $\psi'(\eta)$ in the sample for the continuity of electromagnetic fields.

1. *Band-gap structure*

Forbidden zones in the $\Omega(q)$ dependence, or so-called "photonic crystal" [116], can occur when the electromagnetic wave propagates through a periodically modulated structure, e.g., through the JV lattice. The dimensionless spatial period $\Delta \eta$ of the structure considered here is $1/\sqrt{h_{ab}}$. Following the usual band-theory approach, we obtain the solution of Eq. (90) in the form of the Bloch wave,

$$\psi(\eta) = u(\eta, q) \exp(iq\eta), \tag{94}$$

where $u(\eta, q)$ is a periodic function of $\eta$ with the period $1/\sqrt{h_{ab}}$, and the dimensionless wave vector $q$ is within the first Brillouin zone, $-\pi\sqrt{h_{ab}} < q < \pi\sqrt{h_{ab}}$. The solution of the linear equation (90) within $j$th elementary cell, $\eta_j < \eta < \eta_j + 1/\sqrt{h_{ab}}$ (see Fig. 10), either within the JV core, $\eta_j < \eta < \eta_j + 1$, or outside the core, $\eta_j + 1 < \eta < \eta_j + 1/\sqrt{h_{ab}}$, is a sum of exponential terms multiplied by constants $C_j$. Using the continuity of $\psi$ and $\psi'$ and the periodicity of the Bloch functions $u(\eta, q)$, we obtain a set of homogeneous linear equations for $C_j$. The non-trivial solution of these equations exists only if the determinant of the set of these equations is zero. From this, Ref. 122 obtains the dispersion equation for $\Omega(q)$ in the form,

$$\cos(\kappa_1 b)\cos(\kappa_2) - \frac{\kappa_1^2 + \kappa_2^2}{2\kappa_1\kappa_2}\sin(\kappa_1 b)\sin(\kappa_2)$$
$$= \cos[q(b+1)], \tag{95}$$

where $b = 1/\sqrt{h_{ab}} - 1$ and

$$\kappa_1 = \kappa_0 \left(\Omega^2 - 1\right)^{1/2}, \quad \kappa_2 = \kappa_0 \left(\Omega^2 + \frac{3}{2}\sqrt{h_{ab}} - 1\right)^{1/2}. \tag{96}$$

This spectrum is shown in Fig. 11 for two different values of the transverse wave vector $k$. Two particular features of the spectrum $\Omega(q)$ should be emphasized. First, in the presence of the in-plane magnetic field, the propagation of the Josephson plasma waves is possible at frequencies lower than $\omega_J$, due to the suppression of the Josephson current in the cores of the JVs. Second, the gap in the spectrum, or forbidden frequency band, arises for sufficiently high values of the transverse wave vector $k$ and the field $h_{ab}$. The width $\Delta\Omega$ of the forbidden band is gradually suppressed when decreasing $H_{ab}$ or $k$.

### 2. Tunable transparency

The THz photonic crystal discussed above indicates that the JV lattice can significantly affect the transparency of the medium. Here we calculate the transmission and reflection coefficients for two cases:

- JPW is emitted inside a sample, e.g., by the moving JVs (see Fig. 12a);

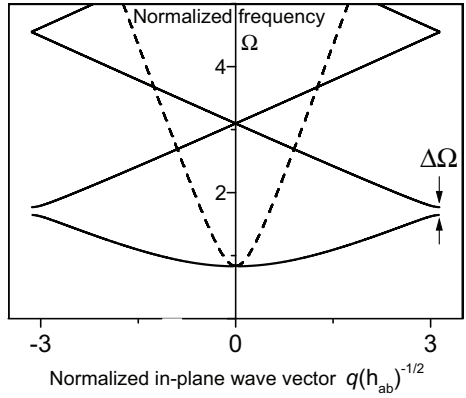

FIG. 11: Band-gap structure of the spectrum of JPWs propagating in a layered superconductor with the JV lattice calculated for $h_{ab} = 0.2$ (From Ref. 115). The values of the parameter $kD$ are $0.3\pi$ (solid line) and $kD = 0.05\pi$ (dashed line). The frequency gap between the first and second zones is marked as $\Delta\Omega$. Here we use $D = 15$ Å, $\lambda_{ab} = 2000$ Å, $\gamma = 600$.

- JPW is excited by the extrnal electromagnetic wave (see Fig. 12b).

*JPW inside a sample*— In the first case, the solution of Eq. (90) for the $j$th cell of the magnetic structure can be expressed in the vector form,

$$\vec{\psi}_\alpha^j = \{C_{1\alpha}^j \exp(i\kappa_\alpha x); \ C_{2\alpha}^j \exp(-i\kappa_\alpha x)\},$$

where $\alpha$ takes integer values either 1 or 2, $\kappa_\alpha$ is defined by Eq. (96), and $C_{i\alpha}^j$ are constants.

The requirement of continuity of $\psi$ and $\psi'$ at any point of discontinuity of the function $\tilde{\omega}^2(\eta)$ gives a set of linear equations relating $\vec{\psi}_\alpha^{j-1}$ and $\vec{\psi}_\alpha^j$. The solution of these equations can be presented in a symbolic form $\vec{\psi}_\alpha^j = \hat{L}\vec{\psi}_\alpha^{j-1}$, where $\hat{L}$ is a $2 \times 2$ matrix. Then, we use a linear non-degenerate transformation $\hat{G}$ that diagonalizes $\hat{L}$. By applying $N$ times such a procedure, we find the linear transformation

$$\vec{\psi}_\alpha^N = \hat{G}^{-1} \left(\hat{G}\hat{L}\hat{G}^{-1}\right)^N \hat{G}\vec{\psi}_\alpha^0 \qquad (97)$$

that propagates the solution from the zeroth to $N$th elementary cell.

Here we consider the case of frequencies higher than the plasma frequency. We denote the amplitude of the incident wave $C_{11}^0$ as 1, the amplitude of the reflected wave $C_{12}^0 = r$, and the amplitude of the transmitted wave $C_{11}^N$ as $\tau$. Using Eq. (97), we obtain two linear equations for two independent variables, $r$ and $\tau$; since $C_{12}^N = 0$. By solving these equations, we find

$$r = \beta_2 \frac{M_2^N - M_1^N}{M_1^N - \beta_2\beta_1 M_2^N} \qquad (98)$$

where

$$\begin{aligned}
M_{1,2} &= \left[\cos\kappa_2(b-1) \pm i\frac{\kappa_1^2 + \kappa_2^2}{2\kappa_1\kappa_2}\sin\kappa_2(b-1)\right] \\
&\quad \times \exp(\mp i(b-1)\kappa_1), \qquad (99)\\
\beta_{1,2} &= \pm\frac{M_2 - M_1 + \sqrt{(M_2 - M_1)^2 + 4L_1L_2}}{2L_{1,2}}, \qquad (100)\\
L_{1,2} &= \pm i\frac{\kappa_1^2 - \kappa_2^2}{2\kappa_1\kappa_2}\sin\kappa_2(b-1)\exp(\pm i(b-1)\kappa_1).
\end{aligned}$$

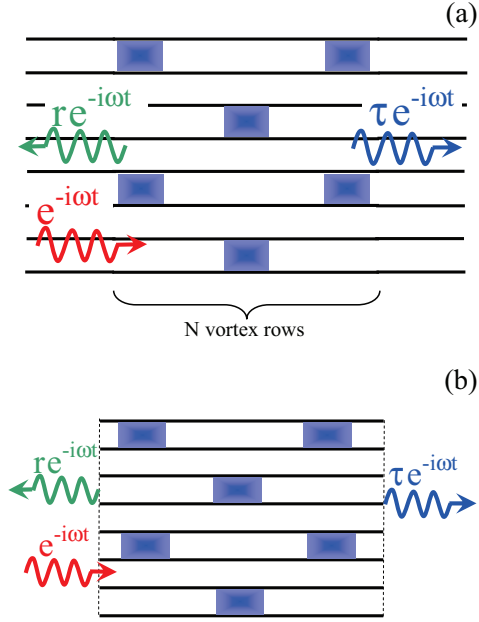

FIG. 12: (Color online) Reflection and transmission of electromagnetic waves from internal (a) and external (b) sources (From Ref. 122).

The frequency dependence of the reflection coefficient $R = |r|^2$ is shown in Fig. 13 for different magnetic fields $H_{ab}$ and the transverse wave vectors $k$. The transparency (transmission $T$) of the crystal increases when increasing the frequency $\Omega$ and when decreasing either the sample length $l = \gamma DN/\sqrt{h_{ab}}$ or $H_{ab}$, due to the decrease of the number of scattering layers. The frequency dependence of the reflection $R$ or transmission $T$ coefficients is much more interesting for large $k$, when the interaction of the electromagnetic wave and the JVs becomes stronger. In this case, the oscillations in the frequency dependence of $R(\Omega)$ and $T(\Omega) = |\tau|^2$ are obtained due to the interference of the transmitting and reflecting waves [55]. Moreover, close to the forbidden frequency zones, the dependence of $R$ and $T$ versus $\Omega$ has several characteristic deep and narrow peaks. At $k = 0$, the corresponding functions are monotonous. Varying the applied magnetic field $H_{ab}$ tunes the reflection at a given frequency from 0 to 1. In a long sample, this tuning remains significant even at small $k$, due to cumulative effect of a large number of weak scatterers.

### 3. Reflection from the sample boundaries

Now we consider the case of irradiation of a sample by a wave from the vacuum along the **ab**-planes. The value of $kD$ is small for THz-range radiation since in vacuum $k^2 + q^2 = \omega^2/c^2$, while $D$ is in the nanometer range. For $\omega/2\pi = 1$ THz and $D = 2$ nm, we find the estimate $kD \leq 4.19 \cdot 10^{-5}$. Imposing the continuity of both $H$ and $E_z$ at the sample surface and using Eq. (97), we find the expression for the amplitude $r$ of the reflected wave,

$$r = \frac{1 + Z(\Omega)S(\Omega)\exp(-2i\kappa_1 b)}{Z(\Omega) + D(\Omega)\exp(-2i\kappa_1 b)},$$
$$S = \frac{\beta_1(Z + \beta_2)M_2^N - (1 + \beta_1 Z)M_1^N}{(Z + \beta_2)M_2^N - \beta_2(1 + \beta_1 Z)M_1^N}, \tag{101}$$

where $Z = (\kappa_1 - i\Omega g)/(\kappa_1 + i\Omega g)$ and $g = D/\left(\sqrt{\varepsilon}\lambda_{ab}\sqrt{1 - c^2k^2/\omega^2}\right)$. Here we assume that two flux-free zones with thickness $b$ exist near the sample edges. Note, that Eq. (98) corresponds to $Z = 0$ in Eq. (101).

The calculated frequency dependence of the reflection coefficient is shown in Fig. 14 for $k = 0$ at different magnetic fields and different sample lengths. The transparency increases when increasing the frequency and when decreasing the number of scattering layers, due to a decrease of the magnetic field $H_{ab}$ or due to a decrease of the sample length. The oscillations in the transition and reflection coefficients occur due to the interference of the scattered and transmitted waves on JVs and sample boundaries. These frequency windows can be easily tuned by the in-plane magnetic field $H_{ab}$.

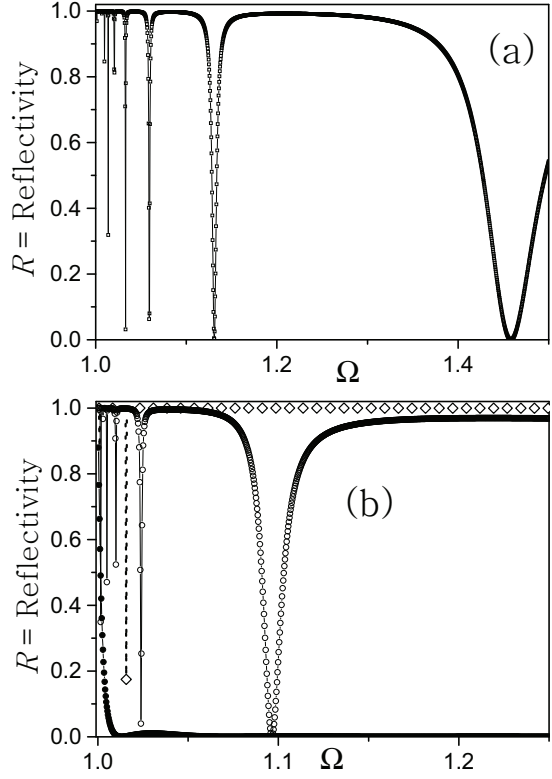

FIG. 13: Internal reflection (From Ref. 115): electromagnetic wave emitted (e.g., by a moving JV) inside a sample reflects back with intensity $R = |r|^2$ and transmits with intensity $T = |\tau|^2 = 1 - R$. The reflection coefficient $R$ versus the electromagnetic wave frequency $\Omega$ for a sample with length $l = 100\gamma D$; (a) for $h_{ab} = 0.2$, $kD = 0.3\pi$, and (b) for $kD = 0.05\pi$, $h_{ab} = 0.2$ (diamond), $h_{ab} = 0.05$ (open circles), $h_{ab} = 0$ (solid circles). Other parameters are the same as in Fig. 11.

Varying $H_{ab}$, one can easily change by an order of magnitude both the transmission, $T$, and reflection, $R = 1 - T$, coefficients of the electromagnetic wave. Thus, a layered superconducting sample can operate as a THz-frequency filter tuned by the applied magnetic field $H_{ab}$. The manufacturing of artificial layered systems, with periodically modulated properties along the layers, would magnify the predicted effects. However, in this case the properties of the system would be not so easily tunable. In general, the transmitted wave should be partially polarized since only waves with a magnetic field along the **ab** plane can propagate through the layered system. These features are potentially useful for THz filters.

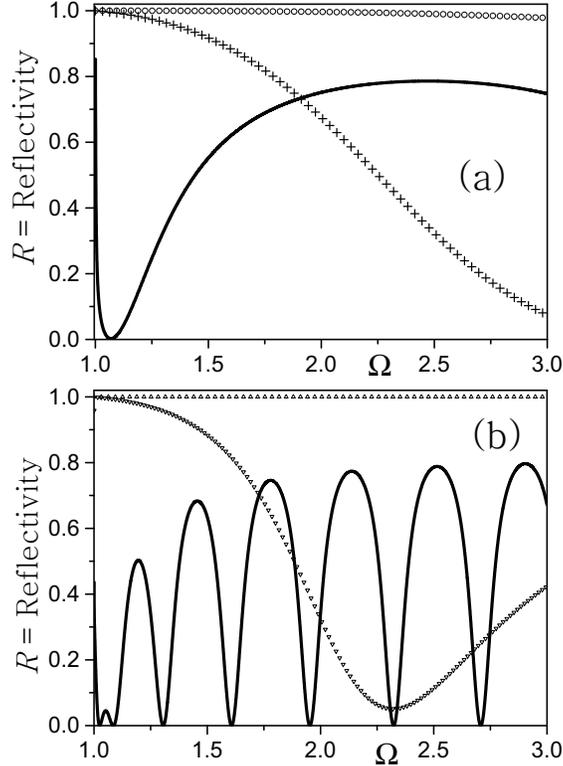

FIG. 14: (Color online) Reflection of electromagnetic waves from the sample, or THz "filter" (From Ref. 122): the reflection coefficient $R$ versus $\Omega$ for an electromagnetic wave incident from the vacuum at $k = 0$, (a) for $l = 100\gamma D$, $h_{ab} = 0.1$ (open circles), $h_{ab} = 0.01$ (crosses), $h_{ab} = 0$ (solid line); and (b) for $l = 1000\gamma D$, $h_{ab} = 0.1$ (up triangles), $h_{ab} = 0.001$ (down triangles), $h_{ab} = 0$ (solid line). Other parameters are the same as in Fig. 11. The deep minima for the solid line in (b) is due to the reflection at the sample boundaries.

### F. Josephson plasma waves localized on the Josephson vortices

In the absence of an external magnetic field, weak (linear) Josephson plasma waves can propagate only if their frequency is above the plasma frequency $\omega_J$. When an external magnetic field is applied along the planes, Josephson vortices can penetrate the junctions. The presence of a vortex locally suppresses the critical current density $J_c$ and, thus, the Josephson plasma frequency, since $\omega_J \propto J_c^{1/2}$. This affects the propagation of JPWs in layered superconductors. For instance, this can provide a tunable photonic crystal, as discussed above. Moreover, the local suppression of the Josephson plasma frequency can support the propagation of JPWs along the vortices *below the plasma frequency.* In other words, the Josephson vortices can serve as specific waveguides for the low-frequency Josephson plasma waves. Such waves, for a single Josephson junction in the presence of an *array* of Josephson vortices, were predicted more than forty years ago in Refs. 123, 124. Here we will not consider arrays, but focus on Josephson plasma waves localized on a single Josephson vortex.

Following Ref. 125, consider a long and wide Josephson junction located in the $xy$-plane, i.e., the $z$-axis is perpendicular to the junction plane. The sine-Gordon equation for the gauge-invariant phase difference $\varphi$ reads,

$$\frac{\partial^2 \varphi}{\partial t^2} + \sin\varphi - \frac{\partial^2 \varphi}{\partial x^2} - \frac{\partial^2 \varphi}{\partial y^2} = 0. \tag{102}$$

Here, the coordinates $x$ and $y$ are normalized by the Josephson length $\lambda_J$ and time $t$ is normalized by $\omega_J^{-1}$. We seek a solution of Eq. (102) in the form

$$\varphi = \varphi_{JV} + \psi,$$

where

$$\varphi_{JV} = 4\arctan[\exp(x)]$$

is a stationary phase distribution produced by a fixed JV and $|\psi| \ll 1$. The perturbation $\psi$ corresponds to Josephson plasma waves propagating along the JV and has the form

$$\psi = \exp(i\omega t - iky)\chi(x).$$

If keeping only the linear terms in $\psi$, the wave amplitude $\chi$ obeys the equation analogous to the 1D Schrödinger equation with reflectionless potential $-1/\cosh^2 x$,

$$\frac{d^2\chi}{dx^2} + 2\left\{\frac{\omega^2 - k^2 - 1}{2} + \frac{1}{\cosh^2 x}\right\}\chi = 0. \tag{103}$$

The solution

$$\chi_{\text{loc}}(x) = \frac{a}{\cosh x} \tag{104}$$

corresponds to a localized wave, running along the vortex. In dimensional variables we obtain the linear dispersion law $\omega = c_{\text{sw}} k$, where $c_{\text{sw}} = \lambda_J \omega_J$ is the Swihart velocity. This branch of the spectrum, shown by the solid straight line in Fig. 15, is gapless, which is unusual for conventional JPWs.

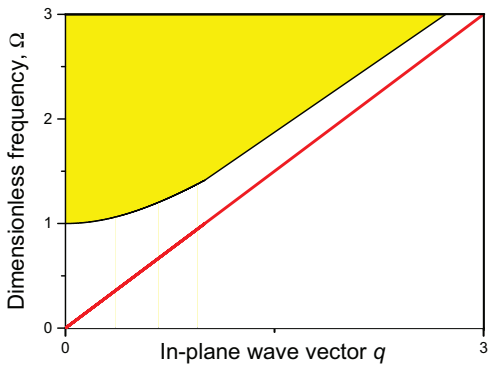

FIG. 15: (Color online) Spectrum of the Josephson plasma waves (From Ref. 125). Solid straight line shows the gapless spectrum of the wave localized on a Josephson vortex. The shaded region corresponds to the delocalized waves within the continuum spectrum.

Obviously, similar localized modes below the plasma frequency can exist in layered superconductors due to the same origin, namely, the suppression of the Josephson current near the vortices.

Localized modes can be employed to guide and control the propagation of THz waves. Since no waves with $\omega < \omega_J$ can propagate without a Josephson vortex, the JPWs will always follow the vortex lines. Because the arrangements of Josephson vortices can be controlled by an external magnetic field or electric currents, it is possible to change the direction of propagation of plasma waves by tuning external parameters. An example of such a device was suggested in Ref. 125. In that paper, numerical simulations were performed on the 2D sine-Gordon equation for a hexagonal splitter attached by three transmission lines (Fig. 16). A magnetic field applied to an edge of the hexagon forces vortices to penetrate a hexagon linking the two neighboring edges, see Fig. 16. The propagation of linear Josephson plasma waves through each vortex arrangement was simulated. To model the transmission line, a long Josephson junction with a plasma frequency $\tilde{\omega}_J$ lower than $\omega_J$ was used. This allows the propagation of waves of frequency $\omega$ ($\tilde{\omega}_J < \omega < \omega_J$) in the transmission line, but not in the hexagon itself if the vortices are absent. The continuity boundary conditions link the electromagnetic field in the hexagon and transmission lines. One can see in Fig. 16 that the distribution of the electric field component in the hexagon tends to localize on each vortex.

So far, many works focused on designing sources and detectors of THz radiation [20] including optical lasers, quantum cascade lasers, solid state and superconducting devices.

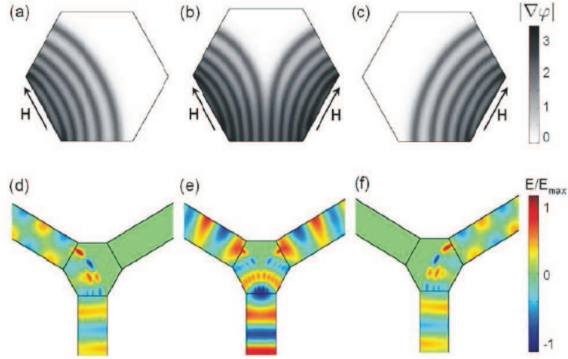

FIG. 16: (Color online) (a-c) Vortex configurations in a hexagonal Josephson junction with each side equal to 20 Josephson penetration lengths $\lambda_J$ (From Ref. 125). The color scale denotes the magnetic field density given by the gradient of the superconducting phase difference $|\nabla\varphi|$ and normalized to $\lambda_J J_c$. (d-f) Propagation of Josephson plasma waves with frequency $\omega = 0.32\,\omega_J$. The plasma frequency $\tilde\omega_J$ in the Josephson transmission lines is 5 times smaller than in the hexagon junction, $\tilde\omega_J = 0.2\,\omega_J$. The width of the attached Josephson transmission lines is $20\lambda_J$. The damping is absent. (d), (e) and (f) correspond to the vortex configurations (a), (b) and (c). The color scale depicts the relative amplitude of the plasma waves as a ratio of the transverse electric field $E$ to its maximal value $E_{\max}$ for each simulation. (d) and (e) show mode transfers from plane-like waves to alternating wave mode in the upper branches. Numerical solutions showing the spatiotemporal evolution of the waves are available online at http://dml.riken.jp/waveguides.

However, less studies have been devoted to the guiding of THz waves. Metal tubes [126, 127] and wires [128], plastic ribbons [129] and dielectric fibers [130] were proposed to guide THz waves. The proposals described here may lead to new type of waveguides for THz waves that can be controlled by magnetic field or electric currents.

## III.  NONLINEAR JOSEPHSON PLASMA WAVES

It is well-known from optics that nonlinearity results in a number of phenomena, including harmonic mixing, self-induced transparency, and self-focusing, which are both of fundamental interest and also important for many practical applications [131, 132]. These nonlinear effects can be derived from the electric or magnetic field dependence of the refraction coefficient. In contrast to optics, the nonlinearity in Josephson media is due to the nonlinear dependence, $J = J_c \sin\varphi$, of the tunnelling supercurrent on the gauge-invariant phase differ-ence, which determines the electromagnetic fields in the system. In the strongly nonlinear regime ($\varphi \sim \pi$), the sine-Gordon equation possesses soliton and breather solutions [78, 79]. However, the nonlinearity becomes crucial even at small wave amplitudes, at $|\varphi| \ll 1$, due to a gap in the spectrum of JPWs. In this section, following Refs. 80, 81, 83, we discuss such phenomena. Some of these (e.g., JPWs self-focusing effects, the pumping of weaker waves by stronger one, the nonlinear plasma resonance [80], and nonlinear surface and waveguide propagation [81]) have analogues in traditional nonlinear optics. In addition, we discuss an unusual stop-light phenomenon caused by both nonlinearity and dissipation [80].

The profound analogy of the nonlinear effects in layered superconductors with several nonlinear optical phenomena could open new avenues in the study of THz plasma waves in superconductors, providing a program for future research in this fast growing field. The close analogy between nonlinear JPWs and nonlinear optics is shown in Table 2.

## A. Nonlinear plane wave below plasma frequency; light slowing-down

Here we focus on weakly nonlinear ($\sin\varphi \approx \varphi - \varphi^3/6$, $|\varphi| \ll 1$) waves at frequencies around $\omega_J$. In the long-wavelength limit (compared to the interlayer spacing), the phase difference $\varphi$ obeys

$$\left(1 - \frac{\partial^2}{\partial z^2}\right)\left(\frac{\partial^2 \varphi}{\partial t^2} + r\Omega\frac{\partial \varphi}{\partial t} + \varphi - \frac{\varphi^3}{6}\right) - \frac{\partial^2 \varphi}{\partial x^2} = 0. \tag{105}$$

We use the dimensionless coordinates and time, $x \to x/\lambda_c$, $z \to z/\lambda_{ab}$, $t \to \omega_J t$.

We employ the asymptotic expansion method to obtain periodic solutions of the nonlinear Eq. (105) in the form,

$$\varphi = \sum_{n=0}^{\infty} A_{2n+1}(x,z) \sin\left[(2n+1)\Omega t - \eta_{2n+1}(x,z)\right]. \tag{106}$$

We consider JPWs with frequencies close to $\omega_J$, i.e., $|1 - \Omega^2| \ll 1$, and the amplitudes $A_1 \sim |1 - \Omega^2|^{1/2}$. For these waves, the nonlinear term $\varphi^3$ in Eq. (105) is of the same order as the linear one, $\partial^2\varphi/\partial t^2 + \varphi$, and even a weak nonlinearity plays a key role in the wave propagation.

For plane-waves propagating along the $x$-axis, the asymptotic expansion of Eq. (105), with respect to $(1 - \Omega^2)$, produces a set of ordinary differential equations for $A_1(x)$, $A_3(x)$,... and $\eta_1(x)$, $\eta_3(x)$,...:

$$A_1'' - [1 - \Omega^2 + (\eta_1')^2]A_1 + \frac{A_1^3}{8} = 0, \tag{107}$$

$$r\Omega A_1 + 2A_1'\eta_1' + A_1\eta_1'' = 0, \qquad (108)$$

$$A_3'' - [1 - 9\Omega^2 + (\eta_3')^2]A_3 + \frac{A_1^3 \cos(\eta_3 - 3\eta_1)}{24} = 0, \qquad (109)$$

$$\frac{A_1^3}{24}\sin(\eta_3 - 3\eta_1) = 0. \qquad (110)$$

Here the prime denotes the derivative with respect to $x$. Due to the nonlinearity, also described in Table 2, the propagating JPW includes higher harmonics with decreasing amplitudes

$$A_{2n+1} \propto |1 - \Omega^2|^{n+1/2}.$$

At $r = 0$, the set of Eqs. (107)-(110) has a solution with constant amplitudes and wave vectors. The amplitude $A_1$ and the wave vector $q = \eta_1'$ of the first harmonics are related by a well-known equation,

$$q = \sqrt{\frac{A_1^2}{8} - 1 + \Omega^2}. \qquad (111)$$

From Eqs. (106) and (111) we conclude that the nonlinear JPW can propagate even below Josephson plasma frequency if its amplitude is strong enough,

$$A_1^2 > A_c^2 = 8(1 - \Omega^2). \qquad (112)$$

This result (confirmed by numerical simulations [80], as shown in Fig. 17) is very unusual for any conducting media where plasma waves propagate only with frequencies above the plasma resonance. Wave packets formed by the nonlinear plane JPW exhibit very weak spreading if their frequency widths are much less than $(1 - \Omega)$.

The dissipation ($r \neq 0$) damps the wave, and the amplitude $A_1$ decays with $x$. To describe this decay, consider the case when the parameter $r$ satisfies the following inequalities:

$$(1 - \Omega^2)^2 \ll r \ll \left(1 - \Omega^2\right). \qquad (113)$$

Due to the right inequality in Eq. (113), nonlinear JPWs decay on a scale much longer than the wavelength, whereas the left inequality allows us to neglect the higher harmonics.

When propagating along the $x$-axis, the wave damps due to dissipation. At some $x = x_0$, the amplitude $A_1(x)$ achieves the critical value $A_c$. At this point, the wave vector $q$ and the group velocity

$$v_g = \frac{\partial \Omega}{\partial q} \propto (A_1 - A_c)^{1/2}$$

formally vanish according to Eq. (111). In other words, the stop-light phenomenon occurs.

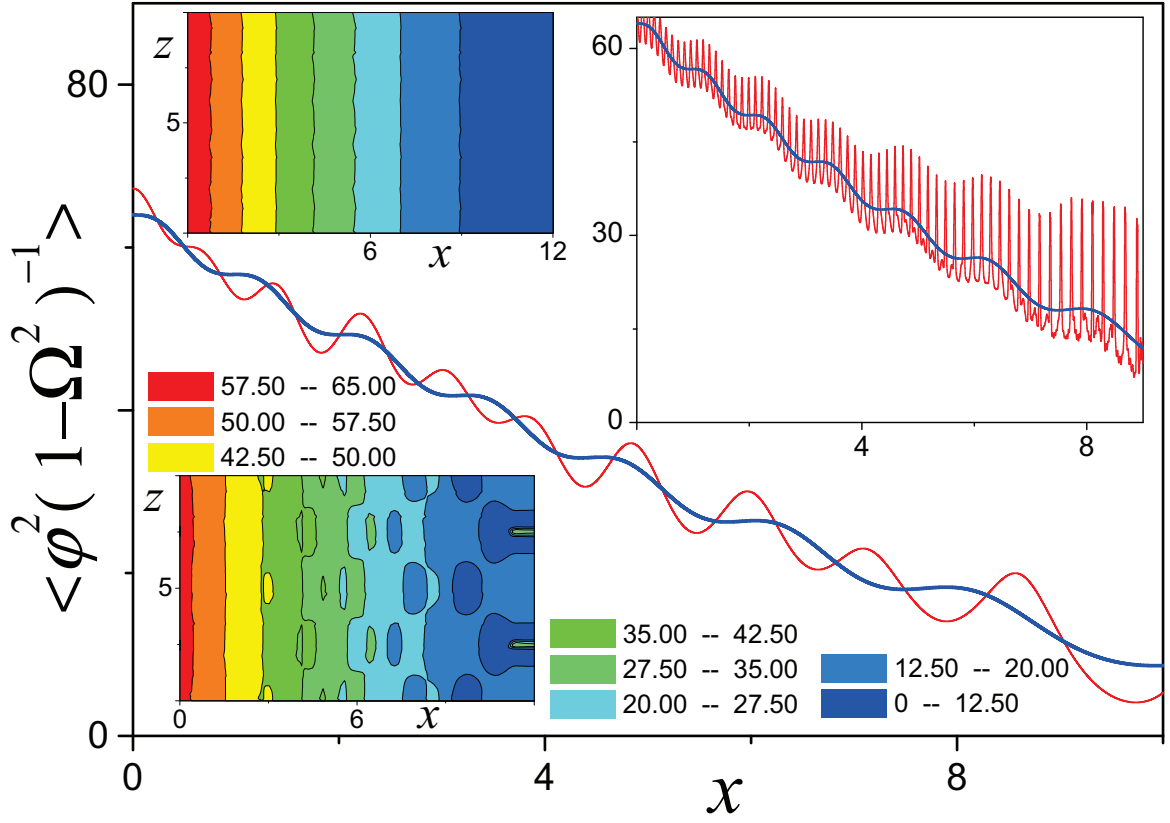

FIG. 17: (Color online) Numerically obtained in Ref. 80 self-induced transparency and pumping: the spatial dependence of the normalized time average of $\varphi^2(x, z = 0)$ for a single nonlinear JPW with $A_1(0) = 8(1 - \Omega^2)^{1/2}$, $r = 0.5$ (blue line) and for a mixture (red line) of nonlinear JPW with the same parameters and a weaker wave with amplitude $0.2 \cos(kz)$ at the surface. Here, $k = 0.4\pi$ ($k = 4\pi$) for the red line in the main panel (right top inset). Left top (bottom) inset shows a 2D contour plot for the blue (red) line in the main panel.

More detailed analysis of Eqs. (107)-(110) yields an estimate for the minimum of $v_g$:

$$v_g^{\min} \sim \frac{r}{\sqrt{1 - \Omega^2}}. \tag{114}$$

Far enough from $x = x_0$ (deeper in the sample), the nonlinearity becomes irrelevant and the JPW decays on a scale $1/(1 - \Omega^2)^{1/2}$, practically without oscillations.

Note that the nonlinear waves considered here, with frequencies below $\omega_J$, are unstable with respect to small fluctuations, and the modulating instability results in formation of the breathers [78, 79]. However, the left inequality in Eq. (113) allows to neglect the modulating instability. The stop-light phenomenon occurs before the formation of breathers.

### B. Self-induced transparency

A further analysis, done in Ref. 80, shows that the nonlinear plane wave with $\Omega < 1$ is stable with respect to small fluctuations of the form,

$$\delta\varphi \propto \exp(ikz + ipx - i\Omega t). \tag{115}$$

The dispersion equation for $p$ has only real solutions. For example, at $r = 0$,

$$p = \pm\sqrt{(1+k^2)[2(1-\omega^2)+3q^2] - q^2}. \tag{116}$$

This indicates that the nonlinear wave assists the propagation of linear waves of the same frequency and wave vector $p$, which could not propagate by themselves because of $\Omega < 1$. This effect is an analog to the self-induced transparency in nonlinear optics.

### C. Nonlinear pumping of a weak wave by a strong one

We have shown above that a running nonlinear wave allows to propagate weak linear waves below the plasma frequency. More interestingly, that decaying nonlinear wave with the amplitude $a_1$ below the critical value $a_c$ pumps weak waves with a large transverse wave number $k$. This occurs if either the amplitude of the running nonlinear wave drops below $a_c$ due to weak dissipation or the amplitude of the incident wave is below the propagating threshold.

For simplicity we consider the latter case. When $A_1 < A_c = (8(1-\Omega^2))^{1/2}$ the strong wave,

$$\varphi(x,t) \approx A_1(x)\sin(qx - \Omega t),$$

decays on a scale $(1-\Omega^2)^{-1/2}$. A weak wave,

$$\varphi_k = A_k(x)\exp(ikz)\sin(\Omega t),$$

interacting with the strong one is described by the equation

$$A_k'' + (1+k^2)\left[\frac{3A_1^2(x)}{8} - 1 + \Omega^2\right]A_k = 0, \tag{117}$$

which can be easily derived from Eq. (105). It is readily seen that when $A_1^2(x) > 8(1-\Omega^2)/3$, the last equation describes a non-decaying wave. In the WKB approximation, for $k \gg 1$, we find from Eq. (117)

$$A_k(x) = A_k(0)\frac{\sqrt{3A_1^2(0) - 8(1-\Omega^2)}}{\sqrt{3A_1^2(x) - 8(1-\Omega^2)}} \tag{118}$$

$$\times \cos\left[\sqrt{1+k^2}\int_0^x dx' \sqrt{\frac{3A_1^2(x')}{8} - 1 + \Omega^2}\right].$$

The amplitude of the weak wave increases when the strong wave approaches a "turning point" $x = x_1$ where $A_1^2(x_1) = 8(1-\omega^2)/3$. This indicates the pumping of the weak waves (with short wave-length along the $z$-axis) by the strong plane wave, as is shown in Fig. 17.

### D. Nonlinear plasma resonance

Consider an electromagnetic wave with frequency $\Omega > 1$ and wave vector $\mathbf{k} = (q_v, k)$, incident from the vacuum, $x < -l$, at the edge of a slab of a layered superconductor, $-l < x < l$, Fig. 18. For the wave in vacuum at $x < -l$, we can write down equations for the electric and magnetic fields,

$$H = H_i \exp(iq_v x) + H_r \exp(-iq_v x), \tag{119}$$

$$E_x = -\frac{k\lambda_{ab}\sqrt{\varepsilon}}{\lambda_c\Omega}H, \tag{120}$$

$$E_z = -\frac{q_v\sqrt{\varepsilon}}{\Omega}\left[H_i \exp(iq_v x) - H_r \exp(-iq_v x)\right], \tag{121}$$

where $H_i$ and $H_r$ are the amplitudes of the incident and reflected waves. To simplify notations, in this subsection we omit the multiplier $\exp(-i\Omega t + ikz)$. In dimensionless units, the dispersion law for the wave in vacuum has the form

$$\left(\frac{k\lambda_{ab}}{\lambda_c}\right)^2 + q_v^2 = \frac{\Omega^2}{\varepsilon}.$$

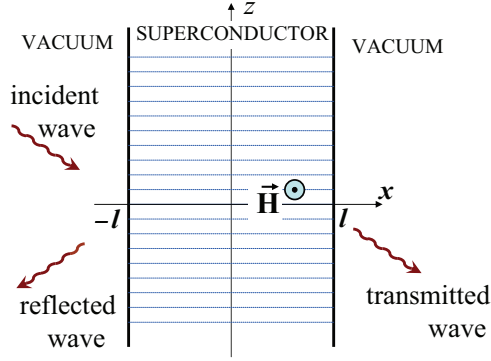

FIG. 18: (Color online) Geometry of the problem (From Ref. 83). A THz wave incident onto a surface of a slab of layered superconductor that occupies the region $-l < x < l$.

In the half-space $x > l$, there exists only a transmitted wave with

$$H = H_t \exp(iq_v x), \ E_x = -H_t \frac{k\lambda_{ab}\sqrt{\varepsilon}}{\lambda_c \Omega} \exp(iq_v x), \tag{122}$$

$$E_z = H_t \frac{q_v \sqrt{\varepsilon}}{\Omega} \exp(iq_v x). \tag{123}$$

1. *Linear approximation*

For the layered superconductor ($-l < x < l$), Eq. (105) can be solved by a standard perturbation approach. First, we examine this problem in the linear approximation and seek a solution for JPWs in the sample in the form of a sum of waves propagating forward and backward,

$$H_s = A \exp(iq_s x) + B \exp(-iq_s x), \tag{124}$$

$$\varphi = a \exp(iq_s x) + b \exp(-iq_s x), \tag{125}$$

with the dispersion law

$$q_s^2 = \left(\Omega^2 - 1\right)\left(k^2 + 1\right). \tag{126}$$

From Eqs. (15)–(17) we find:
$$\frac{\partial H_s}{\partial x} = \mathcal{H}_0(\Omega^2 - 1)\varphi, \tag{127}$$

the relations between the amplitudes $A, B$ and $a, b$,
$$a = \frac{iq_s A}{\mathcal{H}_0(\Omega^2 - 1)}, \qquad b = -\frac{iq_x B}{\mathcal{H}_0(\Omega^2 - 1)}, \tag{128}$$

and the expression for the electric field $E_z$,
$$E_z = -\frac{q_s \Omega}{\sqrt{\varepsilon}(\Omega^2 - 1)} \left[ A \exp(iq_s x) - B \exp(-iq_s x) \right]. \tag{129}$$

Now we should match the solutions in the vacuum and in the sample requiring the continuity of the magnetic field and the tangential component of the electric field, $E_z$, at the sample boundaries. This yields

$$H_i \exp(-iq_v l) + H_r \exp(iq_v l) = A \exp(-iq_s l) + B \exp(iq_s l),$$

$$H_i \exp(-iq_v l) - H_r \exp(iq_v l)$$
$$= Q(\Omega) \left[ A \exp(-iq_s l) - B \exp(iq_s l) \right],$$

$$H_t \exp(iq_v l) = A \exp(iq_s l) + B \exp(-iq_s l), \tag{130}$$

$$H_t \exp(iq_v l) = Q(\Omega) \left[ A \exp(iq_s l) - B \exp(-iq_s l) \right],$$

where
$$Q(\Omega) = \frac{q_s \Omega^2}{q_v \varepsilon(\Omega^2 - 1)}.$$

Solving these algebraic equations, we express all the amplitudes via the amplitude $H_i$ of the incident wave,
$$A = \frac{2(1+Q)H_i \exp(-i(q_v + q_s)l)}{(1+Q)^2 \exp(-2iq_s l) - (1-Q)^2 \exp(2iq_s l)}, \tag{131}$$

$$B = -\frac{2(1-Q)H_i \exp(-i(q_v - q_s)l)}{(1+Q)^2 \exp(-2iq_s l) - (1-Q)^2 \exp(2iq_s l)}, \tag{132}$$

$$H_r = -i(1-Q)\sin(2q_s l) A \exp(-i(q_v - q_s)l), \tag{133}$$

$$H_t = \frac{2Q}{1+Q} A \exp(-i(q_v - q_s)l). \tag{134}$$

One can see that the reflected wave disappears and the amplitudes of the JPW in the sample increase under the resonance condition
$$q_s l = \pi n/2, \tag{135}$$

where $n$ is an integer. For this case, we derive from Eqs. (254)–(134)

$$A = i^n \frac{1+Q}{2Q} H_i \exp(-iq_v l), \tag{136}$$

$$B = (-i)^{n+1} \frac{1-Q}{2Q} H_i \exp(-iq_v l), \tag{137}$$

$H_r = 0$, and $|H_t| = |H_i|$. The values of $Q_n$ for the resonance conditions are

$$Q_n \approx \frac{2\Omega l \left(1+k^2\right)}{\sqrt{\varepsilon}\pi n}. \tag{138}$$

Here we take into account that $q_v \approx \Omega/\varepsilon^{1/2}$. If $Q_n \gg 1$ the resonance amplitudes of the wave in the sample, Eqs. (136) and (137), are much higher than far from the resonance. Thus, the electromagnetic energy stored in a sample of length $2L = 2l\lambda_c$ increases, under resonance conditions, by a factor of about

$$Q_n^2 \sim \frac{4L^2}{\pi^2 \varepsilon \lambda_c^2 n^2}$$

in dimensional units. For a sample with $L = 1$ cm, $\lambda_c = 10^{-2}$ cm, and $\varepsilon = 16$, this value is about 200, for $n = 1$.

2. *Beyond the linear approximation*

Now we take into account the nonlinearity in Eq. (105) which gives rise to corrections to the wave amplitudes and the dispersion law (126). The main idea of the following calculations is analogous to that in the case of nonlinear oscillators [133]. It is of interest to study the system behavior near the first $(n = 1)$ resonance at $\Omega$ close to 1, when $Q_1 \equiv Q \gg 1$ and the power of the JPW in the sample is maximum.

The nonlinearity results in a shift in the dispersion law, which we describe by replacing $q_s \to q_s + \delta q$, where $\delta q$ is a function of the wave amplitude. So, we rewrite the denominator $\Delta$ in Eqs. (254) and (132) as

$$\Delta = (1+Q)^2 \exp(-2iq_s l) - (1-Q)^2 \exp(2iq_s l)$$

$$= -2i(1+Q^2)\sin(2q_s l) + 4Q\cos(2q_s l).$$

In the vicinity of the first resonance, $2q_s l = \pi + 2\delta q l$ and $Q \gg 1$, we obtain in first approximation

$$\Delta = -4Q(1 - iQ\delta q l).$$

Substituting the last expression in Eqs. (254) and (132) we obtain

$$A = -B = \frac{iH_i \exp(-iq_v l)}{2(1 - iQ\delta q l)}. \tag{139}$$

Correspondingly, by means of Eqs. (136), (137), and (139) we derive

$$a = b = -\frac{H_i}{\mathcal{H}_0} \frac{Q\sqrt{\varepsilon} \exp(-iq_v l)}{2(1 - iQ\delta q l)}. \tag{140}$$

Taking the real part of Eq. (125), we can write down

$$\varphi^{(0)} = a_0[\cos(\Omega t - q_s x - kz - \eta)$$

$$+ \cos(\Omega t + q_s x - kz - \eta)], \tag{141}$$

for the phase difference in the first order approximation. Here

$$a_0 = \frac{QH_i}{2\mathcal{H}_0} \frac{\sqrt{\varepsilon}}{\sqrt{1 + Q^2\delta q^2 l^2}}, \quad \eta = -q_v l + \tan^{-1}(Q\delta q l). \tag{142}$$

We seek a solution of Eq. (105) in the form $\varphi = \varphi^{(0)} + \varphi^{(1)}$. Substituting this into Eq. (105) we find, in the first order approximation in $\varphi^{(1)}$ and $\delta q$,

$$[(1 + k^2)(1 - \Omega^2) + q_s^2]\varphi^{(1)}$$

$$= \left(1 - \frac{\partial^2}{\partial z^2}\right)\frac{\varphi^{(0)3}}{6} - 2q_s\delta q \varphi^{(0)}. \tag{143}$$

Following Ref. 133, we choose the value of $\delta q$ to eliminate the first harmonics (resonance terms containing $\cos(\Omega t \pm q_s x - kz - \eta)$) in the right-hand side of Eq. (143). Substituting Eq. (141) into Eq. (143) we obtain

$$\delta q = \frac{3(1 + k^2)}{16 q_s} a_0^2. \tag{144}$$

As in the case of ordinary nonlinear oscillators [133], the shift of the dispersion law is proportional to $a_0^2$.

Let us now consider the case when the frequency of the wave is not exactly equal to the resonance frequency and differs from $\Omega_{\text{res}}$ by a small and slowly-varying value $\Omega_{\text{det}}(t)$, i.e., $\Omega = \Omega_{\text{res}} + \Omega_{\text{det}}(t)$. In this case, the variation of the wave vector,

$$\delta q_\gamma = \frac{\partial q_s}{\partial \Omega}\Omega_{\text{det}} = \left(\frac{1 + k^2}{\Omega_{\text{res}}^2 - 1}\right)^{1/2} \Omega_{\text{det}}, \tag{145}$$

should be added to (144),

$$\delta q = \frac{2(1+k^2)l}{\pi}\left(\frac{3}{16}a_0^2 + \Omega_{\text{det}}\right). \tag{146}$$

Here we take into account that $\Omega_{\text{res}} \approx 1$ and $q_s = \pi/2l$. Substituting this relation into Eq. (142) we derive a self-consistency condition for the wave amplitude

$$f(a_0^2, \Omega_{\text{det}}) = a_0^2\left[1 + \alpha^2\left(\frac{3a_0^2}{16} + \Omega_{\text{det}}\right)^2\right] - h^2 = 0, \tag{147}$$

where

$$h = \frac{(1+k^2)lH_i}{\pi\mathcal{H}_0}, \quad \alpha = \frac{4(1+k^2)^2 l^3}{\pi^2\sqrt{\varepsilon}}. \tag{148}$$

Equation (147) defines the dependence of the wave amplitude $a_0$, near the resonance, on the frequency detuning $\Omega_{\text{det}}$ and the amplitude $H_i$ of the incident wave. The function $a_0(\Omega_{\text{det}})$ is shown in Fig. 19 for different $H_i$. One can see that this dependence is single-valued if $H_i$ is smaller than some critical value $H_{\text{cr}}$. At $H_i > H_{\text{cr}}$, there arises an interval of frequencies where the function $a_0(\Omega_{\text{det}})$ has three branches. As usual, the intermediate branch is unstable while the lower and upper branches are stable. These stable branches can be reached when $\Omega_{\text{det}}(t)$ either increases or decreases. As a result, a hysteresis in the $a_0(\Omega_{\text{det}})$ dependence can be observed if $H_i > H_{\text{cr}}$. Obviously, to observe the hysteretic jumps in $a_0(t)$, the magnitude of the frequency change should exceed a critical value, $\Omega_{\text{det}} > |\Omega_{\text{det}}^{\text{cr}}|$.

The critical values $H_{\text{cr}}$ and $\Omega_{\text{det}}^{\text{cr}}$ can be derived from Eq. (147) and the condition $\partial^2 f/\partial a_0^2 = 0$. The latter condition results in

$$3\alpha^2\left(\frac{3}{16}\right)^2 a_0^4 + \frac{3}{4}\alpha^2\Omega_{\text{det}}a_0^2 + 1 + \alpha^2\Omega_{\text{det}}^2 = 0. \tag{149}$$

This is a quadratic equation with respect to $a_0^2$. It has real roots if its discriminant $\mathcal{D}(\Omega_{\text{det}})$ is positive. Thus, the threshold frequency deviation $\Omega_{\text{det}}^{\text{cr}}$ is defined by the evident condition $\mathcal{D}(\Omega_{\text{det}}^{\text{cr}}) = 0$. From this we obtain

$$\Omega_{\text{det}}^{\text{cr}} = -\frac{\sqrt{3}}{\alpha}. \tag{150}$$

From Eqs. (149) and (150) we calculate the critical amplitude $a_{\text{cr}}$ that corresponds to $\Omega_{\text{det}}^{\text{cr}}$,

$$a_{\text{cr}} = \frac{2^{5/2}}{3^{3/4}\alpha^{1/2}}. \tag{151}$$

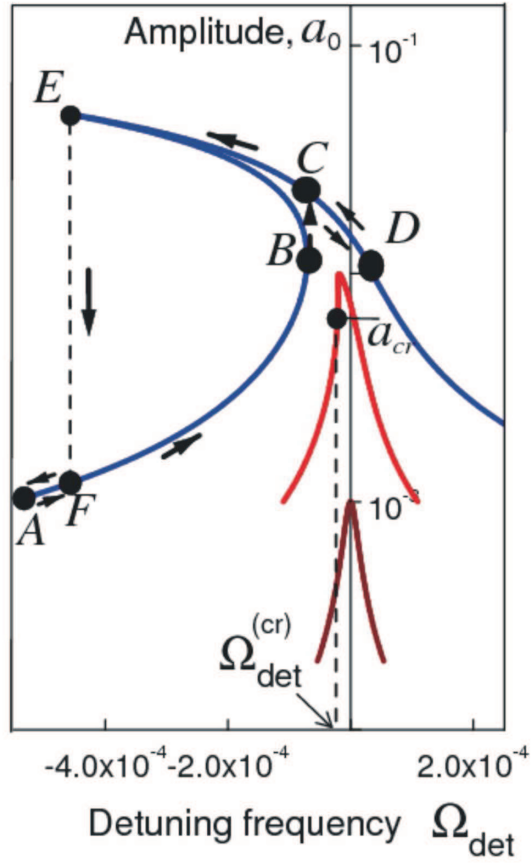

FIG. 19: The dependence of the amplitude $a_0$ of the nonlinear Josephson plasma wave inside the sample on the frequency detuning $\Omega_{\text{det}}$ near a resonance, calculated in Ref. 83 for different incident wave amplitudes: brown (lower) curve corresponds to $H_i/\mathcal{H}_0 = 5 \cdot 10^{-5}$, red curve (critical) for $H_i/\mathcal{H}_0 = 10^{-4}$, and blue curve (upper) for $H_i/\mathcal{H}_0 = 5 \cdot 10^{-4}$. The values of other parameters used here are: $l/\lambda_c = 100$, $\varepsilon=16$, $k = 0.25$.

Finally, by means of Eq. (147), we obtain the inequality to determine the minimal incident wave amplitude

$$h > h_{\text{cr}} = \frac{8\sqrt{2}|\Omega_{\text{det}}^{\text{cr}}|^{1/2}}{3}$$

necessary to observe the hysteresis effect. In dimensional units, this produces the condition

$$H_i > H_{\text{cr}} = \frac{2^{5/2}\pi}{3^{3/4}} \frac{\varepsilon^{1/4}}{(1+k^2)l^{5/2}}\mathcal{H}_0. \tag{152}$$

Using the same values as for the estimate after Eq. (138), we obtain $\Omega_{\text{det}}^{\text{cr}} \approx -10^{-5}$. Thus, the critical frequency detuning is approximately $-5$ MHz for Josephson plasma frequency $\sim 0.5$ THz. The estimate for the critical amplitude gives the ratio $H_{\rm cr}/\mathcal{H}_0 \approx 10^{-4}$. If $D = 1.5$ nm, then $\mathcal{H}_0 \approx 21$ Oe and $H_{\rm cr} \approx 2 \cdot 10^{-3}$ Oe.

The hysteresis of the wave amplitude can result in an interesting phenomenon of transformation of the continuous THz radiation to a set of short bursts with amplitudes significantly higher than the amplitude of the incident wave (see animation at http://dml.riken.go.jp/). Indeed, while the frequency of the incident wave increases approaching $\Omega_{\rm res}$ (route ABCD in Fig. 19), the energy of the electromagnetic wave is accumulated in the sample. When the frequency is decreased (route DCEFA in Fig. 19), the amplitude of the wave abruptly decreases (jump E→F) and a significant part of the stored energy is released in the form of a short THz pulse.

### E. Localized THz beam

Another example of nonlinear effects in layered superconductors is the possible formation (below the plasma frequency $\omega_J$) of plasma waves localized across the layers. The existence of such localized beams can be understood by means of a simple analysis of the coupled sine-Gordon equations Eq. (105) and the dispersion law Eq. (126) for the linear plasma waves. The tails of the localized beams can be considered as linear waves. They can propagate along the $x$-direction with $\omega < \omega_J$ due to the concave profile of $\varphi(z)$. Indeed, Eq. (126) shows that the $x$-component of the wave vector, $q$, can be real for waves with $\omega < \omega_J$ only in the case of imaginary $k$ with $k^2 + 1 < 0$. In other words, the tails of the beam (parts of the THz beam located far enough from the beam center $z = 0$) should have a form

$$\varphi \propto \exp(iqx - i\Omega t \pm \kappa z)$$

with real $\kappa$. Note that such a concave profile of $\varphi(z)$ also describes surface Josephson plasma waves localized near the sample boundary. The center part (the "peak") of the beam cannot have the concave profile of $\varphi(z)$. However, this part of the beam *can* propagate when $\Omega < 1$ *due to the nonlinearity*. Indeed, in the nonlinear regime, the cubic term $\varphi^3$ in Eq. (105) can change the sign of the sum in the second bracket, if the wave amplitude exceeds the threshold value Eq. (112). Thus, we can imagine the localized (in the $z$ direction) beam consisting of two "linear tails" decaying as $\exp(-\kappa|z|)$, with $\kappa > 1$ when $|z| \to \infty$, that are connected with each other via the nonlinear "peak", where the amplitude of the wave

exceeds the threshold value.

We seek a solution of Eq. (105) using the asymptotic expansion Eq. (106), where we only keep only the first harmonics with small amplitude,

$$A_1 \sim (1 - \Omega^2)^{1/2} \ll 1.$$

The equation for $A_1$ has a form,

$$\left(1 - \frac{d^2}{dz^2}\right)\left[(1 - \Omega^2)A_1 - \frac{A_1^3}{8}\right] + k^2 A_1 = 0, \quad (153)$$

with boundary conditions

$$A_1(\pm\infty) = 0 \quad (154)$$

corresponding to a localized solution. Introducing the new variables,

$$a = A_1/(1 - \Omega^2)^{1/2}, \quad \kappa = q/(1 - \Omega^2)^{1/2}, \quad \xi = \kappa z, \quad (155)$$

we rewrite Eq. (153) in the form

$$\left[1 - \kappa^2 \frac{d^2}{d\xi^2}\right]\left(a - \frac{a^3}{8}\right) + \kappa^2 a = 0. \quad (156)$$

Using Eqs. (15)–(17) we obtain the relation between the phase amplitude $a(\xi)$ and the components of the electromagnetic field of the beam:

$$H = H(\xi) \cos(\Omega t - qx), \quad (157)$$

$$H(\xi) = -\mathcal{H}_0 \frac{(1-\Omega^2)}{\kappa} h(\xi), \quad h(\xi) = a(\xi) - \frac{a^3(\xi)}{8}, \quad (158)$$

$$E_x = E_x(\xi) \sin(\Omega t - qx), \quad (159)$$

$$E_x(\xi) = \mathcal{H}_0 \frac{\lambda_{ab}}{\sqrt{\varepsilon}\lambda_c}\left(1 - \Omega^2\right) h'(\xi), \quad (160)$$

$$E_z = -E_z(\xi) \cos(\Omega t - qx), \quad (161)$$

$$E_z(\xi) = \mathcal{H}_0 \left(1 - \Omega^2\right)^{1/2} \frac{1}{\sqrt{\varepsilon}} a(\xi). \quad (162)$$

Equation (153) has a first integral:

$$\left(\frac{da}{d\xi}\right)^2 = \frac{C + a^6 - 12a^4\left(\kappa^2 + \frac{4}{3}\right) + 64a^2\left(\kappa^2 + 1\right)}{\kappa^2\left(8 - 3a^2\right)^2}. \quad (163)$$

Using this equation we can construct the phase diagram in the $(a, a')$-plane (here prime denotes the derivative with respect to $\xi$). For simplicity, we now restrict our analysis to the case when $\kappa \gg 1$, since this simplification does not change the results qualitatively. In this limit, Eq. (163) yields

$$(a')^2 = -\frac{4}{3} + \frac{G}{(8-3a^2)^2}. \tag{164}$$

The phase trajectories $a'(a)$ Eq. (164) that correspond to the localized beam are shown in Fig. 20. According to the boundary conditions Eq. (275), the point (0,0) in Fig. 20 corresponds to $|z| = \infty$. Thus, the black solid lines at $|a(\xi)| < (8/3)^{1/2}$ correspond to the tails of the beam. As it follows from Eq. (158), the value of $a$ is negative at the tails if we demand the positiveness of the magnetic field amplitudes. The peak of the beam is described by the green dashed line in Fig. 20 where the point with $a' = 0$ and $a'' < 0$ (the point of beam maximum) exists. At the beam peak $a(\xi) > 8^{1/2}$, as it also follows from Eq. (158). Obviously, the transition from the tails to peak of the beam is possible only through the jumps between the phase trajectories as is indicated by the dashed arrows in Fig. 20. Such jumps are allowed since the conditions of continuity for the magnetic $h(\xi)$ and electric $E_x(\xi)$ fields can be satisfied.

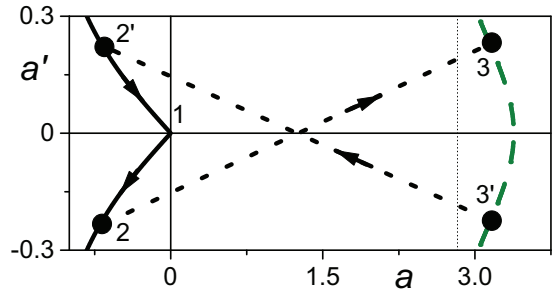

FIG. 20: (Color online) The phase trajectories $a'(a)$ Eq. (164) that correspond to the localized beam (From Ref. 83).

It is convenient to illustrate the beam behavior in the plot $h(a)$, Eq. (158), shown in Fig. 21. The point $(0,0)$ in this plot corresponds to $\xi = \pm\infty$. When increasing $\xi$ from $-\infty$, the values of $a$ and $h$ decrease (see the route from point 1 to point 2 shown by the arrow in Fig. 21). This movement corresponds to the left tail of the beam. At $a = a_{J1} < 0$, the transition from tail to peak of the beam occurs. This transition is shown in Fig. 21 by the horizontal arrow from point 2 to point 3 with $a = a_{J2} > 2\sqrt{2}$. With further increase of $\xi$, the value of $a$ increases while $h$ decreases, and motion from point 3 to point 4 occurs along the $h(a)$-trajectory. Point 4 corresponds to the beam maximum, $a(\xi = 0) = a_m$ and $h(\xi = 0) = h_m$. At $\xi > 0$ we follow the same route, 4-3-2-1, in the reverse direction since the beam is symmetric with respect to $\xi = 0$. The magnetic field is evidently continuous at the points $\xi = \pm \xi_J$ of the jumps. The condition of continuity of the electric field $E_x$ determines the positions $\pm \xi_J$ of the jumps $a_{J1}$ as well as the values $a_{J2}$.

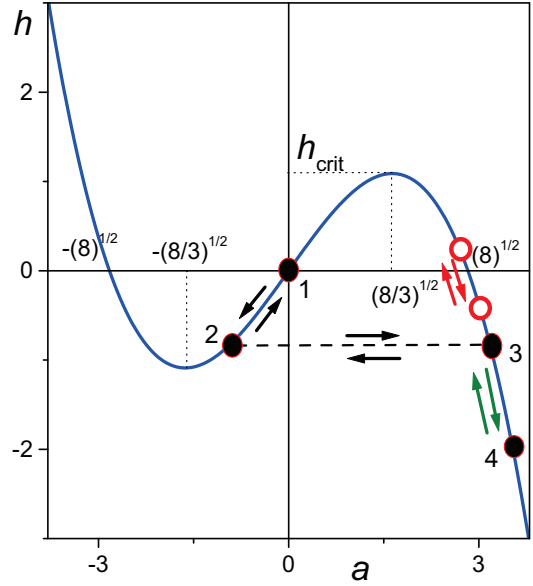

FIG. 21: The non-monotonous dependence of $h$ versus $a$, Eq. (158) (From Ref. 83). The route 1-2-3-4-3-2-1 corresponds to the localized beam profile when the coordinate $\xi$ varies from $-\infty$ to $\infty$. The red arrows between the open circles schematically show the change of $h$ and $a$ for a strongly nonlinear waveguide mode (discussed below in Subsection F) inside the superconducting plate, $-d/2 < z < d/2$.

Integrating Eq. (164) we derive the form of the beam for the case $\kappa \gg 1$. For the peak of the beam, the constant $G$ is determined from the condition $a'(0) = 0$, that is, $G = 4(8 - 3a_m^2)^2/3$. So, the peak of the beam is described by the implicit expression,

$$\int_{a(\xi)}^{a_m} \frac{du\,(3u^2 - 8)}{\sqrt{3(a_m^4 - u^4) - 16(a_m^2 - u^2)}} = 2\kappa|\xi|,\ |\xi| \leq \xi_J. \tag{165}$$

This equation taken at the point $\xi = \xi_J$,

$$\int_{a_{J2}}^{a_m} \frac{du\,(3u^2 - 8)}{\sqrt{3(a_m^4 - u^4) - 16(a_m^2 - u^2)}} = 2\kappa\xi_J, \tag{166}$$

relates the position of the jump with $a_{J2}$. For the tails of the beam, $G = 256/3$ since $a' = a = 0$ at $|\xi| = \infty$. Thus, we have from Eq. (164)

$$\int_{a_{J1}}^{a(\xi)} \frac{du\,(8 - 3u^2)}{u\sqrt{16 - 3u^2}} = 2\kappa\,(|\xi| - \xi_J),\ |\xi| > \xi_J. \tag{167}$$

The asymptotics of the $a(\xi)$ dependence at $\xi \to \infty$, $a \propto \exp(-\kappa\xi)$, coincides with the $z$-coordinate behavior of linear surface waves.

The continuity conditions for $h(\xi)$ and $E_x(\xi)$ give two equations for $a_{J1}$ and $a_{J2}$:

$$a_{J2}\left(1 - \frac{a_{J2}^2}{8}\right) = a_{J1}\left(1 - \frac{a_{J1}^2}{8}\right),$$

$$a_{J1}^2 - a_{J2}^2 + a_m^2 = \frac{3}{16}\left(a_{J1}^4 - a_{J2}^4 + a_m^4\right).$$

Thus, the form of the beam, positions of the jumps, and the values $a_{J1}$, $a_{J2}$ depend on the parameters $\Omega$, $\kappa$, and $a_m$. The form of the beam is illustrated in Fig. 22. The dependences $a(\xi)$ and $h(\xi)$ are shown by the blue solid and red dashed lines, respectively.

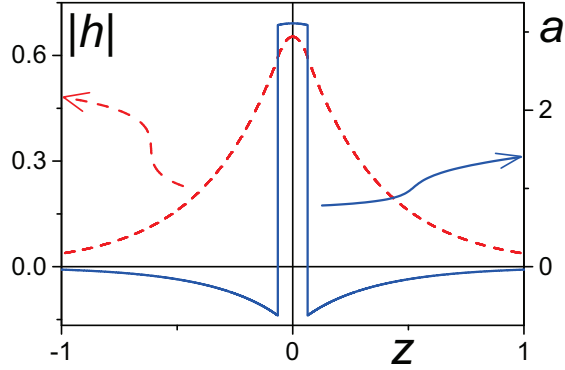

FIG. 22: (Color online) The profile of the localized beam (From Ref. 83): blue solid line shows $a(\xi)$ and red dashed line shows $|h(\xi)|$. The parameters used here are: $\omega = 0.9$, $\kappa = 10$, and $a_m = 3.5$.

### F. Nonlinear waveguide modes

In this subsection we discuss self-sustained JPWs propagating along a thin slab ($-d/2 < z < d/2$) of a layered superconductor (both symmetric and antisymmetric with respect to the middle, $z = 0$, of the sample). The geometry of the problem considered here is shown in the inset of Fig. 23. Weakly nonlinear waves exist in slabs of arbitrary thickness $d$, and coincide with linear surface waves for $d \to \infty$. For *thin* slabs ($d \lesssim \lambda_{ab}$), nonlinear waveguide modes (NWGMs) can be excited in the sample. Surprisingly, even though the magnetic field $H$ for NWGMs can be very small, the electric field $E$ remains strong. Moreover, the magnetic field of the NWGM at the sample surface can be much weaker than the one in the middle of the slab. For this case, the wave amplitude significantly affects the dispersion properties of the NWGMs. The dispersion relation $\omega(q)$ for this wave mode is nonmonotonous. As a result, the stop-light phenomenon, $\partial\omega(q,H)/\partial q = 0$, controlled by the magnetic field amplitude $H$ could be observed.

The Maxwell equations for NWGMs in vacuum ($z > d/2$) determine the distributions of the magnetic (directed along the $y$-axis) and electric fields,

$$H(x, z > d/2, t), \; E_z(x, z > d/2, t)$$

$$\propto \exp[-k_v(z - d/2)]\cos(qx - \omega t); \tag{168}$$

also,

$$E_x(x, z > d/2, t) \propto \exp[-k_v(z - d/2)]\sin(qx - \omega t), \tag{169}$$

with the spatial decrement $k_v = (q^2 - \omega^2/c^2)^{1/2}$. The impedance ratio,

$$\left.\frac{E_x}{H}\right|_{z=d/2} = -\sqrt{\frac{c^2 q^2}{\omega^2} - 1}, \tag{170}$$

should match the one obtained for the superconducting slab at the interface $z = d/2$.

A spatial distribution of the gauge-invariant phase difference

$$\varphi = A_1(z)\cos(qx - \omega t) \tag{171}$$

inside a layered superconductor is defined by Eqs. (155), (157)–(162), and (164) with $\kappa \gg 1$. For symmetric and antisymmetric solutions we use the boundary conditions $a'(0) = 0$ and $a(0) = 0$, respectively, in the middle of the sample. The matching of the impedance (continuity of $E_x(z)$ and $H(z)$) at the sample surface $z = d/2$ results in the dispersion relation for the NWGMs:

$$\left(\frac{q^2 c^2}{\omega_J^2} - \Omega^2\right)^{1/2} = \frac{\lambda_{ab}}{\sqrt{\varepsilon}\lambda_c}\Omega^2 \kappa f_s(\Omega, \kappa, H/H_0). \tag{172}$$

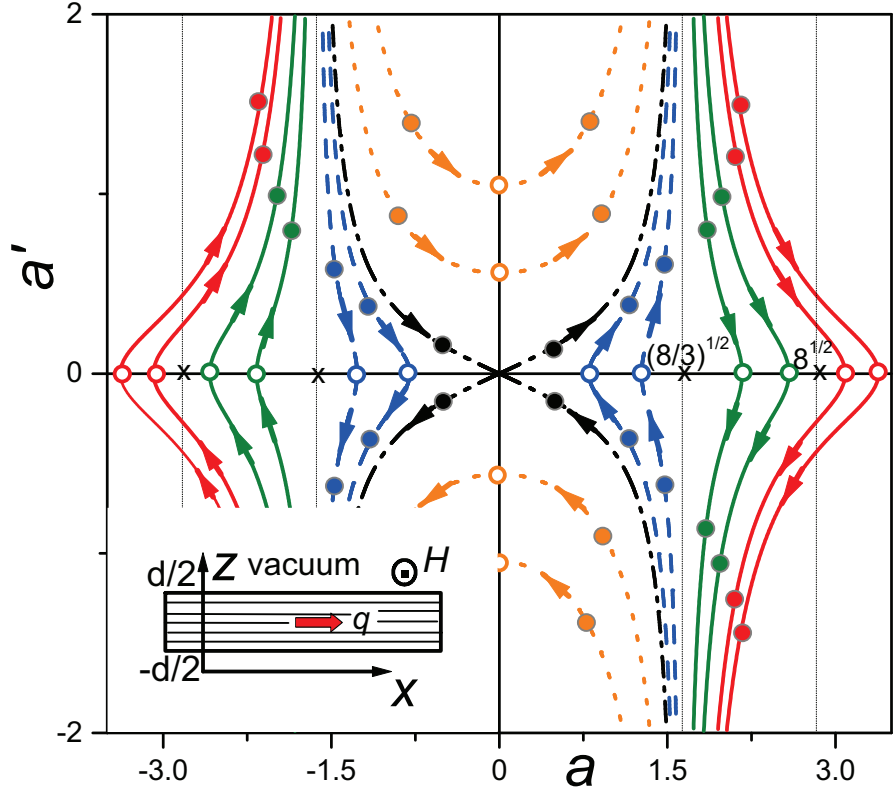

FIG. 23: (Color online) Phase diagram $a'(a)$ (From Ref. 81). Moving along trajectories between solid circles corresponds to the change of coordinate $z$ inside the sample ($-d/2 < z < d/2$). The blue dashed and orange dotted trajectories mark the symmetric and antisymmetric quasi-linear self-sustained modes; the black dash-dotted curve separates symmetric and antisymmetric solutions and corresponds to the nonlinear surface wave; the green and red solid trajectories describe the symmetric strongly nonlinear waveguide modes. The inset shows the geometry of the problem.

The factor

$$f_s = \left.\frac{h'}{h}\right|_{\xi = \kappa d/2\lambda_{ab}} \tag{173}$$

provides the amplitude dependence of the spectrum of the self-sustained waves. This factor has to be obtained by solving Eq. (164). In Eq. (172) we denote $H = H(z = d/2)$.

The phase diagram of Eq. (164), i.e., the set of $a'(a)$ curves for different constants $G$, is shown in Fig. 23. Different phase trajectories correspond to different types of self-sustained waves in the superconducting slabs. Solid circles mark the sample boundaries, while open circles indicate the middle of the slab, arrows show the direction of motion along the trajectories when $\xi$ increases. In order to match the vacuum-superconductor boundary conditions, the starting and ending points of trajectories should be within the interval from $-8^{1/2}$ to $8^{1/2}$. Trajectories confined between $\pm(8/3)^{1/2}$ are weak-amplitude modes, called below quasi-linear ones. For these modes, the effective magnetic field $h$ increases with $a$ (Fig. 21) according to Eq. (157). The quasi-linear waves can be both symmetric and antisymmetric and transform to linear surface waves, when approaching the point (0,0) in Fig. 23. The trajectories with $|a| > (8/3)^{1/2}$ represent symmetric strong-amplitude NWGMs with "reverse" dependence $h(a)$ (see Fig. 21), i.e., $h$ decreases when increasing $a$. This is responsible for the unusual properties of high-amplitude NWGMs: the electric field amplitude $E_z$ can increase inside the sample, while the magnetic field amplitude $H$ decreases. There are no strongly nonlinear self-consistent antisymmetric NWGMs.

1. *Quasi-linear waves*

The blue dashed trajectories in Fig. 23 describe the symmetric waves having the spectrum in Eq. (172), with

$$f_s = \tanh \delta \cdot \left[1 + \frac{h_s^2}{64}\left(7\tanh^2 \delta - 5 + \frac{3\delta}{\sinh \delta \cosh^3 \delta}\right)\right], \tag{174}$$

and orange dotted lines correspond to antisymmetric waves with

$$f_s = \coth \delta \cdot \left[1 + \frac{h_s^2}{64}\left(5 - 3\coth^2 \delta + \frac{3\delta}{\sinh^3 \delta \cosh \delta}\right)\right]. \tag{175}$$

Here we assume

$$h_s = h(\xi = \delta) \ll 1, \quad \delta = \kappa d/2\lambda_{ab}. \tag{176}$$

For thick slabs, $d \to \infty$, the trajectories for both symmetric and antisymmetric waves tend to the black dash-dotted trajectory corresponding to the nonlinear surface wave. For $h \to 0$, the spectrum (174) coincides with the spectrum of linear surface waves. For the parameters corresponding to the $Bi_2Sr_2CaCu_2O_{8+\delta}$ compounds, the spectrum of symmetric quasi-linear waves is located close to the "vacuum light line", $\omega = cq$, and deviates from this line only at very small values of $(1 - \Omega^2) \sim \lambda_{ab}^2/\varepsilon\lambda_c^2$. Thus, these waves are unlikely to be excited in $Bi_2Sr_2CaCu_2O_{8+\delta}$ compounds. However, for artificial superconducting multi-layers, or other compounds like $YBa_2Cu_3O_{7-\delta}$, the conditions for symmetric quasi-linear wave excitations could be satisfied.

Concerning the antisymmetric quasi-linear waves, their spectrum shifts far from the "vacuum light line" for thin slabs, $d \ll \lambda_c$ (see Fig. 24). The electromagnetic field of these waves has a very simple, almost linear distribution inside the sample (inset in Fig. 24) and decays in the vacuum over a short enough (sub-millimeter) distance. Due to the latter, layered $Bi_2Sr_2CaCu_2O_{8+\delta}$ superconductors can act as a waveguide for the antisymmetric THz modes. Also, nonlinear antisymmetric waves can produce Wood's-like anomalies in both amplitude and angular dependence of the reflectivity, transmissivity and absorptivity coefficients. Similar properties are also inherent for the symmetric strongly nonlinear waves considered below.

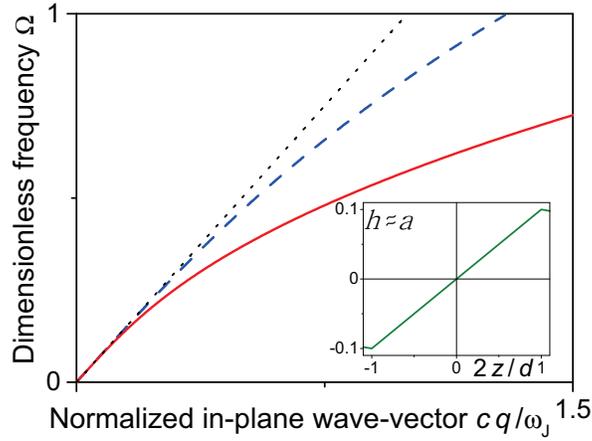

FIG. 24: (Color online) Dispersion relation, $\Omega(cq/\omega_J)$, for the antisymmetric waveguide mode (From Ref. 81). Parameters are $\lambda_c/\lambda_{ab} = 200$, $\varepsilon = 16$, $d/\lambda_{ab} = 0.1$ and 0.3 for solid red and dashed blue curves, respectively. The black dash-dotted line corresponds to the "vacuum light line". Inset: schematics of the spatial distribution of the dimensionless magnetic field amplitude $h(2z/d) \approx a(2z/d)$ inside the sample.

## 2. Symmetric strong-amplitude NWGMs

The function $f_s$ in Eq. (172), describing the deviation of the spectrum of the NWGMs from the "vacuum light line", has a very complicated structure with asymptotics:

$$f_s \propto \frac{\kappa d}{h_s \lambda_{ab}} \quad \text{for} \quad \frac{\kappa^2 d^2}{\lambda_{ab}^2} \ll 1 - h_s \tag{177}$$

and

$$f_s \propto \frac{\kappa^2 d^2}{h_s \lambda_{ab}^2} \quad \text{for} \quad \frac{\kappa d}{\lambda_{ab}} \gg 1. \tag{178}$$

This allows to construct a simple interpolation of the dispersion relation

$$1 - \Omega^2 = \frac{d^2}{3\varepsilon \lambda_{ab}^2} \left[ \left(\frac{H}{H_t}\right)^2 \left(1 - \frac{\omega_J^2}{c^2 k^2}\right) - \frac{c^2 k^2}{4\omega_J^2} \right] \tag{179}$$

where the threshold amplitude

$$H_t \approx \frac{0.8 \mathcal{H}_0 d^2}{\varepsilon^{3/2} \lambda_{ab} \lambda_c} \tag{180}$$

defines the lowest value of the magnetic field amplitude at the sample surface: at lower fields the predicted NWGMs do not exist. The interpolation formulae Eq. (179) is in perfect agreement with numerical results (see Fig. 25) obtained by the integration of Eq. (164).

The spectrum of the strong-amplitude NWGMs is non-monotonic (Fig. 25) and $\Omega(q)$ reaches the minimal value

$$\Omega_{\min} = \left\{ 1 - \frac{d^2 H}{3\varepsilon \lambda_{ab}^2 H_t} \left(\frac{H}{H_t} - 1\right) \right\}^{1/2} \tag{181}$$

at

$$q = \frac{\omega_J}{c} \sqrt{\frac{2H}{H_t}}. \tag{182}$$

Thus, the stop-light phenomenon, $\partial \omega(q, H)/\partial q = 0$, occurs in the THz superconducting waveguide. This stop-light effect can be easily controlled by the magnetic field amplitude.

It is interesting to note that the spectrum of the NWGMs is located between $(q_s, \Omega_s)$ and $(q_f, \Omega_f)$. At these peculiar points of the spectrum, the value of the dimensionless magnetic field amplitude $h$ achieves its critical value

$$h_{\text{crit}} = h\left(a = \sqrt{8/3}\right) = \sqrt{32/27} \tag{183}$$

(see Fig. 21). At the sample edges $z = d/2$, $a'$ tends to infinity according to Eq. (164), but $h'$ is not singular.

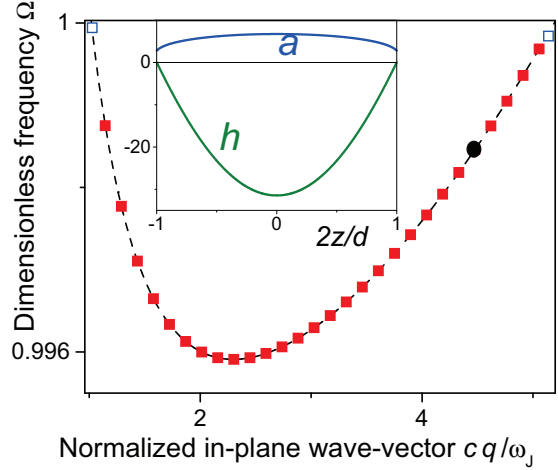

FIG. 25: (Color online) Dispersion relation, $\Omega(cq/\omega_J)$, for the strongly nonlinear waveguide mode (From Ref. 81): the solid red squares present the result of the numerical simulation using Eq. (164); the dashed black line is obtained by interpolating between two asymptotics. The simulations and interpolation perfectly coincide. Here we use the following set of parameters: $d/\lambda_{ab} = 0.3$, $\lambda_c/\lambda_{ab} = 200$, $\varepsilon = 16$, $H/\mathcal{H}_0 = 1.5 \cdot 10^{-5}$. Inset: the spatial distribution of the dimensionless magnetic field amplitude $h(2z/d)$ and the amplitude of the gauge-invariant phase difference $a(2z/d)$ inside the sample, for the same set of parameters as in the main panel and for $\Omega$ and $cq/\omega_J$ marked by the solid circle in the main panel. The open blue squares (located at the top left and top right corners) mark the starting and ending points of the spectrum.

For the strong-amplitude NWGMs, the magnetic field amplitude at the sample surface is less than inside the slab, while the phase $a(\xi)$ and, thus, the electric field $E_z$ do not significantly change in the sample (see inset of Fig. 25). Due to this feature, i.e., $H(0)/H(d/2) \gg 1$, the spectrum of the NWGMs is remarkably far from the $\omega = cq$ line despite the smallness of the parameter $\lambda_{ab}/\sqrt{\varepsilon}\lambda_c$ in Eq. (172).

The numerical analysis of Eqs. (153) and (172) [or (179)] shows that strong-amplitude NWGMs exist for sample thicknesses $d$ smaller than some critical value $d_c$, because of the instability of NWGMs for thick samples (see the next section). This threshold thickness $d_c$ depends on the sample parameters (in particular, the ratio $\lambda_c/\lambda_{ab} \gg 1$ and $\varepsilon$) as well as the NWGM frequency and wave vector. However, $d_c$ is about several $\lambda_{ab}$ in any realistic case. For given parameters of the incident EMW and material characteristics (e.g., $\lambda_c/\lambda_{ab}$, $\varepsilon$), the amplitude of the magnetic field and the gauge-invariant phase oscillations increase in the middle of the sample when the sample thickness $d$ grows. When $d > d_c$, the large-amplitude NWGMs suddenly become unstable. However, the large-amplitude NWGMs are stable with respect to small perturbations when $d$ is smaller than $d_c$.

The nonlinear waveguide modes can be excited in a superconducting slab using two dielectric prisms. As a result of the NWGM excitation, a resonance increase of the electromagnetic absorptivity can be observed if the ac amplitude, frequency, and wave vector satisfy the dispersion relation Eq. (179). Let us now consider the plane monochromatic electromagnetic wave incident from the dielectric prisms through the vacuum interlayers onto a plate of layered superconductor, from both of its sides (see Fig. 26).

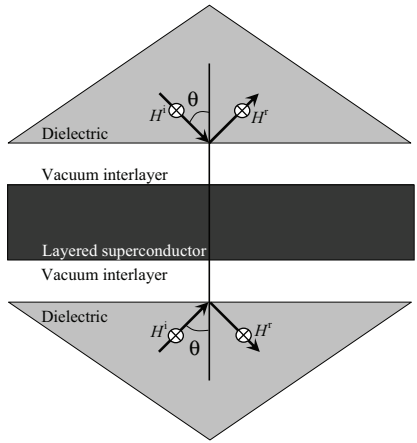

FIG. 26: A geometry, proposed in Ref. 81, for an experiment probing the resonance excitations of NWGMs in a layered superconductor sandwiched between two dielectric prisms.

This experimental configuration corresponds to the prism method of excitation of surface waves with attenuated total wave reflection (see Subsection II C). Usually, the incident angle $\theta$ is varied in one-sided or unilateral experiments, and the resonance suppression of the wave reflection is observed if the wave vector satisfy the dispersion relation for the surface wave in a conductor.

Two essential additional features of this proposed experiment are important for our case. First, the superconducting plate is excited from both its sides, resulting in the magnetic field of the incident waves to be symmetric with respect to the middle of the superconducting plate. Second, the considered waveguide mode is nonlinear. This offers a novel possibility to observe the anomalies in the reflection coefficient and absorptivity as a function of the *amplitude* of the incident wave with given frequency and incident angle. This allows to distinguish the predicted nonlinear waveguide modes from linear ones (for which there is no amplitude anomaly). Namely, if the sum of the magnetic fields of the incident and reflected waves at the sample surfaces takes the resonance value $H_{\text{res}}$,

$$H_{\text{res}} = H_t \left\{ \frac{\sin^2(\theta)\epsilon}{\sin^2(\theta)\epsilon - 1} \left[ \frac{3\varepsilon\lambda_{ab}^2(1 - \Omega^2)}{d^2} + \frac{\sin^2(\theta)\epsilon}{4} \right] \right\}^{1/2} \quad (184)$$

and

$$\sin^2(\theta) > \frac{1}{\epsilon}, \quad (185)$$

a sharp decrease of the reflection coefficient and increase of the electromagnetic absorption should be observed. Here $\epsilon$ is the dielectric constant of a prism.

The dependence of the dispersion relation on the wave amplitude is the main feature that can be used to experimentally distinguish the predicted NWGMs from ordinary plasma waves. The excitation of NWGMs produces an increase in the EMW transparency of the sample near the plasma frequency, if the amplitude of the incident wave $H$ exceeds the threshold value $H_t$, Eq. (180). Using characteristic values for BSCCO ($\lambda_{ab} = 200$ nm, $\lambda_c/\lambda_{ab} = 200$, $\varepsilon = 16$, and $D = 15$ nm) and assuming that the sample thickness $d$ is equal to $\lambda_{ab}$, we obtain $H_t \approx 2 \cdot 10^{-3}$ Oe, which corresponds to a power of the incident (from the vacuum) EMWs of the order of 1 mW/cm$^2$.

## IV. RADIATION OF THE JOSEPHSON PLASMA WAVES BY MOVING JOSEPHSON VORTICES

If an external magnetic field is applied parallel to the **ab**-plane of a layered superconductor, the Josephson vortices in between the superconducting layers enter the sample. Since Josephson vortices do not have a normal core, the vortex pinning force and the friction force, that slow down vortex motion, are very weak. Thus, vortices can move very fast when a driving current along the **c**-axis is applied. The vortex velocity can reach the phase velocity of plasma waves. In particular, very fast-moving Josephson vortices were observed in annular tunnel Josephson junctions [134, 135]. Therefore, the generation of Josephson plasma waves by moving vortices due to the Cherenkov mechanism could occur under certain conditions. The main problem is if the maximal possible vortex velocity can *exceed* the phase velocity of plasma waves. In this context, we consider three different Josephson systems: (i) a single vortex in a single Josephson junction; (ii) a single vortex in a layered superconductor; and (iii) a vortex lattice in a layered superconductor.

### A. Cherenkov radiation in a long Josephson junction. Nonlocal electrodynamics of Josephson vortices

Following Mints and Snapiro [56], we consider a Josephson vortex in a long Josephson junction of thickness $d$ with critical current density $J_c$ and dielectric constant $\varepsilon$. The dynamics of the vortex is described by the sine-Gordon equation for the gauge-invariant phase difference $\varphi(x,t)$ (the $x$- and $y$-axes are oriented in the plane of junction, the vortex is parallel to the $y$-axis),

$$\frac{1}{\omega_J^2}\frac{\partial^2 \varphi}{\partial t^2} - \lambda_J^2 \frac{\partial^2 \varphi}{\partial x^2} + \sin\varphi = 0, \tag{186}$$

where

$$\lambda_J = \sqrt{\frac{c\Phi_0}{16\pi^2 \lambda J_c}}, \quad \omega_J = \sqrt{\frac{8\pi e d J_c}{\hbar \varepsilon}} \tag{187}$$

are the Josephson penetration length and the Josephson plasma frequency. We neglect the dissipation and driving terms in Eq. (186), assuming them to be small. The well-known solution of Eq. (186),

$$\varphi_0(x,t) = 4\tan^{-1}\left[\exp\left(\frac{x - Vt}{\lambda_J \sqrt{1 - V^2/c_{\rm sw}^2}}\right)\right], \tag{188}$$

describes the uniform motion of a Josephson vortex with a certain velocity $V$. It follows from Eq. (188) that a Josephson vortex moves in a long junction similarly to a relativistic particle with the highest possible velocity $c_{\rm sw} = \lambda_J \omega_J$ (Swihart velocity) [68].

Now we consider JPWs in a long Josephson junction [136],

$$\varphi(x,t) = \varphi_a \exp(-i\omega t + iqx), \quad |\varphi_a| \ll 1. \tag{189}$$

The relation between $\omega$ and $q$ is given by the formula [68],

$$\omega = \omega_J \sqrt{1 + \lambda_J^2 q^2}. \tag{190}$$

Correspondingly, the phase velocity of the wave is

$$v_{\rm ph} = \frac{\omega}{q} = c_{\rm sw}\sqrt{1 + \frac{1}{q^2\lambda_J^2}}. \tag{191}$$

This velocity is obviously higher than the Swihart velocity $c_{\rm sw}$ for any $q$ and, at a first glance, the Cherenkov radiation of JPWs by the vortex moving in a long Josephson junction is impossible.

However, as noted in Ref. 56, the above consideration is valid only within the framework of the local theory. In Ref. 69, Gurevich proved that the evolution of the phase difference $\phi$ is governed by the *nonlocal* sine-Gordon equation,

$$\frac{1}{\omega_J^2}\frac{\partial^2 \varphi}{\partial t^2} - \frac{\lambda_J^2}{\pi\lambda}\int_{-\infty}^{\infty} K_0\left(\frac{|x-u|}{\lambda}\right)\frac{\partial^2 \varphi}{\partial u^2}du$$
$$+ \sin\varphi = 0, \tag{192}$$

where $K_0(x)$ is the zero order modified Bessel function. Only when the spatial variation of $\varphi(x,t)$ is smooth in space, i.e., if $\lambda \ll \lambda_J$, Eq. (192) is reduced to the local form Eq. (186). However, the spectrum of the electromagnetic wave differs significantly from Eq. (190) at high enough values of the wave vector $q \geq 1/\lambda$. Namely, the dispersion relation, accounting nonlocality, is

$$\omega = \omega_J\sqrt{1 + \frac{q^2\lambda_J^2}{\sqrt{1+q^2\lambda^2}}}. \tag{193}$$

Correspondingly, the phase velocity,

$$v_{\rm ph} = \frac{\omega}{q} = c_{\rm sw}\sqrt{\frac{1}{\sqrt{1+q^2\lambda^2}} + \frac{1}{q^2\lambda_J^2}}, \tag{194}$$

tends to zero when $q \to \infty$ and, therefore, becomes less than the vortex velocity. This means, that the Cherenkov radiation of the plasma waves with high values of the wave number could exist in a long Josephson junction at any value of the Josephson vortex velocity [56].

### B. Cherenkov out-of-plane radiation by a moving Josephson vortex in layered superconductors.

The Cherenkov radiation propagating along the vortex velocity caused by the nonlocal effects should exist not only in a single Josephson junction but in layered superconductors as well. The principal question arises if the Cherenkov radiation can propagate in other directions similarly to the Cherenkov radiation of a relativistic particle. In other words, can the Cherenkov cone of the radiated waves be observed in layered superconductors? As shown in Refs. 70–72, the answer this question is certainly "yes" if one of the junctions differs significantly from others in a stack of Josephson junctions.

The geometry of the problem is shown in Fig. 27b. Two layered superconducting half-spaces are joined by a "weak" junction parallel to the layers. The $x$-axis is directed along the layers, the $z$-axis is perpendicular to them, and the magnetic field $\vec{H}(x,z,t)$ is oriented parallel to the $y$-axis. The weak junction is situated at $z=0$ and has a thickness $D^*$ and the critical current density $J_c^*$. The corresponding values for the junctions in the layered superconductors are $D$ and $J_c$, as in previous sections.

The gauge-invariant phase difference, $\varphi^{l+1,l}(x,z,t)$, in both superconducting half-spaces is described by the coupled sine-Gordon equations, Eq. (12). We assume that the main phase difference is across the weak contact, where a JV is located, whereas $\varphi^{l+1,l}$ is small across other junctions. Thus, for the waves emitted by the vortex, we can linearize Eq. (12).

We seek a travelling solution of Eq. (12),

$$\varphi^{l+1,l} = \varphi^{l+1,l}(x - Vt), \tag{195}$$

which corresponds to a Josephson vortex moving with the constant velocity $V$. Using the Fourier transform,

$$\varphi^{l+1,l}(q) = \int_{-\infty}^{\infty} d\zeta \exp(-iq\zeta)\varphi^{l+1,l}(\zeta), \tag{196}$$

we obtain, from linearizing Eq. (12), the solution

$$\varphi^{l+1,l}(q) = \varphi^{1,0}(q) \cdot \exp[i\,\text{sign}(q)\,k(q)D|l|], \tag{197}$$

where $\text{sign}(q) = 1$ if $q > 0$, and $-1$ if $q < 0$, the wave number $k(q)$ is defined by the relation

$$\sin^2\left(\frac{k(q)D}{2}\right) \approx \frac{D^2}{4\lambda_{ab}^2} \frac{\omega_J^2 + c^2 q^2/\varepsilon}{q^2 V^2 - \omega_J^2}. \tag{198}$$

Obviously, the transverse wave numbers defined by Eq. (198) correspond to Josephson plasma waves propagating not only in the $x$ direction but across the layers as well, if $0 < \sin^2(kD/2) < 1$. Specifically, in this situation the Josephson vortex emits the plasma waves in a wide range of angles, i.e., the Cherenkov cone can be observed. The analysis of Eq. (198)

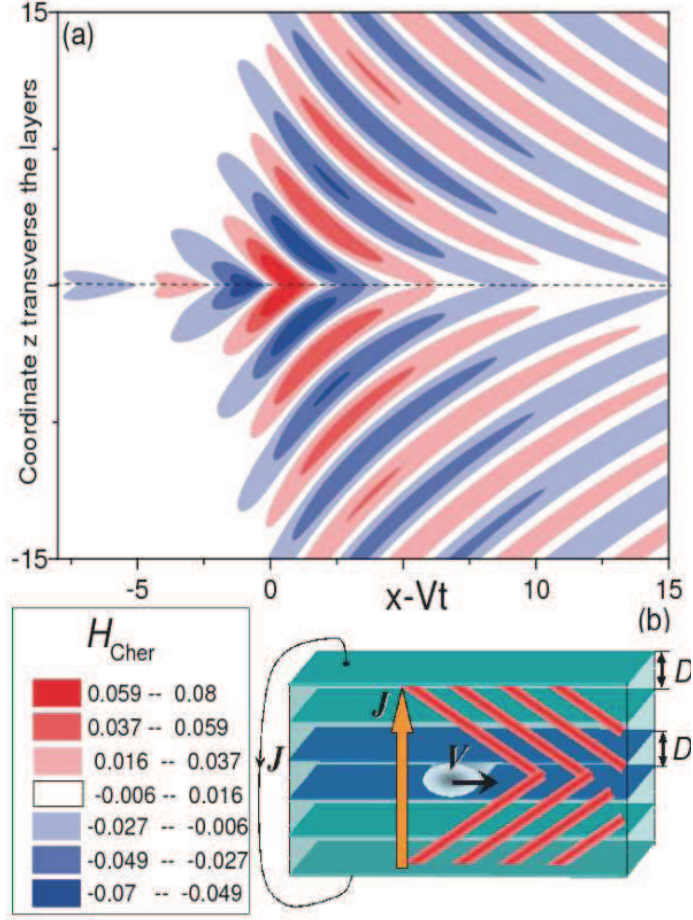

FIG. 27: (Color online) Cherenkov radiation generated by a fast Josephson vortex (located at $x = Vt$) moving in a weaker junction with the critical current $J_c^* < J_c$ and the junction thickness $D^*$ (From Ref. 70). (a) Magnetic field distribution $H(x - Vt, z)$ in units of $\Phi_0/2\pi\lambda_c\lambda_{ab}$ for $J_c^*/J_c = 0.2$, $D^*/D = 1.2$, $V/V_{\max} = 0.9$. The "running" coordinate, $x - Vt$, is measured in units of $\lambda_c D/(\pi\lambda_{ab}\sqrt{v^2\beta^2 - 1})$, while the out-of-plane coordinate $z$ is normalized by $D/(\pi\sqrt{v^2\beta^2 - 1})$, where $\beta = J_c D^*/J_c^* D$ and $v = V/V_{\min}$. The moving vortex emits radiation propagating forward. This radiation forms a cone determined by the vortex velocity $V$. (b) Geometry of the problem: in a weaker junction (located between the two blue superconducting planes), an **c**-axis current $J_{\|c}$ drives a Josephson vortex with velocity $V$, which is higher than the minimum phase velocity $V_{\min}$ of the propagating electromagnetic waves. Red strips in (b) schematically show out-of-plane Cherenkov radiation.

shows that this is the case if the vortex velocity exceeds the minimal value $V_{\min}$,

$$V > V_{\min} = \frac{cD}{2\sqrt{\varepsilon}\lambda_{ab}}, \tag{199}$$

and the wave numbers $q$ is high enough,

$$|q| > q_{\min} = \frac{\omega_J}{\sqrt{V^2 - V_{\min}^2}}. \tag{200}$$

When $|q| < q_{\min}$, the waves Eq. (197) propagate along the $x$-axis and decay with distance from the weak junction.

The magnetic field of the emitted waves can be expressed in terms of the phase difference $\varphi^{l+1,l}$. As it follows from Eq. (16),

$$H^{l+1,l}(q) = -\frac{4\pi i J_c}{cq}\left(1 - \frac{q^2 V^2}{\omega_J^2}\right)\varphi^{l+1,l}(q). \tag{201}$$

Thus, the dependence of the magnetic field on the layer number obeys the same law as the phase difference $\varphi^{l+1,l}$,

$$H^{l+1,l}(q) = H^{1,0}(q) \cdot \exp[i\,\text{sign}(q)\,k(q)D|l|]. \tag{202}$$

Now we should find the relation connecting the amplitude of the emitted waves and the phase difference $\phi$ in the weak junction where the vortex is located. We use the following relation [68]:

$$\phi'(\zeta) = -\frac{8\pi^2\lambda^2}{c\Phi_0}\{J_x(\zeta, l = 0) - J_x(\zeta, l = -1)\}, \tag{203}$$

where $\zeta = x - Vt$, the prime denotes differentiation with respect to $\zeta$, and $J_x(\zeta, l = 0, 1)$ are the values of the $x$-components of the current density taken at the top and bottom edges of the weak contact. Equation (203) is valid if $D^* \ll 2\lambda_{ab}$, that is, if the magnetic flux through the weak junction is small compared with $\Phi_0$. In other words, the gradient of the gauge-invariant phase difference $\phi$ along the central junction occurs due to the gradient of the phase of the order parameter in the layers forming the weak junction rather than the trapped magnetic flux.

Using Eq. (203) and the Maxwell equation Eq. (6), we obtain the desired formula in the form

$$H^{1,0}(q) = -\frac{i\Phi_0 Dq}{4\pi\lambda_{ab}^2(1 - \exp(-ik(q)D))}\phi(q). \tag{204}$$

We would like to pay particular attention to a very important feature of Josephson plasma waves emitted by the moving vortex. According to Eq. (198), the wave numbers $k(q)$ are about $\pi/D$ for $q$ close to the minimal value $q_{\min}$, Eq. (200). It follows from Eq. (6), that the ratio $E_x/H$ for these waves,

$$\frac{E_x}{H} = \frac{-icD}{\lambda_{ab}^2 \omega(1 - \exp(-ik(q)D))} \approx \frac{-icD}{2\lambda_{ab}^2 \omega_J}, \tag{205}$$

appears to be about unity in Bi- or Tl-systems at such conditions. This fact seems to be quite unexpected because the effective conductivity $\sigma_{\text{ef}}$ of these materials along the superconducting layers is high. In usual conductors, the ratio $E_x/H \sim \omega\delta/c$ is always very small. Here $\delta \sim c/(\omega\sigma_{\text{ef}})^{1/2}$ is the skin depth. However, for JPWs in layered superconductors, contrary to usual conductors, the characteristic length of the magnetic field variations along the $z$-direction is much smaller than the skin depth $\delta$. It is not defined by the effective conductivity but is governed by the plasma wave spectrum. Specifically, due to very short spatial scale of the magnetic field change (about $D$), the amplitudes of the magnetic field and the $x$-component of electric field in the plasma wave are of the same order.

1. *Nonlocal sine-Gordon equation for the Josephson vortex*

In order to obtain the equation describing the phase difference generated by the Josephson vortex in the weak junction, we follow Ref. 69. First, using Eqs. (201), (202), and (204), we express the current normal to the layers in terms of the phase difference $\phi$ in the weak junction. Performing a reverse Fourier transform, we obtain

$$J_z^{l+1,l}(\zeta) = -\frac{ic\Phi_0 D}{16\pi^2 \lambda_{ab}^2}$$

$$\times \int_{-\infty}^{\infty} \frac{q^2 dq}{2\pi} \frac{\exp(iq\zeta + i\,\text{sign}(q)\, k(q) D |l|)}{1 - \exp(-ik(q)D)} \phi(q). \tag{206}$$

Below we consider the case where the main contribution to the integral in Eq. (206) comes from the region

$$q^2 \ll \omega_J^2/V^2. \tag{207}$$

It will be shown below that this inequality is valid if the vortex velocity satisfies a certain limitation. In this region of $q$, the value of $k(q)D$ is much less than unity, and we obtain from Eq. (198) the asymptotic expression for $k(q)$,

$$k(q) = \frac{1}{\lambda_{ab}} \left( \frac{\omega_J^2 + c^2 q^2/\varepsilon}{q^2 V^2 - \omega_J^2} \right)^{1/2}. \tag{208}$$

Substituting Eq. (208) into Eq. (206), we find the current $J_y$ through the weak junction, $l = 0$. Using the reverse Fourier transform for $\phi$ and taking into account the inequality (207), one gets

$$J_z^{1,0}(\zeta) = \frac{c\Phi_0}{16\pi^3 \lambda_{ab} \lambda_c} \int_{-\infty}^{\infty} d\zeta' K_0\left(\frac{|\zeta' - \zeta|}{\lambda_c}\right) \frac{\partial^2 \phi}{\partial \zeta'^2}, \tag{209}$$

where $K_0(x)$ is the modified Bessel function. Equating the current Eq. (209) to the sum of Josephson and displacement currents in the weak junction, we obtain the nonlocal sine-Gordon equation for the Josephson vortex in a layered superconductor,

$$\frac{V^2}{\omega_J^{*2}} \frac{\partial^2 \phi}{\partial \zeta^2} + \sin\phi = \frac{\lambda_J^{*2}}{\pi \lambda_c} \int_{-\infty}^{\infty} d\zeta' K_0\left(\frac{|\zeta' - \zeta|}{\lambda_c}\right) \frac{\partial^2 \phi}{\partial \zeta'^2}, \tag{210}$$

where

$$\lambda_J^* = \sqrt{\frac{c\Phi_0}{16\pi^2 \lambda_{ab} J_c^*}}, \qquad \omega_J^* = \sqrt{\frac{8\pi e D^* J_c^*}{\hbar \varepsilon}}. \tag{211}$$

The physical reason for the nonlocal structure of this equation results from the fact that the magnetic flux of the vortex is distributed along the layers over distances $\lambda_c$ much higher than the region of the nonlinearity, $\lambda_J^2/\pi\lambda_c \approx \gamma D$, (the JV core). Therefore, the component Eq. (209) of the current normal to the layers and inflowing into the weak junction at the point $\zeta$ "feels" the phase difference $\phi$ not only at the same point but is defined by the phase distribution over a region of about $\lambda_c$ around the point $\zeta$. A similar situation was considered in [69] for single Josephson junctions with high critical currents, when $\lambda_J$ is much less than the London penetration depth.

Now we use Eq. (210) in order to estimate the maximum velocity of the Josephson vortex in the weak junction and to analyze the soliton solution. If $\lambda_c \ll \lambda_J^*$, the kernel $K_0$ in the integral in Eq. (210) is a sharper function of $\zeta'$ than $\partial^2\phi/\partial\zeta'^2$. In this case, one can take away $\partial^2\phi/\partial\zeta'^2$ from the integral at the point $\zeta' = \zeta$. This reduces Eq. (210) to the usual local sine-Gordon equation. However, the $c$-axis London penetration depth $\lambda_c$ for layered superconductors is usually much higher than $\lambda_J$ [3], i.e.,

$$\lambda_c \gg \lambda_J^*. \tag{212}$$

So, below we consider only this case.

Following the standard procedure, we multiply both parts of Eq. (210) by $\partial\phi/\partial\zeta$ and integrate over $\zeta$ from $-\infty$ to zero,

$$\frac{V^2}{2\omega_J^{*2}}(\phi'(0))^2 + 2$$

$$= \frac{\lambda_J^{*2}}{\pi\lambda_c}\int_{-\infty}^{\infty}d\zeta'\int_{-\infty}^{0}d\zeta K_0\left(\frac{|\zeta-\zeta'|}{\lambda_c}\right)\frac{\partial^2\phi}{\partial\zeta'^2}\frac{\partial\phi}{\partial\zeta}. \quad (213)$$

The function $\phi(\zeta)$ varies on a scale $l$, that is much less than $\lambda_c$. Therefore, the main contribution to the integral in Eq. (213) comes from $|\zeta|$, $|\zeta'| \ll \lambda_c$. This enables us to replace $K_0(x)$ by its expansion at small arguments, $K_0(x) = -\ln x$. Integrating by parts over $\zeta'$ in the right hand side of Eq. (213), we find

$$\frac{V^2}{2\omega_J^{*2}}(\phi'(0))^2 + 2 = \frac{\lambda_J^{*2}}{\pi\lambda_c}\mathcal{P}\int_{-\infty}^{\infty}d\zeta'\frac{\partial\phi}{\partial\zeta'}\int_{-\infty}^{0}d\zeta\frac{1}{\zeta'-\zeta}\frac{\partial\phi}{\partial\zeta}, \quad (214)$$

where $\mathcal{P}$ stands for the principal value of the integral. We assume a simple estimation, $\phi'(0) = 1/l$. Taking into account that $\phi'(\zeta)$ is a sharp function, we can put $\phi'(\zeta') = 2\pi\delta(\zeta')$ in the internal integral in Eq. (214). Estimating the remaining integral over $\zeta$ as $1/l$, we obtain the approximate algebraic equation for the characteristic size of the moving soliton,

$$l^2 - l_0 l + \frac{V^2}{4\omega_J^{*2}} = 0, \qquad l_0 = \frac{\lambda_J^{*2}}{\lambda_c}. \quad (215)$$

Obviously, the soliton size at $V = 0$ is $l_0$. This result coincides with the exact one obtained by Gurevich for isotropic superconductors in the nonlocal mode [69].

The real roots of the quadratic equation (215) exist if the vortex velocity does not exceed the critical value

$$V_c^* = \omega_J^* l_0 = \frac{\omega_J^* \lambda_J^{*2}}{\lambda_c} = c_{\text{sw}}^* \frac{\lambda_J^*}{\lambda_c} \ll c_{\text{sw}}^*, \quad (216)$$

where $c_{\text{sw}}^*$ is the Swihart velocity for the weak junction. We emphasize that the last result is valid not only for the weak junction but for any other Josephson junction in the layered medium. The limiting velocity $V_c$ of the Josephson vortex in layered superconductors is much less than the Swihart velocity, $V_c = c_{\text{sw}}\lambda_J/\lambda_c \ll c_{\text{sw}}$, due to the nonlocal effects. Note that the limitation $V < c_{\text{sw}}$ was found in the paper by Krasnov [67], however, he considered the local limit, $\lambda_c \ll \lambda_J$.

The characteristic size of the soliton decreases with an increase of $V$, as in the local case. However, the value of $l$ remains non-zero even at $V = V_c^*$. As was assumed before, $l$ is always much less than $\lambda_c$ if the inequality (212) is fulfilled.

Now we can clarify the physical meaning of the limitation (207). Since the integral in Eq. (206) defines the soliton structure, the main contribution to it comes from $q \sim l^{-1}$. Thus, we find the validity conditions of our results in the form:

$$V^2 \ll \omega_J^2 l^2 = V_c^{*2} \frac{J_c D}{J_c^* D^*}. \tag{217}$$

If the critical current in the weak junction is much less than in other ones, the last inequality is valid at any vortex velocity up to $V = V_c^*$.

### 2. Cherenkov out-of-plane radiation

The magnetic field of the Josephson plasma wave running away from the weak junction can be found by means of the reverse Fourier transformation of Eq. (202). From Eqs. (202) and (204) we get

$$H^{l+1,l}(x,t) = i\mathcal{H}_0 \frac{D^2 \lambda_c}{4\pi \lambda_{ab}^2} \int_{-\infty}^{\infty} \frac{q\,dq\,\phi(q)}{1 - \exp(-ik(q)D)}$$
$$\times \exp[iq(x - vt) + i\,\mathrm{sign}(q)\,k(q)D|l|], \tag{218}$$

where $\phi(q)$ is the Fourier transform of the solution of Eq. (210). The interval $|q| < q_{\min}$ of integration in Eq. (218) corresponds to the complex values of $k(q)$. The contribution to the integral coming from this region decays from the weak junction. The out-of-plane Cherenkov radiation is provided by the contribution coming from the region $|q| > q_{\min}$. We write this term in the form,

$$H_{\mathrm{Ch}}^{l+1,l}(x,t) = i\mathcal{H}_0 \frac{D^2 \lambda_c}{4\pi \lambda_{ab}^2} \int_{q_{\min}}^{\infty} \frac{q\,dq}{1 - \exp(-ik(q)D)}$$
$$\times \{\phi(q) \exp[iq(x - vt) + ik(q)D|l|]$$
$$-\phi(-q) \exp[-iq(x - vt) - ik(q)D|l|]\}. \tag{219}$$

Here we stress again that the emitted wave can exist only in the case when the vortex speed $V$ is higher than the minimal value $V_{\min}$ (see Eq. (200)). This fact could be readily seen from Eqs. (219), (200). Indeed, the interval of integration in Eq. (219) diminishes at $V \to V_{\min}$. Thus, using Eqs. (199), (216), and (217), we can write down the conditions for the out-of-plane Cherenkov radiation as

$$\frac{J_c^* D}{J_c D^*} < \frac{V^2}{V_c^{*2}} < \frac{J_c D}{J_c^* D^*}. \tag{220}$$

These conditions can be readily satisfied if there exists a weak junction with the critical current $J_c^* \ll J_c$.

Unfortunately, we cannot find the explicit solution of Eq. (210) for the soliton profile $\phi(\zeta)$. As an example, let us now use the profile of a stationary vortex [69],

$$\phi(\zeta) = \pi + 2\tan^{-1}[\zeta/l(V)], \tag{221}$$

with $l$ dependent on $V$ according to Eq. (215),

$$l(V) = \frac{l_0}{2}\left(1 + \sqrt{1 - \frac{V^2}{V_c^{*2}}}\right). \tag{222}$$

The Fourier transform of (221) is

$$\phi(q) = -\frac{2\pi i}{q}\exp[-|q|l(V)]. \tag{223}$$

Substituting this expression into Eq. (219) we find

$$H_{\text{Ch}}^{l+1,l}(x,t) = i\mathcal{H}_0 \frac{D^2 \lambda_c}{\lambda_{ab}^2} \int_{q_{\min}}^{\infty} \frac{dq}{1 - \exp[-ik(q)D]}$$
$$\times \exp[-ql(V)]\sin[q(x-Vt) + k(q)D|l|]. \tag{224}$$

Due to the rather unusual dispersion relation (20), i.e., the decrease of $k(q, \omega = qV)$ when increasing $q$, as well as due to the spatial extension of a vortex, the generated electromagnetic waves are located *outside* the Cherenkov cone (Fig. g-1), which is drastically different from the Cherenkov radiation of a fast (point-like) relativistic particle. The new type of radiation predicted here could be called *outside-the-cone* Cherenkov radiation.

The frequency spectrum of the Cherenkov radiation has a sharp edge at $\omega = q_{\min}V \approx \omega_J$. At lower frequencies the radiation is absent, while it exponentially decreases at higher frequencies due to the exponential factor $\exp[-|q|l(V)]$ in Eq. (224). Thus, the moving vortex emits the out-of-plane radiation mainly at the frequency $\omega \approx \omega_J$. Note that the Cherenkov radiation propagates mainly in the direction transverse to the layers, since $k(q_{\min}) = \pi/D \gg q_{\min}$.

The Cherenkov radiation is exponentially small if $V \ll V_c^*$, and grows with $V$ approaching $V_c^*$. If the vortex velocity becomes comparable with the critical value $V_c^*$, the exponential factor in the integral Eq. (224) is about $\exp[-(J_c/J_c^*)^{1/2}]$.

As was mentioned above, the ratio $E_x/H$ is about unity for the emitted waves. This means that contrary to the longitudinal propagation of Cherenkov radiation, where the ratio $E_z/H \sim V/c$ is very small, *there is no the impedance mismatch for the out-of-plane radiation.* However, the out-of-plane Cherenkov radiation can never pass through the sample boundary that is parallel to the velocity of moving particle. This is because of the large longitudinal wave vector $q = \omega/V \gg \omega/c$ of radiation. As was shown in Refs. 70–72, the longitudinal wave vector can be decreased by means of the spatial modulations of the critical interlayer current, resulting in emission of the Cherenkov radiation from the top and bottom sample surfaces.

### C. Transition radiation of Josephson plasma waves

In the previous section we have shown that the vortex moving in a weak junction produces the out-of-plane Cherenkov radiation if the vortex velocity is high enough. However, the out-of-plane radiation can be emitted even at small vortex velocity if the junction, where the vortex moves, is non-uniform [72]. In this case, it is not necessary to assume that this junction differs significantly from others. *We only assume that the dielectric constant $\varepsilon$ (or the critical current $J_c$) of the junction is periodically modulated. This is an analogue of the transition radiation.* We predicted this in Ref. 72.

It is convenient to present the modulated dielectric constant in the form,

$$\varepsilon(x) = \varepsilon \left(1 + \mu \sum_{n=-\infty}^{\infty} F(x - na)\right), \tag{225}$$

where

$$F(\xi) = \begin{cases} f(\xi), & |\xi| < a/2; \\ 0, & |\xi| > a/2. \end{cases} \tag{226}$$

For simplicity, we assume that $f(\xi)$ is an even function with a maximum $f(0) = 1$. We consider the case $\mu \ll 1$, in order to derive an explicit formula for radiation.

Contrary to the previous studies, the problem under consideration does not have a travelling-wave solution since the sample is not homogeneous along the vortex motion. Performing the Fourier transformation of Eq. (12) over $t$ and $x$ and neglecting the dissipation term, we obtain

$$\left(1 + \frac{q^2 \lambda_c^2}{1 - \omega^2/\omega_J^2}\right) \varphi^{l+1,l} - \frac{\lambda_{ab}^2}{D^2} \partial_l^2 \varphi^{l+1,l} = 0. \tag{227}$$

Following the procedure used in Subsection IV B, we obtain the modified expression for the

Fourier transform of the magnetic field,

$$H^{l+1,l}(\omega,q) = H^{1,0}(\omega,q) \cdot \exp[i\,\text{sign}(\omega)\,k(\omega,q)D|l|] \tag{228}$$

with

$$\sin^2\left(\frac{k(q)D}{2}\right) \approx \frac{D^2}{4\lambda_{ab}^2} \frac{\omega_J^2 + c^2 q^2/\varepsilon}{\omega^2 - \omega_J^2}. \tag{229}$$

The relation between the magnetic field $H^{1,0}(\omega,q)$ and the phase difference $\phi(\omega,q)$ in the weak junction is given by Eq. (204), where $k(q)$ should be replaced by $k(\omega,q)$ from Eq. (229). The equation for the phase difference $\phi(x,y,t)$ now has a form

$$\frac{\varepsilon(x)}{\varepsilon\omega_J^{*2}}\frac{\partial^2\phi}{\partial t^2} + \sin\phi = \frac{\lambda_J^{*2}}{\pi\lambda_c}\int_{-\infty}^{\infty} dx' K_0\left(\frac{|x'-x|}{\lambda_c}\right)\frac{\partial^2\phi}{\partial x'^2} \tag{230}$$

with $\omega_J^*$ taken for the unperturbed case. This equation can be solved by an iterative procedure with respect to $\mu \ll 1$. We seek a solution of Eq. (230) as a sum of the travelling wave $\phi(x-Vt)$ for the unperturbed soliton moving with the constant velocity $V$ and a small perturbation $\psi(x,t)$. The function $\phi(x-Vt)$ is the solution of Eq. (210). To first order approximation in $\psi$, we have

$$\frac{1}{\omega_J^{*2}}\frac{\partial^2\psi}{\partial t^2} + \frac{\mu}{\omega_J^{*2}}\sum_{n=-\infty}^{\infty} F(x-na)\frac{\partial^2\phi}{\partial t^2} + \psi\cos\phi$$

$$= \frac{\lambda_J^{*2}}{\pi\lambda_c}\int_{-\infty}^{\infty} dx' K_0\left(\frac{|x'-x|}{\lambda_c}\right)\frac{\partial^2\psi}{\partial x'^2}. \tag{231}$$

We can neglect the last term in the left hand side of Eq. (231) since we are interested in the high frequency radiation with $\omega \gg \omega_J^*$. Solving Eq. (231) by means of a Fourier transform and neglecting terms of the order of $1/q^2\lambda_c^2 \ll 1$ (corresponding to high frequency and strongly nonlocal modes) we obtain $\psi(\omega,q)$

$$\psi(\omega,q) = \frac{\mu\omega^2}{|q|V_c^*\omega_J^* - \omega^2}$$

$$\times \int_{-\infty}^{\infty} dx \sum_{n=-\infty}^{\infty} F(x-na)\phi(\omega,x)\exp(-iqx), \tag{232}$$

where $\phi(\omega,x)$ is the Fourier transform of $\phi(x-Vt)$ from Eq. (221) over time:

$$\phi(\omega,x) = -\frac{2\pi i}{\omega}\exp\left(\frac{i\omega x - |\omega|l(V)}{V}\right). \tag{233}$$

Now using Eqs. (226), (228), (229), (232), (233), and (204), where $k(q)$ is replaced by $k(\omega, q)$, we obtain, after cumbersome but straightforward algebra, the expression for the magnetic field of the transition radiation,

$$H_{\text{tr}}^{l+1,l} = \frac{i\mu \mathcal{H}_0 D^2 \lambda_c}{\lambda_{ab}^2 a}$$

$$\times \sum_{m=-\infty}^{\infty} \tilde{f}_m \int_{\omega_J}^{\infty} d\omega \frac{\omega q_m}{1 - \exp(-ik(\omega, q_m)D)}$$

$$\times \frac{\exp(-|\omega|l(V)/V)}{|q_m| V_c^* \omega_J^* - \omega^2} \sin[q_m x + k(\omega, q_m)D|l| - \omega t]. \tag{234}$$

Here

$$\tilde{f}_m = 2 \int_0^{a/2} dx f(x) \cos\left(\frac{2\pi mx}{a}\right), \tag{235}$$

$$q_m = \frac{\omega}{V} - \frac{2\pi m}{a}. \tag{236}$$

We also take into account that only real values of $k(\omega, q_m)$ correspond to the emitted waves. Thus the integration in Eq. (234) is performed from $\omega_J$. Additional limitations on the summation and integration regions should be imposed since $\sin^2(k(\omega, q_m)D/2) < 1$ in Eq. (229).

In the general case, the analysis of Eq. (234) is rather cumbersome. So, below we only discuss the case of low vortex velocities, $V < V_{\min}$ (see Eq. (199)). The condition that the right-hand side in Eq. (229) cannot exceed one produces a quadratic inequality for the value of $\omega$, as it follows from Eq. (236). At $v = V/V_{\min} < 1$, the allowed frequencies lie between the values $\omega_{1,2}(m)$, which are defined by the roots of the corresponding quadratic equation,

$$\omega_{1,2}(m) = \frac{2\pi m V_{\min}}{a} \frac{v^2}{1 - v^2}$$

$$\times \left[\frac{1}{v} \pm \sqrt{1 - \omega_J^2 \left(\frac{a}{2\pi m V_{\min}}\right)^2 \frac{1 - v^2}{v^2}}\right]. \tag{237}$$

Evidently, we should choose only $m > m_{\min}$,

$$m_{\min} = \frac{a\omega_J}{2\pi V_{\min}} \frac{\sqrt{1 - v^2}}{v}, \tag{238}$$

since the values of $\omega$ are real and positive. A simple analysis shows that both frequencies $\omega_{1,2}(m)$ are higher than $\omega_J$ if $m > m_{\min}$. Thus, we should keep only the terms with $m > m_{\min}$ in the sum in Eq. (234) and perform integration from $\omega_1$ to $\omega_2$,

$$H_{\text{tr}}^{l+1,l} = \frac{i\mu \mathcal{H}_0 D^2 \lambda_c}{\lambda_{ab}^2 a}$$

$$\times \sum_{m=[m_{\min}]+1}^{\infty} \tilde{f}_m \int_{\omega_1(m)}^{\omega_2(m)} d\omega \frac{\omega q_m}{1 - \exp(-ik(\omega, q_m)D)}$$

$$\times \frac{\exp\left(-|\omega|l(V)/V\right)}{|q_m|V_c^* \omega_J^* - \omega^2} \sin\left[q_m x + k(\omega, q_m)D|l| - \omega t\right], \tag{239}$$

where $[m_{\min}]$ denotes the integer part of $m_{\min}$.

In contrast to the outside-the-cone Cherenkov-like radiation (Fig. 27), the transition radiation (Fig. 28a) propagates both *forward and backward* in space. Indeed, the in-plane component $k_x$ of the wave vector and, thus, the corresponding phase velocity both change sign (Fig. 28b,c). Also, the waves running backward can be directly seen from the magnetic field distribution shown in Fig. 28a. For fast Josephson vortices moving with velocity close to $V_{\min}$, the frequency zones overlap for different zone number $m$ (Fig. 28b). However, for slower speeds, we obtain the forbidden frequency ranges $\omega_{\max}(m) < \omega < \omega_{\min}(m+1)$ (see, Fig. 28c) of radiated electromagnetic waves.

The radiation is more intensive for $m_{\min} \leq 1$ since the value of $\tilde{f}_m$ in Eq. (239) drops with increasing $m$. This means that the modulation period $a$ should not be too large. On the other hand, the radiation intensity decays if the modulation period is small compared with the soliton size $l$. Indeed, the exponential factor in Eq. (239) decreases sharply at $a \ll l$. Thus the optimal value $a$ is of the order of $l_0$ since $l(V)$ does not change drastically with the fluxon velocity $V$. In this case, the minimal frequency $\omega_1$ of the emitted radiation is about $\omega_J$.

Different frequency bands $\Delta\omega_m = \omega_2(m) - \omega_1(m)$ correspond to each $m$th term in the sum in Eq. (239). The neighboring bands can be divided by the gaps of forbidden frequencies under appropriate choice of parameters. For example, at $a = 2\pi V_{\min}/\omega_J$ and $v^2 = 1/2$, when $a$ is about $l_0$ and $m_{\min} = 1$, the gap $(\sqrt{3} - \sqrt{2})\omega_J$ between the first and second bands exists.

Note that the transition radiation could arise for junction inhomogeneities of different origin, e.g., for junctions with a modulated critical current density $J_c(x)$ [72].

## V.  SUPERRADIANCE OF THZ WAVES IN LAYERED SUPERCONDUCTORS

The possibility of generating Cherenkov and transition radiations produced by a moving *single* Josephson vortex promises, at first glance, a way for layered superconductors to generate THz electromagnetic waves. However, several attempts to produce Cherenkov radiation in layered superconductors by a moving *lattice* of Josephson vortices [58–62] has uncovered an important problem that is very difficult to solve. For example, the far-field radiation power obtained experimentally from BSCCO is limited to the pW range [137], which is very small. To generate radiation having a higher intensity, JPWs induced by the moving lattice have to be in phase in different layers, which can be realized only if vortices form a *rectangular lattice*. However, such a rectangular lattice is usually unstable [53, 138, 139]. In other words, the vortex-vortex interaction favors the triangular lattice that can produce only a weak non-coherent radiation with the intensity proportional to the total number $N$ of layers in a superconductor. Thus, the problem of coherent superradiance in layered superconductors, with an intensity proportional to $N^2$, is of importance. In this subsection, we describe a way to achieve superradiance, recently reported in Ref. 140 and also studied in Ref. 141. References 140, 141 considered the synchronization, by the radiated field, of the Josephson plasma oscillations for the simplest case when the dc magnetic field is not applied, i.e., when there are no vortices in a superconductor. Only JPWs themselves produce the in-plane gradients of the gauge-invariant phase difference $\varphi$. In this case, the **c**-axis bias-current works as a source of Josephson coherent radiation.

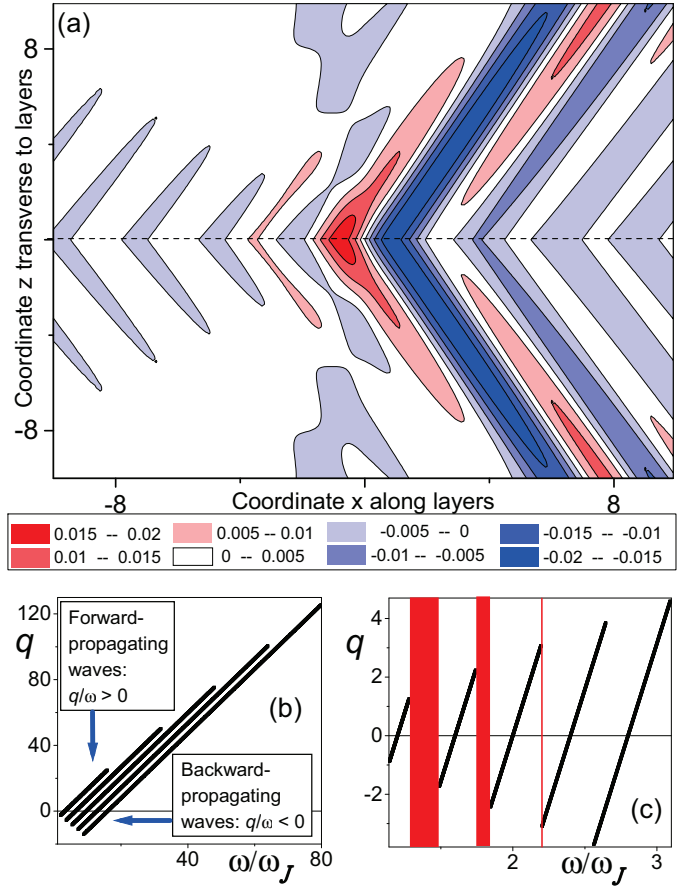

FIG. 28: (Color online) Transition radiation emitted by a Josephson vortex (located at $x = 0$) moving through a spatially modulated (along the **ab**-plane) layered superconducting sample (From Ref. 72). (a) Magnetic field distribution $H(x,z)$ (at a certain time, say $t = 0$) in units of $\mu \mathcal{H}_0 D^2 \lambda_c / \lambda_{ab}^2 a$ for $V/V_{\min} = 0.8$, $a/l(V=0) = 1$. The in-plane and out-of-plane coordinates $x$ and $z$ are normalized by the core size $l$ of a static Josephson vortex and $2D$, respectively. (b,c) The $x$-component $q$ of the wave vector of the radiation versus frequency. In contrast to the Cherenkov radiation (Fig. 27), the phase velocity $q/\omega$ of the transition radiation could be positive or negative (b), resulting in waves propagating both forward and backward with respect to the Josephson vortex motion. The radiation frequency has forbidden zones, shown by red strips in (c), when the vortex moves relatively slow.

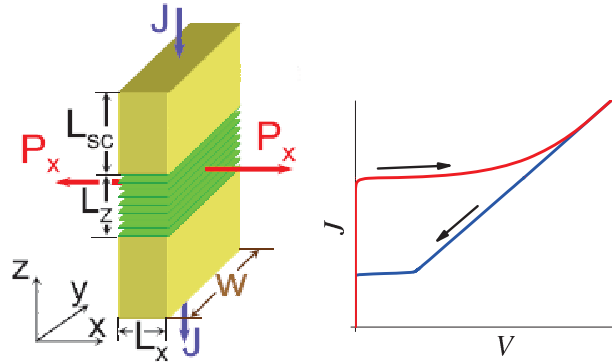

FIG. 29: (Color online) Left: Schematics of a layered superconductor placed in between leads serving as screens (From Ref. 141). The directions of the dc current $J$ and of the radiation Poynting vectors $P_x$ are shown. Right: Schematics of a hysteretic current-voltage characteristics of a SJJ.

### A. The main idea and experimental realization

The main idea of the experiment made in Ref. 140 is as follows. Consider the simple geometry shown in the left panel in Fig. 29: a rectangular SJJ with sizes $L_x \ll W$, $L_z$, $N = L_z/D \gg 1$, and current $J$ flowing in the $z$-direction. The external magnetic field is zero. When a SJJ is in the resistive state, the phase $\varphi$ oscillates at the Josephson frequency, $\omega_{\mathrm{osc}} = 2e\Delta V/\hbar$, where the voltage $\Delta V$ between the neighboring layers is induced by the DC current $J$ flowing across the sample. The phase oscillations excite the JPW with the same frequency. For identical junctions in the stack, the voltage $\Delta V$ is the same in all junctions (except, possibly, the top and bottom junctions in the SJJ) because the same current flows between all layers.

The current-voltage characteristics (CVC) of the Josephson system is characterized by a strong hysteresis (see the right panel in Fig. 29). Moving along the bias-decreasing branch of the CVC, the frequency $\omega_{\mathrm{osc}}$ can be tuned and, for a definite voltage, we attain the resonance conditions under which the high-amplitude standing JPW arises in the sample. The oscillations of the phase $\varphi$ in the SJJ are synchronized by this standing electromagnetic wave, as in a laser [140, 141]. That is, the higher amplitude of the JPW, the more stable is the superradiance against disturbances, which destroy the uniform oscillations in the SJJ.

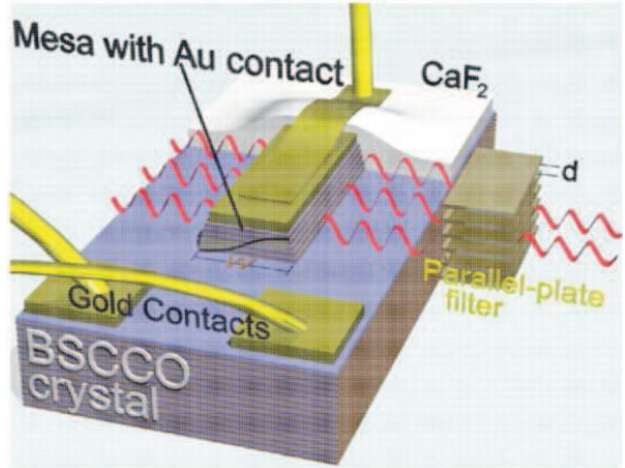

FIG. 30: (Color online) Schematic diagram of the BSCCO mesas (From Ref. 140). The applied **c**-axis current excites the fundamental cavity mode on the width $W$ of the mesa, and high-frequency electromagnetic radiation is emitted from the side faces (red waves), whose polarization and frequency are analyzed with parallel-plate filters.

In the experiment [140], a series of BSCCO samples in the form of mesas (Fig. 30) with sizes $L_x = 40$–$100$ $\mu$m, $W \approx 300$ $\mu$m, and $L_z \approx 1$ $\mu$m or $N > 500$ was used. A continuous radiation power, up to 0.5 $\mu$W at frequencies $f = \omega/2\pi$ up to 0.85 THz, was observed. The emission persisted up to temperatures around 50 K.

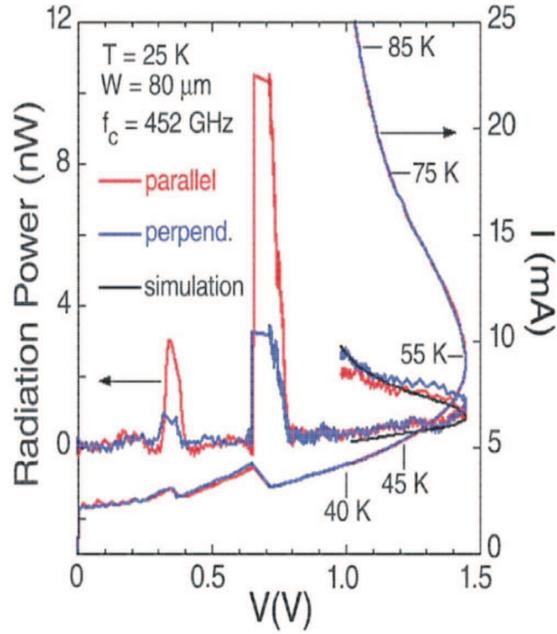

FIG. 31: (Color online) CVC and radiation power for a mesa structure with $L_x = 80$ $\mu$m (From Ref. 140). The voltage dependence of the current (shown in the right $y$-axis) and of the radiation power (left $y$-axis) at $T = 25$ K for parallel and perpendicular settings of the filter, with a cut-off frequency 0.452 THz, are shown for decreasing bias voltage. A polarized Josephson emission occurs near 0.71 and 0.37 V, and unpolarized thermal radiation occurs at higher biases.

Experimental results are shown in Figs. 31 and 32. The figures show the CVC and the radiation power as functions of decreasing bias voltage for the parallel and perpendicular settings of a parallel-plate cut-off filter. The radiation near the peaks is polarized with the electric field perpendicular to the $CuO_2$-planes, while unpolarized radiation was observed for high currents and high voltage biases. The former is identified as Josephson radiation, whereas the latter is thermal radiation. The estimated mesa temperatures $T$ along the CVC are also shown in Fig. 31. Figure 32 shows a significant change of the CVC in the vicinity of the radiation peaks.

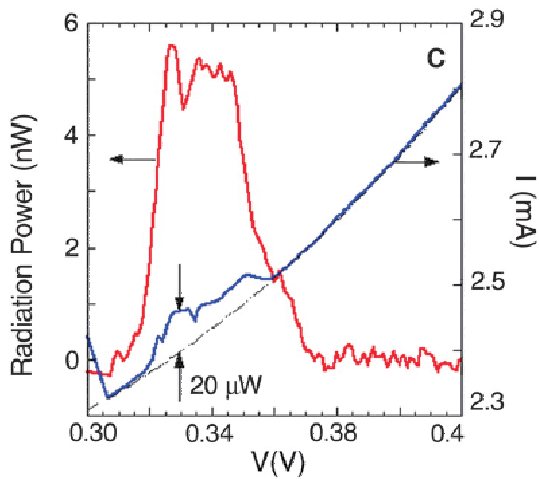

FIG. 32: (Color online) A more detailed plot of one of the emission peaks and the CVC for the mesa with $L_x = 80$ $\mu$m at $T = 20$ K (From Ref. 140).

## B. Theory of superradiance in the resistive mode

We consider here the simplest sample geometry shown in Fig. 29. We assume that the SJJ is bounded by superconducting contacts with the same lateral sizes as the stack, which extend in the $z$-direction over a distance $L_{\rm sc} \gg 1/k_\omega = c/\omega_{\rm osc}$. Such contacts serve as screens, restricting the radiation to the half-spaces $|x| > L_x/2$. The impedance of the contacts is negligibly small. We assume also that $L_x k_\omega, L_z k_\omega < 1$ and $W k_\omega \gg 1$, which is true for the experiment [140], and allows us to disregard the $y$- and $z$-coordinate dependence of the fields in the SJJ. In optimally-doped BSCCO, $\omega_J/2\pi \approx 0.15$ GHz [142], while in experiments [140], $\omega_{\rm osc}/2\pi \approx 0.5$–1 GHz. Thus, we consider here the case $\Omega = \omega_{\rm osc}/\omega_J \gg 1$. In general, the Josephson plasma frequency depends on the current $J$, but if $J$ is significantly lower than the critical value, this dependence is not of importance and we will use here the former expression (13) for $\omega_J$. For the geometry considered here, the electromagnetic fields in the vacuum have the components $\mathbf{H} = (0, H_y, 0)$, $\mathbf{E} = (E_x, 0, E_z)$, and only outgoing waves propagate, that is, $A \propto \exp{(iq|x| + ikz - i\omega t)}$ at $|x| > L_x/2$. Here $q = \sqrt{k_\omega^2 - k^2}$, when $k_\omega^2 > k^2$, and $q = i\sqrt{k^2 - k_\omega^2}$, when $k_\omega^2 < k^2$.

We now seek a uniform, in the $z$-direction, solution of Eq. (12). Normalizing the coordi-nate $x$ to $\lambda_c$ and time $t$ to $1/\omega_J$, we can rewrite Eq. (12) in the form,

$$\frac{\partial^2 \varphi}{\partial t^2} + r\frac{\partial \varphi}{\partial t} + \sin\varphi = \frac{\partial^2 \varphi}{\partial x^2}. \tag{240}$$

We denote the dimensionless SJJ width as $l_x = L_x/\lambda_c$ and assume that $l_x < 1$. We present the solution of the latter equation as a sum $\varphi_0 + \psi$, where $\varphi_0$ describes the Josephson oscillations in the SJJ due to applied DC current and a small term $\psi$ describes radiation, $|\psi| \ll 1$. The boundary conditions to Eq. (240) follow from the continuity of the $y$-component of the magnetic field and the relation between the magnetic field and the phase difference. For the geometry considered, the Maxwell equations and Eq. (15) give

$$H_y = \mathcal{H}_0 \frac{\partial \varphi}{\partial x}, \tag{241}$$

and, therefore,

$$\frac{\partial \varphi_0}{\partial x} = \pm \frac{H_I}{\mathcal{H}_0} \quad \text{at} \quad x = \pm \frac{l_x}{2}. \tag{242}$$

Here $H_I = 2\pi J L_x/c$ is the self-field produced by the DC current at the SJJ edges. We assume that $H_I/\mathcal{H}_0 = l_x(J/J_c) \ll 1$ since both $J/J_c$ and $l_x$ are smaller than one. For this case, we derive $\varphi_0 \approx \Omega t$.

We present the radiation term as follows:

$$\psi = \text{Re}\left[\psi_\omega(x) e^{-i\Omega t}\right]. \tag{243}$$

Taking into account that $\sin x = \text{Re}[i\exp(-ix)]$ and $\Omega \gg 1$, we derive from Eq. (240)

$$\frac{\partial^2 \psi_\omega}{\partial x^2} + \kappa_r^2 \psi_\omega = i, \tag{244}$$

with $\kappa_r = \sqrt{\Omega^2 + i\Omega r}$. A detailed derivation of the boundary conditions to the latter equation is presented in Refs. 141, 143. Under the assumptions formulated above, these conditions can be written in the form,

$$\psi'_\omega = \pm i\zeta \psi_\omega \quad \text{at} \quad x = \pm l_x/2, \tag{245}$$

where

$$\zeta = \frac{L_z \Omega}{2\varepsilon \lambda_c}\left[|\Omega| - \frac{2i\Omega}{\pi}\ln\frac{5.03\sqrt{\varepsilon}\lambda_c}{|\Omega|L_z}\right]. \tag{246}$$

For the BSCCO mesas with $N \lesssim 10^3$, $|\zeta|$ is a small parameter, $|\zeta| \ll 1$. The solution of Eq. (244) with boundary conditions (245) is

$$\psi_\omega = \psi_{\text{osc}} + \psi_{\text{rad}}$$

$$= \frac{i}{\kappa_r^2} + \frac{\zeta \cos(\kappa_r x)}{\kappa_r^2 \left(\kappa_r \sin \frac{\kappa_r l_x}{2} + i\zeta \cos \frac{\kappa_r l_x}{2}\right)}. \tag{247}$$

Here the term $\psi_{\mathrm{osc}}$ is the correction to the Josephson oscillations and $\psi_{\mathrm{rad}}$ describes electromagnetic waves propagating inside the layered superconductor [141]. They are generated at the boundaries $x = \pm l_x/2$ due to the radiation field.

Now we can calculate the $x$-component of the Poynting vector

$$\mathcal{P}_{\mathrm{rad}} = \frac{c}{4\pi} E_z H_y,$$

i.e., the radiation power from one side of the sample along the $x$-direction. Using Eqs. (15), (241), (243), (245), and (246), we obtain, according Refs. 141, 143,

$$\frac{1}{W}\mathcal{P}_{\mathrm{rad}}\left(\pm\frac{l_x}{2}\right) = \frac{\Phi_0^2 N^2 \omega_{\mathrm{osc}}^3}{64\pi^3 c^2} \left|\psi_\omega\left(\pm\frac{l_x}{2}\right)\right|^2. \tag{248}$$

In the case considered, of uniform junctions,

$$\mathcal{P}_{\mathrm{rad}}(l_x/2) = \mathcal{P}_{\mathrm{rad}}(-l_x/2) = \mathcal{P}_{\mathrm{rad}}.$$

The dimensional factor in Eq. (248) can be estimated as,

$$\frac{\Phi_0^2 N^2 \omega_{\mathrm{osc}}^3}{64\pi^3 c^2} = \frac{1}{W}\mathcal{P}_0 \approx 0.6\,\frac{\mathrm{W}}{\mathrm{cm}}, \tag{249}$$

for $N = 1000$ and $\omega_{\mathrm{osc}}/2\pi = 10^{12}$ Hz. The frequency dependence of the dimensionless radiation $|\psi_\omega|^2$ is shown in Fig. 33. This dependence has resonance peaks at $\Omega l_x \approx 2\pi n$ with amplitudes rapidly decreasing with $n$. Expanding the value $\psi_{\mathrm{rad}}$ in Eq. (247) near the resonance, we find the dependence of the THz radiation on the number of junctions in the SJJ and the thickness $L_x$, which can be conveniently presented in the form [141],

$$\frac{\mathcal{P}_{\mathrm{rad}}}{W} \approx \frac{\Phi_0^2 \omega_J^4 N^2}{64\pi^3 c^2 \omega_{\mathrm{osc}}} \mathcal{L}(a), \quad a = \frac{L_x}{L_z}, \tag{250}$$

$$\mathcal{L}(a) = \frac{a^2 \varepsilon^2}{(a\varepsilon + \mathcal{L}_\omega)^2 + 1}, \quad \mathcal{L}_\omega = \frac{2}{\pi}\ln\left(\frac{5.03c}{L_z \omega_{\mathrm{osc}}}\right).$$

One can see that the energy flux is proportional to $N^2$ for samples with $L_z \ll L_x \varepsilon$ that corresponds to the superradiance [141]. The function $\mathcal{P}(N)$ saturates for higher $N$, when $L_z \gg L_x \varepsilon$. When analyzing the dependence of the radiation on $L_x$, we should take into account that $\omega_{\mathrm{osc}} \propto 1/L_x$. Using Eq. (249) and the results shown in Fig. 33, we obtain for the highest peak $\mathcal{P}_{\mathrm{rad}} \sim 1\,\mu\mathrm{W}$, that is of the order of the experimentally observed value [140].

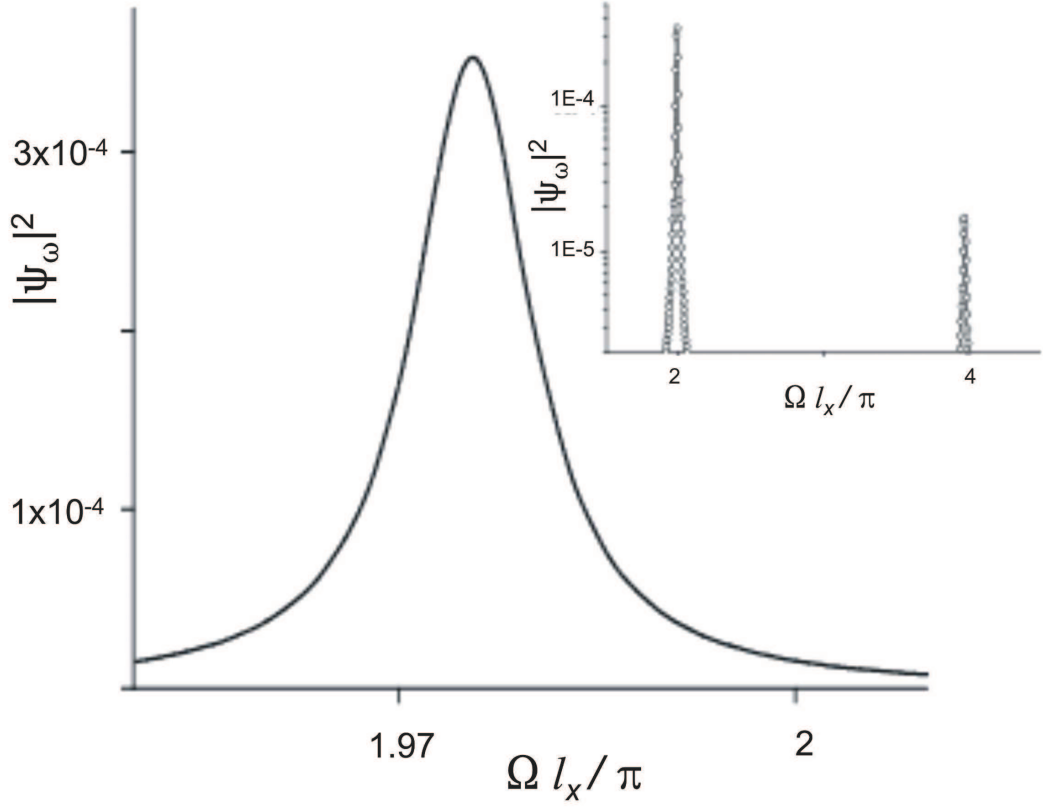

FIG. 33: Frequency dependence of the dimensionless radiation power for $D = 1.56$ nm, $\lambda_c = 100$ $\mu$m, $r = 2 \cdot 10^{-3}$, $\omega_{\mathrm{osc}}/2\pi = 10^{12}$ Hz, $N = 1000$, $\varepsilon = 12$, and $L_x = 0.5\,\lambda_c$; these values are characteristic of BSCCO mesas [140, 141]. Inset shows the results in a wider frequency range.

### C. The ways to increase the superradiance

Here we analyze methods to increase the radiated power proposed in Refs. 143 and 144. Both approaches exploit the same physical idea to increase the value of $|\psi_{\mathrm{rad}}|$ using the SJJ with modulated critical Josephson current. This modulation provides the excitation of the in-phase Fiske mode when the Josephson frequency matches the Fiske-resonance frequency, which is defined by the stack lateral size. Reference 143 propose the use of artificially-prepared inhomogeneous samples while the application of the moderate DC magnetic field is suggested in Ref. 144 for the same purpose.

### 1. Superradiance from inhomogeneous SJJ

Following Ref. 143, we assume that the critical Josephson current is modulated as $J_c(x) = g(x)J_c$ and the values $\lambda_c$ and $\omega_J$ are defined at the reference point, where $g(x) = 1$. In this case, we have instead of Eq. (240)

$$\frac{\partial^2 \varphi}{\partial t^2} + r\frac{\partial \varphi}{\partial t} + g(x)\sin\varphi = \frac{\partial^2 \varphi}{\partial x^2}. \tag{251}$$

Similarly to the approach used in the previous subsection, here we present the solution of this equation as a sum $\varphi = \varphi_0 + \psi$ and derive

$$\frac{\partial^2 \psi_\omega}{\partial x^2} + \kappa_r^2 \psi_\omega = ig(x), \tag{252}$$

where $\psi_\omega$ is defined by Eq. (243) and obeys the boundary conditions (245). In Ref. 143, different types of modulations are analyzed. Here we consider only the case of linear modulation, that is, $g(x) = 1 - 2g_0 x/l_x$. This means that the the critical current density $J_c$ at the left edge of the SJJ is larger by the factor $(1 + g_0)/(1 - g_0)$ than $J_c$ at the right edge.

The solution of Eq. (252) with the boundary conditions (245) can be found in the explicit form [143]. It is convenient to present it as a sum of symmetric and antisymmetric parts, $\psi_\omega = \psi^s + \psi^a$, where

$$\psi^s = \frac{i}{\kappa_r^2} + \frac{\zeta \cos(\kappa_r x)}{\kappa_r^2 \left[\kappa_r \sin\left(\frac{\kappa_r l_x}{2}\right) + i\zeta \cos\left(\frac{\kappa_r l_x}{2}\right)\right]}, \tag{253}$$

$$\psi^a = \frac{ig_0(2x/l_x - 1)}{\kappa_r^2}$$

$$+ \frac{g_0(2i/l_x + \zeta)\sin(\kappa_r x)}{\kappa_r^2 \left[\kappa_r \cos\left(\frac{\kappa_r l_x}{2}\right) - i\zeta \sin\left(\frac{\kappa_r l_x}{2}\right)\right]}. \tag{254}$$

Comparing this solution with Eq. (247) for the uniform SJJ, we find that $\psi^s$ coincides with $\psi$ in the sample without modulations. The antisymmetric term $\psi^a$ has resonance peaks with approximately twice lower frequencies, $\Omega l_x \approx \pi n$, and the height of the first peak can be much larger than in the uniform case if $g_0 \sim 1$ (since $|\zeta| \ll 1$). In the case of symmetric (say, parabolic) modulation, a higher peak has the symmetric part of $\psi$ and the position of this peak is close to that in the case of homogeneous SJJ [143].

The radiation power can be calculated using Eq. (248). The dependence of the dimensionless radiation power $|\psi_\omega(l_x/2)|^2$ on the frequency is shown in Fig. 34. The value $|\psi_\omega(-l_x/2)|^2$

is slightly shifted in frequency compared to $|\psi_\omega(l_x/2)|^2$ [143]. Thus, the radiation power can be increased by 3–4 orders of magnitude by modulating the critical current in the SJJ by a factor of about two. Note that the appearance of a peak at $\Omega l_x \approx 2\pi$ is practically independent of the modulation. We should emphasized that the results presented here are valid if $\Omega \gg 1$, $l_x < 1$, and $|\psi_\omega(l_x/2)|^2 < 1$.

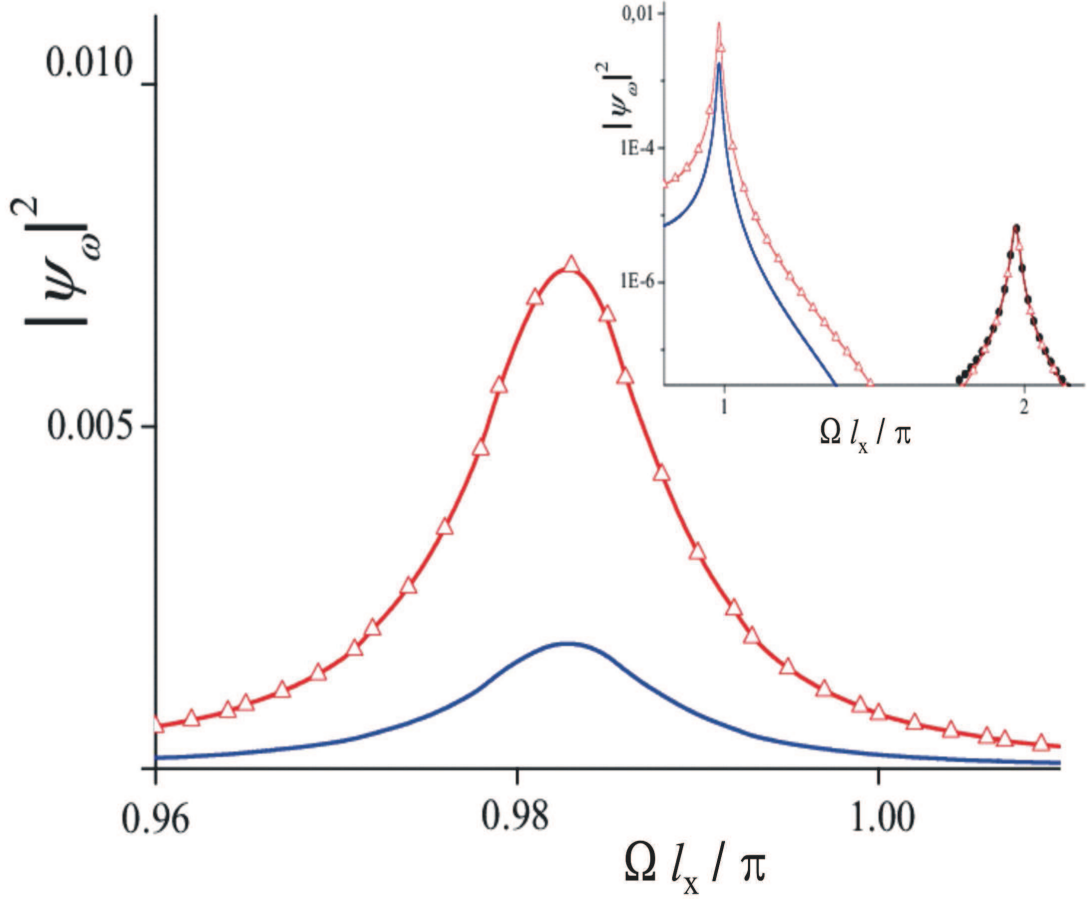

FIG. 34: (Color online) Frequency dependence of the dimensionless radiation power for $D = 1.56$ nm, $\lambda_c = 100$ $\mu$m, $r = 2 \cdot 10^{-3}$, $\omega_{\text{osc}}/2\pi = 10^{12}$ Hz, $N = 1000$, $\varepsilon = 12$, and $L_x = 0.2\lambda_c$. Red curve with triangles and blue solid curve correspond to the current-modulation parameters $g_0 = 1/3$ and $g_0 = 1/6$, respectively. The inset shows the results in a wider frequency range. Black circles correspond to $g_0 = 0$.

## 2. Effect of a DC magnetic field

As mentioned before, at the beginning of this Section, the moving triangular lattice of the Josephson vortices destroys the synchronization of the JPWs in different junctions, and the superradiance in the SJJ disappears. However, the SJJ size is usually smaller than $\lambda_c$, and the lower critical magnetic field for such objects is known to be much larger than for samples with large dimensions, $L_x \gg \lambda_c$. As a result, a DC magnetic field, $H_{\rm DC}$, applied in the $y$-direction, can produce a considerable modulation of the critical current before entering the vortex lattice into the sample. Here we study such a way to increase the superradiance [144]. As before, we seek the solution of Eq. (240) as a sum

$$\varphi = \varphi_0 + \psi, \quad |\psi|^2 \ll 1,$$

but now, the boundary conditions (242) for $\varphi_0$ should be replaced by

$$\frac{\partial \varphi_0}{\partial x} = h \pm \frac{H_I}{\mathcal{H}_0}, \qquad x = \pm \frac{l_x}{2}, \tag{255}$$

where $h = H_{\rm DC}/\mathcal{H}_0$. As in previous subsections, we assume that $L_x < \lambda_c$, $\Omega > 1$, and $H_I/\mathcal{H}_0 \ll 1$. This allows us to neglect the self-field effect and present $\varphi$ in the form

$$\varphi = \Omega t + hx + \psi,$$

where $\psi$ obeys the boundary conditions (245). Using the definition (243), we find that $\psi_\omega$ satisfies Eq. (252) with $g(x) = \exp(-ihx)$. In this case, the solution $\psi$ can be obtained explicitly,

$$\psi_\omega = A_1 e^{i\kappa_r x} + A_2 e^{-i\kappa_r x} + \frac{i \exp(-ihx)}{\kappa_r^2 - h^2}, \tag{256}$$

with

$$A_1 = \frac{1}{Z}\left[-\left(h\kappa_r - \zeta^2\right)\sin\frac{(\kappa_r - h)l_x}{2}\right.$$
$$\left. + i\zeta(\kappa_r - h)\cos\frac{(\kappa_r - h)l_x}{2}\right],$$

$$A_2 = \frac{1}{Z}\left[\left(h\kappa_r + \zeta^2\right)\sin\frac{(\kappa_r + h)l_x}{2}\right.$$
$$\left. + i\zeta(\kappa_r + h)\cos\frac{(\kappa_r + h)l_x}{2}\right],$$

$$Z = 2i\left(\kappa_r^2 - h^2\right)\left[\kappa_r \cos\frac{\kappa_r l_x}{2} - i\zeta \sin\frac{\kappa_r l_x}{2}\right]$$

$$\times \left[\kappa_r \sin\frac{\kappa_r l_x}{2} + i\zeta \cos\frac{\kappa_r l_x}{2}\right].$$

This solution can be presented as a sum of symmetric and antisymmetric, with respect to $x = 0$, terms. A high-resonance peak at $\Omega l_x = \pi$ is related to the antisymmetric term. The radiation from the left and right edges of the sample are equal. The frequency dependence of the dimensionless radiation power $|\psi_\omega(l_x/2)|^2$ is shown in Fig. 35, for different applied magnetic fields. The characteristic value of the field $\mathcal{H}_0$ for BSCCO is about 20 Oe. As follows from the results shown in Fig. 35, the application of the DC field of the order of several Oe gives rise to a significant increase of the radiation power. Our results are only valid for $|\psi_\omega| \ll 1$. This means, in particular, that $hl_x < \pi$ or

$$\frac{hl_x}{\pi} = \frac{2DL_x H_{\mathrm{DC}}}{\Phi_0} \lesssim 1. \tag{257}$$

For the first resonance,

$$\Omega \approx \Omega^{\mathrm{res}} \approx \pi/l_x - 2\mathrm{Im}(\zeta)/\pi,$$

we derive

$$\psi_\omega\left(\pm\frac{l_x}{2}\right) = \frac{-4ihl_x^2 \cos(hl_x/2)}{(\pi^2 - h^2 l_x^2)(\pi r + 4\mathrm{Re}\zeta)}. \tag{258}$$

The peak height increases linearly with the magnetic field if $(hl_x/\pi)^2 \ll 1$. Comparing the maximum resonance powers for a uniform SJJ with and without a DC field, and an artificially modulated SJJ, we find that the application of $H_{\mathrm{DC}}$:

(i) enhances the output radiation power if

$$H_{\mathrm{DC}} > H_e = \frac{3\pi^2 L_z \mathcal{H}_0}{8\varepsilon L_x};$$

(ii) is more effective than the modulation of $J_c$ by a factor

$$g = \frac{1+g_0}{1-g_0} \quad \text{if} \quad H_{\mathrm{DC}} > H_g = 2\mathcal{H}_0 \frac{\lambda_c(g-1)}{L_x(g+1)}.$$

We can estimate $H_e \approx 0.4$ Oe, and $H_g \approx \mathcal{H}_0$ when $g = 1.3$.

For the two configurations considered here, to increase superradiance, the dependence of the radiation power on the number of junctions is illustrated in Fig. 36. The value $\mathcal{P}_{\mathrm{rad}}(N)$ increases with $N$ for low $N$, and saturates for higher $N$. The resonant feature of the radiation power becomes less pronounced with increasing $N$, due to increase of the radiation damping, $|\zeta|^2 \propto N^2$, which results in a saturation of the radiation power.

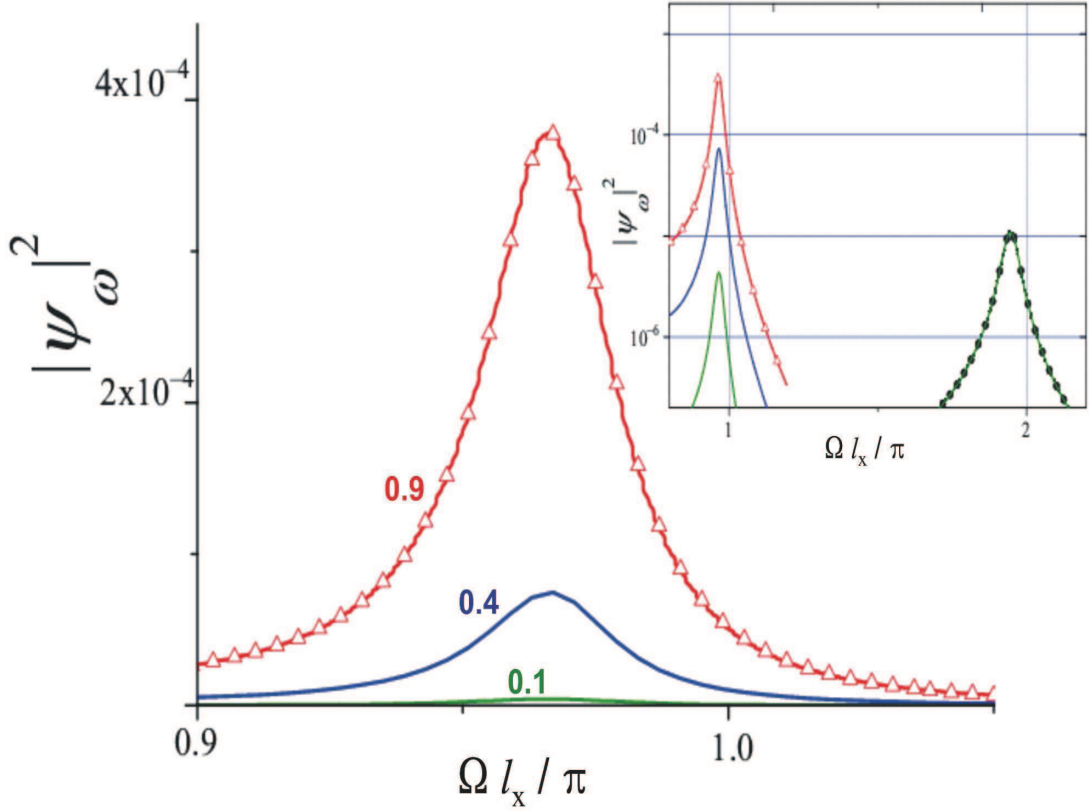

FIG. 35: (Color online) Dimensionless radiation power $|\psi_\omega(l/2)|^2$ versus $\Omega$, for different values of the DC magnetic field $h$ shown near the curves (From Ref. 144). Sample parameters are: $L = 0.25\lambda_c$, $\lambda_c = 100$ $\mu$m, $D = 1.56$ nm, $N = 3 \cdot 10^3$, $\varepsilon = 12$, and $r = 0.002$. Inset shows the results in a wider frequency range. Black circles correspond to $h = 0$.

### D. Stability of the superradiance

In previous parts of this section, we assumed that all contacts of the SJJ radiate in-phase JPWs, giving rise to superradiance. In this subsection, we analyze the stability of such a coherent THz emission with respect to $z$-dependent perturbations. This problem was considered in detail in Refs. 141, 144. Here we consider, as an example, the case when the radiation near the first resonance is increased due to an application of the DC magnetic field. This allows us to outline the general idea of the investigation of the stability and to find limitations on the value of the DC magnetic field.

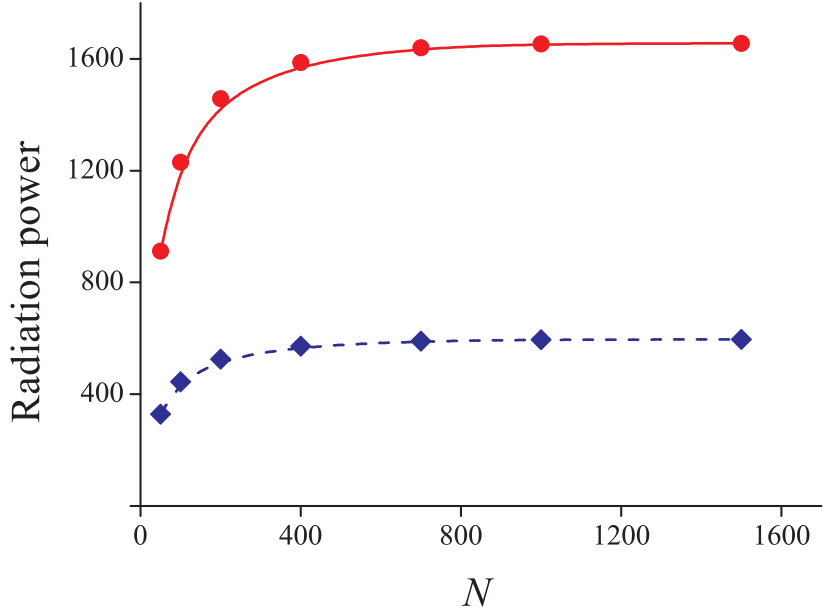

FIG. 36: (Color online) Dependence of the radiation power (in arbitrary units) on the number of junctions $N$: SJJ with linear critical current modulation, $g_0 = 1/12$, (red circles) and uniform SJJ in the DC field $h = 0.5$ (blue diamonds). Other parameters of the samples are the same as in Figs. 34 and 35.

We use here the coupled sine-Gordon equations in the form of Eq. (12) and present the phase difference $\varphi^{l+1,l}$ as

$$\varphi^{l+1,l} = \Omega t + hx + \psi + \psi^{l+1,l}. \qquad (259)$$

The first tree terms here describe a uniform, along the $z$-axis, solution, while $\psi^{l+1,l}$ is an infinitesimally small perturbation, which disturbs this uniformity. In the limit $\Omega \gg 1$, we neglect higher harmonics ($m\Omega$ with $m > 1$) and seek the perturbation $\psi^{l+1,l}$ in the form [141],

$$\psi^{l+1,l} = \sum_p \left[ \psi_p + \psi_p^+ e^{i\Omega t} + \psi_p^- e^{-i\Omega t} \right]$$

$$\times \sin(pl) \exp(-i\nu(p)t), \qquad (260)$$

where $p = \pi n/(N+1)$, $n = 1, 2, ..., N$, $|\nu(p)| \ll 1$. The inequality $\text{Im}(\nu(p)) < 0$ for all $p$ corresponds to the stability of the superradiance. We analyze the case of essentially non-uniform perturbations when $p \gg \pi/N$ and also assume that the number of junctions is high enough, $N \gg \lambda_{ab}/D$. Linearizing the sine-Gordon equations with respect to $\psi^{l+1,l}$, we derive an equation for the perturbation in the form,

$$\frac{\partial^2 \tilde{\psi}}{\partial x^2} = \left(1 + \tilde{p}^2\right)$$

$$\times \left(\frac{\partial^2 \tilde{\psi}}{\partial t^2} + r\frac{\partial \tilde{\psi}}{\partial t} + \cos\left(\Omega t + hx + \psi\right)\tilde{\psi}\right), \tag{261}$$

with $\tilde{p}^2 = 2(\lambda_{ab}/D)^2(1 - \cos p)$ and

$$\tilde{\psi} = \left[\psi_p + \psi_p^+ e^{i\Omega t} + \psi_p^- e^{-i\Omega t}\right] e^{-i\nu t}.$$

We present $\psi$ as $\psi = \psi_r \cos(\Omega t) + \psi_i \sin(\Omega t)$, where $\psi_r = \mathrm{Re}(\psi_\omega)$ and $\psi_i = \mathrm{Im}(\psi_\omega)$. Taking into account that $|\psi| \ll 1$, we obtain

$$\cos\left(\Omega t + hx + \psi\right) = \cos\left(\Omega t + hx\right)$$

$$- \frac{\psi_r}{2}\left[\sin(hx) + \sin(2\Omega t + hx)\right]$$

$$- \frac{\psi_i}{2}\left[\cos(hx) - \cos(2\Omega t + hx)\right].$$

Substituting this expression and Eq. (260) in Eq. (261), neglecting higher harmonics ($m\Omega$ with $m > 1$), and extracting terms with equal frequencies, we derive

$$\frac{1}{1+\tilde{p}^2}\frac{\partial^2 \psi_p^+}{\partial x^2} + F^*(\Omega - \nu)\psi_p^+ = \frac{\psi_p e^{ihx}}{2} - \frac{e^{ihx}}{4i}\psi_\omega^* \psi_p^-,$$

$$\frac{1}{1+\tilde{p}^2}\frac{\partial^2 \psi_p^-}{\partial x^2} + F(\Omega + \nu)\psi_p^- = \frac{\psi_p e^{-ihx}}{2} + \frac{e^{-ihx}}{4i}\psi_\omega \psi_p^+,$$

$$\frac{1}{1+\tilde{p}^2}\frac{\partial^2 \psi_p}{\partial x^2} + F(\nu)\psi_p = \frac{\psi_p^+ e^{-ihx}}{2} + \frac{\psi_p^- e^{ihx}}{2} \tag{262}$$

where

$$F(u) = u^2 + iru + \frac{\psi_r}{2}\sin(hx) + \frac{\psi_i}{2}\cos(hx), \tag{263}$$

and $f^*$ is the complex conjugate of $f$. The boundary conditions to these equations were found in Ref. 141. Using our notation, these can be written as

$$\frac{1}{\psi_p}\frac{\partial \psi_p}{\partial x} = \pm\chi(\nu), \quad \frac{1}{\psi_p^+}\frac{\partial \psi_p^+}{\partial x} = \pm\chi(\Omega - \nu),$$

$$\frac{1}{\psi_p^-}\frac{\partial \psi_p^-}{\partial x} = \pm\chi(\Omega + \nu) \quad \text{at} \quad x = \pm\frac{l_x}{2}, \tag{264}$$

$$\chi(\Omega) = \frac{\Omega^2 \left(1 + \tilde{p}^2\right) D}{\varepsilon p \lambda_c}. \tag{265}$$

These conditions are valid if $k_\omega L_z \ll 1$ and $p \gg 1/\pi N$. We can fix $\chi(\nu) = 0$, since $\nu$ and $D/\varepsilon p \lambda_c$ are small for the disturbances analyzed. We also neglect terms of the order of $|\psi_\omega|$, compared to $\Omega \gg 1$ in Eq. (262).

A characteristic spatial scale of the variation of $\psi_p$ is

$$\frac{1}{|\Omega G(\Omega)|} \approx \tilde{p}/\Omega \gg l_x, \quad G^2 = \Omega^2(1 + \tilde{p}^2),$$

as follows from Eq. (262). Thus, $\psi_q(x)$ is almost constant and we can find solutions of Eq. (262) explicitly. A numerical analysis of these solutions shows that they are practically independent of $h$, if $h^2 \ll \pi^2/l_x^2$. Neglecting $\nu \ll 1$ compared to $\Omega \gg 1$, we obtain

$$\psi_p^- \exp(ihx) \approx \psi_p V(x)$$

with

$$V = \frac{1}{2G^2} \left[ -\frac{\chi(\Omega) \cos Gx}{G \sin \frac{Gl_x}{2} + \chi(\Omega) \cos \frac{Gl_x}{2}} \right] \tag{266}$$

and a similar expression for $\psi_p^+$. Substituting these results into Eq. (262), we have

$$\frac{1}{1 + \tilde{p}^2} \frac{\partial}{\partial \psi_p x} + \left\{ \nu^2 + ir\nu + \frac{\psi_i \sin(hx) + \psi_r \cos(hx)}{2} \right.$$

$$\left. - \operatorname{Re}[V(x)] \right\} \psi_p = 0. \tag{267}$$

Integrating Eq. (267), with boundary conditions (264), we derive a dispersion equation for $\nu(p)$ in the form,

$$\nu^2 + ir\nu \approx -\left[W_1(\omega) + W_2(\omega)\right], \tag{268}$$

where

$$W_1 = \frac{1}{l_x} \operatorname{Im} \left[ A_1 \frac{\sin \frac{(\kappa_r + h)l_x}{2}}{\kappa_r + h} + A_2 \frac{\sin \frac{(\kappa_r - h)l_x}{2}}{\kappa_r - h} \right], \tag{269}$$

$$W_2 = \operatorname{Re} \left\{ \frac{(1 + \tilde{p}^2)\chi(\Omega)}{l_x \Omega^3 \left[ G + \chi(\Omega) \cot\left(\frac{Gl_x}{2}\right) \right]} \right\}. \tag{270}$$

Here we assume that $N \ll \varepsilon \lambda_c/s \approx 5 \cdot 10^5$. If the term $W_1$ is negative, it stabilizes the uniform oscillations due to radiation coupling. The term $W_2$ can result in an instability due to the excitation of the Fiske resonance [141]. Note that dissipation favors stability [141, 144].

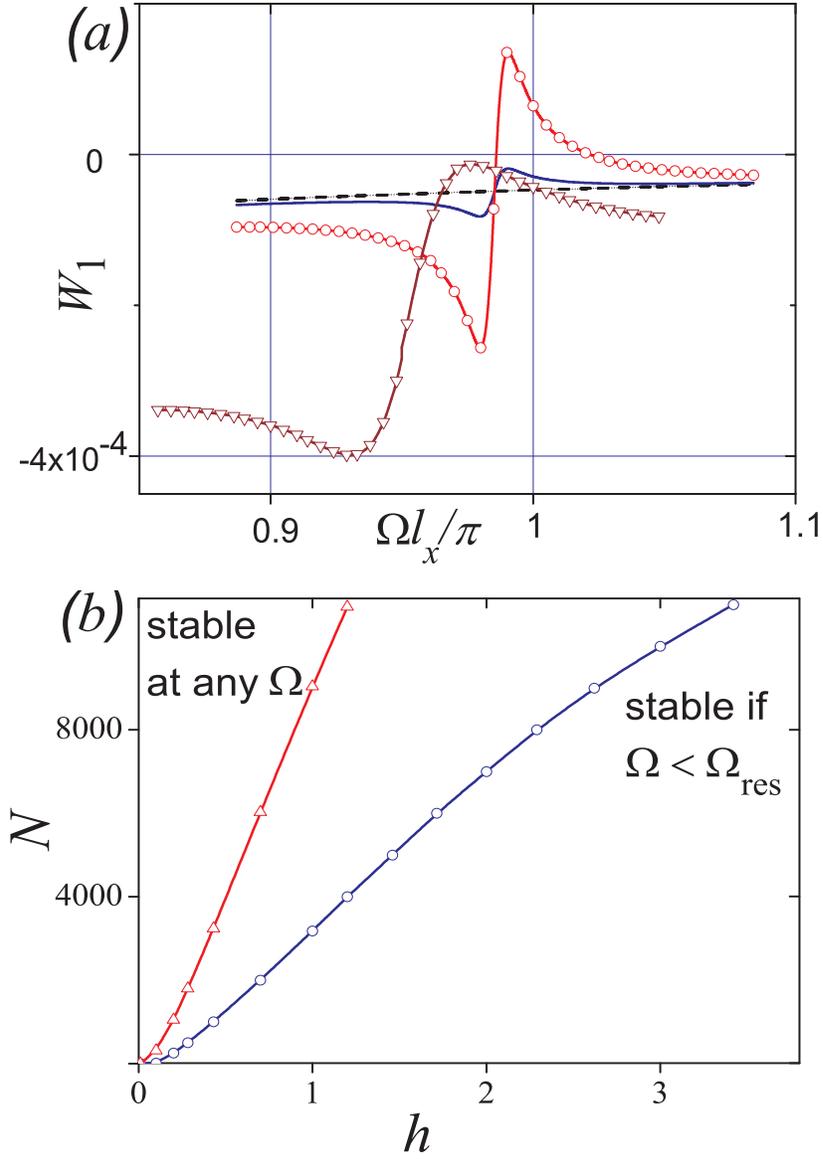

FIG. 37: (Color online) (a) Function $W_1(\omega)$ near resonance (From Ref. 144). Black dashed, blue solid, and red line with circles correspond to $h = 0$, $N = 10^3$, $h = 0.2$, $N = 10^3$, $h = 0.5$, $N = 10^3$, respectively. The wine curve with triangles is obtained for $h = 1$, $N = 4.5 \cdot 10^3$. Other parameters are the same as in Figs. 34 and 35. (b) Stability regions in the plane $(N, h)$ for $l_x = 0.25$ (blue line with circles) and $l_x = 0.4$ (red line with triangles) (From Ref. 144).

The magnetic field does not practically affect the term $W_2$, but significantly reduces the radiation coupling. Actually, the value of $W_1$ changes its sign and becomes zero in the main approximation at the resonance point (see Fig. 37 (a)). So, the stabilization via the radiation coupling can occur only due to the next order terms with respect to $\zeta$. As a result, for sufficiently high magnetic fields and low number of junctions, the resonance radiation is stable at frequencies lower than the resonance frequency and becomes unstable for $\Omega > \Omega^{\text{res}}$. In the range of lower fields and for SJJs with a large number of junctions, the resonance peak is stable for any frequency. The stability regions in the plane $(h, N)$ are shown in Fig. 37 (b) for two different values of $l_x$. The radiation is stable for all frequencies above the lines $N(h)$ and for $\Omega < \Omega^{\text{res}}$ below these curves. The stable region decreases when increasing the ratio $L_x/L_z$. The instability at $\Omega > \Omega^{\text{res}}$ arises due to the negative differential resistivity of the CVC near the peak of the radiation power [140, 141]. Our considerations are invalid if $hl_x \gtrsim \pi$. We can suppose that the synchronization of the oscillations in different junctions will be destroyed at higher $h$, since a vortex lattice enters the sample. However, for the SJJ small thicknesses ($L_x \ll \lambda_c$), this event occurs at magnetic fields much higher than the lower critical field of the bulk layered material.

### E. Concluding remarks

Thus, the effect of the superradiance is observed in SJJ prepared from BSCCO with 500–1000 elementary junctions [140]. The effect could perhaps be enhanced for artificially-modulated samples [143]. The main problem here would be to prepare samples with a number of identically-modulated junctions. An alternative approach, consisting in the application of a moderate DC magnetic field, allows one to overcome this difficulty [144]. However, the stability region of the superradiance regime, for samples in magnetic fields, would be considerably reduced.

As was recently shown in Ref. 145, the modulation can be self-organized in layered superconductors in the absence of an external magnetic field. Reference 145 predicts new states where $(2m+1)\pi$ phase kinks appear around the junction centers, with $m$ being an integer, periodic and, thus, non-uniform in the **c**-direction. Such a state manifests itself in the CVC as current steps occurring at both even and odd cavity modes. Inside the current steps, the plasma oscillations become strong, which generates several harmonics in the frequency spectrum at a given voltage. The superradiance in this case is much stronger than predicted for the uniform case.

It should be also noted that the necessity to work in the resistive state gives rise to a significant dissipation power $VJ \propto \omega_{\rm osc}$. As a result, the problems of overheating and the effectiveness of the power transmission become very important for future THz devices [140, 141, 143].

## VI. QUANTUM JOSEPHSON PLASMA WAVES

In this section, we consider an approach allowing the analysis of the quantum effects in the propagation of JPWs. Such phenomena can be of importance in small samples at low temperatures. A striking example of quantum effects in HTS is the macroscopic quantum tunnelling (MQT) observed at $T < 1$ K in $Bi_2Sr_2CaCu_2O_{8+\delta}$ microbridges, which can be considered as stacks of Josephson junctions [146–149]. Unexpectedly, MQT in such stacks is considerably enhanced compared with single junctions. This observation could open a new avenue for the applicability of stacks of Josephson junctions in quantum electronics [150–152].

Here, for simplicity, we neglect the effect of charge-neutrality-breaking and ignore dissipation, which is negligible for MQT [146–148, 153, 154]. In this section, the in-plane coordinate $x$ is normalized to $\lambda_c$, time $t$ is normalized to $1/\omega_J$; also, $\partial_x = \partial/\partial x$ and $\partial_z f_l = \lambda_{ab}(f_{l+1} - f_l)/D$.

### A. Lagrangian approach and two types of JPWs

The set of coupled sine-Gordon equations, Eq. (12) or Eq. (32), could not be obtained using a Lagrangian formalism. In other words, these equations do not have a Lagrangian. Following the approach used in Refs. [155–157], we introduce an additional gauge-invariant field,

$$p^l \equiv \frac{D}{\lambda_{ab}}\partial_x\chi_l - \frac{2\pi\gamma DA_{xl}}{\Phi_0},$$

which can be considered as the normalized superconducting momentum in the $l$th layer. Then, we introduce the Lagrangian of two interacting classical fields, $\varphi^{l+1,l}$ and $p^l$:

$$\mathcal{L} = \sum_l \int dx \left[ \frac{1}{2} (\dot{\varphi}^{l+1,l})^2 + \frac{1}{2\gamma^2} (\dot{p}^l)^2 \right.$$
$$-\frac{1}{2}(\partial_x \varphi^{l+1,l})^2 - \frac{1}{2}(\partial_z p^l)^2 - \frac{1}{2}(p^l)^2 + \cos(\varphi^{l+1,l})$$
$$\left. + \frac{1}{2} \left( \partial_x p^l \partial_z \varphi^{l+1,l} + \partial_z p^l \partial_x \varphi^{l+1,l} \right) \right]. \tag{271}$$

Varying the action $\mathcal{S} = \int dt \, \mathcal{L}$ produces dynamical equations for the phase difference,

$$\ddot{\varphi}^{l+1,l} - \partial_x^2 \varphi^{l+1,l} + \sin(\varphi^{l+1,l}) + \partial_x \partial_z p^l = 0,$$
$$\frac{1}{\gamma^2} \ddot{p}^l - \partial_z^2 p^l + p^l + \partial_x \partial_z \varphi^{l+1,l} = 0, \tag{272}$$

which reduce to the coupled sine-Gordon equations (12) when $\gamma^2 \to \infty$. Note that the Lagrangian approach for stacks of Josephson junctions can be formulated only for *two* interacting fields, $\varphi$ and $p$. This is because the vector potential has two components, $A_x$ and $A_z$, in stacks of Josephson junctions, in contrast to single Josephson junctions where the electromagnetic field can be described by one component of the vector potential.

Linearizing Eqs. (272) and substituting there $p, \varphi \propto \exp(-i\omega t + iqx + ikz)$, we derive a biquadratic equation,

$$(\omega^2 - q^2 - 1)\left(\frac{\omega^2}{\gamma^2} - k^2 - 1\right) - q^2 k^2 = 0,$$

for the spectrum of JPWs in the continuous limit (i.e., for $kD \ll 1$) and for $\gamma^2 \gg 1$. This equation determines two branches, $\omega = \omega_a(\vec{k})$ and $\omega_b(\vec{k})$, of the JPWs:

$$\omega_a(\vec{k}) = \left(1 + \frac{q^2}{1+k^2}\right)^{1/2}, \quad \omega_b(\vec{k}) = \gamma(k^2 + 1)^{1/2} \tag{273}$$

up to terms of the order of $1/\gamma^2$. The $a$-branch describes Josephson plasma waves propagating both along and perpendicular to the layers. This branch coincides with the dispersion law Eq. (20) in the limit $kD \ll 1$ and $\omega_r = 0$. The $b$-branch describes waves propagating only perpendicular to the layers. The possibility of excitation of the latter JPWs is questionable due to their high frequencies, $\omega_b \geq \gamma \omega_J$. Here we do not consider such waves. However, for samples with not so high values of $\gamma$, as for $Bi_2Sr_2CaCu_2O_{8+\delta}$, the branch $b$ could be observable.

### B. Analogy with quantum electrodynamics

In order to quantize the JPWs, we use the Hamiltonian ($\hbar = 1$),

$$\mathcal{H} = \sum_l \int dx (\Pi_\varphi^l \dot\varphi^{l+1,l} + \Pi_p^l \dot p^l) - \mathcal{L},$$

with the momenta $\Pi_\varphi^l$ and $\Pi_p^l$ of the fields $\varphi^{l+1,l}$ and $p^l$, and require the standard commutation relations,

$$[\varphi^{l'+1,l'}(x'), \Pi_\varphi^l(x)] = i\delta(x-x')\delta_{ll'},$$
$$[p^{l'}(x'), \Pi_p^l(x)] = i\delta(x-x')\delta_{ll'}$$

(all others commutators are zero). Here $\delta$ is either a delta function or a Kronecker symbol. Expanding $\cos\varphi_n = 1 - \varphi_n^2/2 + \varphi_n^4/24 - \ldots$, we can write $\mathcal{H} = \mathcal{H}^0 + \mathcal{H}^{\rm an}$, where we include terms up to $\varphi^2$ in $\mathcal{H}^0$, and the anharmonic terms in $\mathcal{H}^{\rm an}$. Diagonalizing $\mathcal{H}^0$, we obtain the Hamiltonian for the Bosonic free fields $a$ and $b$,

$$\mathcal{H}^0 = \sum_k \int \frac{dq}{2\pi} \left\{ \omega_a(\vec{k})\, a^+ a \;+\; \omega_b(\vec{k})\, b^+ b \right\}.$$

The original fields $\varphi$ and $p$ are related to the free Bosonic fields $a$ and $b$ by

$$\varphi \approx \frac{a^+ + a}{\sqrt{2\omega_a(\mathbf{k})}} - \mathcal{Z}\frac{b^+ + b}{\gamma\sqrt{2\omega_b(\mathbf{k})}},$$
$$p \approx \mathcal{Z}\frac{a^+ + a}{\sqrt{2\omega_a(\mathbf{k})}} + \gamma\frac{b^+ + b}{\sqrt{2\omega_b(\mathbf{k})}},$$

where $\mathcal{Z} = qk/(k^2+1)$. Equation (273) shows that the "mass" of the $a$-quantum equals one, in our dimensionless units, and the "mass" of the heavier $b$-quasi-particle is $\gamma$. The interaction between the $a$ and $b$ fields, including the self-interaction, occurs due to the *anharmonic* terms in

$$\mathcal{H}^{\rm an} \approx -\frac{1}{24}\sum_n \int dx\, \varphi_n^4 + \ldots.$$

In leading order with respect to the bosonic field interactions, an $a$-particle can create either $(a+a)$ or $(a+b)$ pairs. Using Eqs. (273), one can conclude that the amplitudes of these processes have energy thresholds $\omega_a(\vec{k}) = 3$ or $(\gamma+2)$. Note that this is similar to the $2mc^2$ rest energy threshold for $e^- + e^+$ *pair creation in usual quantum electrodynamics*. These can result in resonances in the amplitudes of quantum processes (e.g., decay of the "$a$" quanta).

### C. Enhancement of macroscopic quantum tunnelling

#### 1. Effective Lagrangian

Now we use quantum field theory to describe MQT in layered superconductors [155–158]. We now consider a stack of $N \gg 1$ Josephson junctions having sizes $L_x \times L_y$ in the plane of the junctions, and $L_z = ND$ across them. In the continuous limit, and when $\gamma^2 \gg 1$, Eqs. (272) can be rewritten as

$$\left(1 - \frac{\partial^2}{\partial z^2}\right)\left[\frac{\partial^2 \varphi}{\partial t^2} + \sin \varphi\right] - \frac{\partial^2 \varphi}{\partial x^2} = 0. \tag{274}$$

Here the coordinate $z$ transverse to the layers is normalized to $\lambda_{ab}$. In accordance with an experimental observation of MQT (e.g., Ref. 149), we assume that: (i) the stack of Josephson junctions bridges two bulk superconducting sheets; (ii) the current close to the critical value flows across the stack; (iii) the external magnetic field is zero. We neglect the disturbance produced by the tunnelling fluxon in the bulk superconductors. In this case, the boundary conditions to Eq. (274) are

$$\left.\frac{\partial \varphi}{\partial x}\right|_{x=0,d} = \mp \frac{jd}{2}, \quad \left.\frac{\partial \varphi}{\partial z}\right|_{z=0,L} = 0, \tag{275}$$

where $j$ is the current density normalized by $J_c$, $L = ND/\lambda_{ab}$ and $d = L_x/\lambda_c$. When tunnelling occurs, the phase difference in a junction changes from 0 to $2\pi$, which can be interpreted as the tunnelling of a fluxon through the contact. This process can be safely described within a semiclassical approximation.

#### 2. Probability of quantum tunnelling in the semiclassical limit

The probability of quantum tunnelling in the semiclassical limit is expressed through the classical action of the system in imaginary time [159]. However, the simplified Eq. (274) *has no Lagrangian*. In general, we have to use an action of the form given in Eq. (271) with two interacting bosonic fields, $\varphi$ and $p$, which could produce a rather cumbersome mathematical problem. In order to avoid this difficulty, we follow the approach described in Refs. 155, 156, 158. First, we seek a solution of Eq. (274) in imaginary time $t = i\tau$ in the form

$$\varphi(\tau, x, z) = \varphi_0(x) + \psi(\tau, x, z),$$

where $\varphi_0(x)$ is a steady-state solution corresponding to an energy minimum of the junction's stack; $\varphi_0(x)$ satisfies the equation

$$\frac{\partial^2 \varphi_0}{\partial x^2} = \sin \varphi_0. \qquad (276)$$

We consider short junctions, $d \ll 1$ ($L_x \ll \lambda_c$), corresponding to the experimental conditions (e.g., Ref. 149). In this case, the solution of Eq. (276) with boundary conditions (275) has the form

$$\varphi_0(x) = \arcsin(j) + \frac{j}{2}\left(x - \frac{d}{2}\right)^2 + O(x^4). \qquad (277)$$

Substituting the expansion $\varphi = \varphi_0 + \psi$ into Eq. (274), we derive, in zero approximation with respect to $d \ll 1$, the equation for $\psi(\tau, x, y)$:

$$\left(1 - \frac{\partial^2}{\partial z^2}\right)\left[-\frac{\partial^2 \psi}{\partial \tau^2} - j(1 - \cos\psi) + \sqrt{1-j^2}\sin\psi\right] - \frac{\partial^2 \psi}{\partial x^2} = 0. \qquad (278)$$

We assume that the fluxon tunnels mainly through one of the junctions and linearize Eq. (278) in all junctions except this one. We label this junction as $l$. The linearized equation for $\psi$ is

$$\left(1 - \frac{\partial^2}{\partial z^2}\right)\left[-\frac{\partial^2 \psi}{\partial \tau^2} + \bar{\mu}_0 \psi\right] - \frac{\partial^2 \psi}{\partial x^2} = 0 \qquad (279)$$

where $\bar{\mu}_0 = \sqrt{1-j^2}$. This equation is valid in all junctions except for the $l$-th junction, located at the position $z_0 = L_1 = lD/\lambda_{ab}$. The function $\psi(\tau, x, z)$ satisfies the boundary conditions

$$d\psi/dx = 0 \quad \text{at} \quad x = 0, d$$

and

$$d\psi/dz = 0 \quad \text{at} \quad z = 0, L.$$

The characteristic size of the fluxon is $\gamma s \ll \lambda_c$. So, when $L_x \gtrsim \lambda_J = \gamma D$, the $x$-dependence of $\psi(\tau, x, z)$ is essential.

The solution of Eq. (279) with the boundary conditions specified above, and the continuity condition at $z = L_1$, $\psi(\tau, x, L_1 + 0) = \psi(\tau, x, L_1 - 0)$, can be written as an expansion

$$\psi(\tau, x, z) = \sum_{n=0}^{\infty} \int_0^{\infty} dp\, e^{-p\tau} \cos(k_n x) f_n(p, z)\psi_n(p), \qquad (280)$$

with $k_n = \pi n/d$ and functions

$$f_n(p,z) = \begin{cases} \dfrac{\cosh[\nu_n(p)z]}{\cosh[\nu_n(p)L_1]}, & z < L_1 \\[2ex] \dfrac{\cosh[\nu_n(p)(L-z)]}{\cosh[\nu_n(p)(L-L_1)]}, & z > L_1 \end{cases}, \tag{281}$$

where

$$\nu_n^2 = 1 + \frac{k_n^2}{\bar{\mu}_0 - p^2}.$$

The functions $\psi_n(p)$ in Eq. (280) are derived from the equation for $\psi$ in the $l$th junction, $\bar{\psi}(\tau, x) \equiv \psi(\tau, x, L_1)$. Similarly to subsection IV A, the latter can be derived from the relation between the phase difference and the magnetic field in the $l$th layer. Using Eq. (203) and Maxwell's equations, we obtain

$$\frac{\partial \psi}{\partial x} = \frac{c}{4\pi J_c \gamma D}\left[\left(\frac{\partial H}{\partial z}\bigg|_{L_1+0} - \frac{\partial H}{\partial z}\bigg|_{L_1-0}\right)\right]. \tag{282}$$

We present the $y$-component of the magnetic field in the form $H = H_0 + \bar{H}$, where

$$\frac{\partial H_0}{\partial x} = -\frac{4\pi J_c \lambda_c}{c}\sin\varphi_0.$$

According to Maxwell's equation, the field $\bar{H}$ linearly depends on $\psi$ at $z \neq L_1$, and can be represented in the form,

$$\bar{H}(\tau, x, z) = \sum_{n=1}^{\infty}\int_0^{\infty}dp\, e^{-p\tau}\sin(k_n x)f_n(p,z)h_n(p), \tag{283}$$

where the amplitudes $h_n(p)$ are independent of the coordinates $x$ and $z$. Substituting the expansion Eq. (283) into Eq. (282), we obtain the relation between the functions $h_n(p)$ and $\psi_n(p)$,

$$h_n(p) = \left(\frac{2\pi J_c \gamma s}{c}\right)\frac{k_n \chi_n(p)}{\nu_n(p)}\psi_n(p), \tag{284}$$

with

$$\chi_n = \frac{2\cosh(\nu_n L_1)\cosh(\nu_n(L-L_1))}{\sinh(\nu_n L)}. \tag{285}$$

The Maxwell equation for the contact at $z = L_1$ is non-linear,

$$-\frac{\partial H}{\partial x} = \frac{4\pi J_c \lambda_c}{c}\left[\sin(\varphi_0 + \bar{\psi}) - \frac{\partial^2 \psi}{\partial \tau^2}\right].$$

The continuity of the $z$-component of the current requires the continuity of the derivative $\partial H/\partial x$ at $z = L_1$. Thus, substituting the expansion Eq. (283) into the latter equation, we obtain the following set of equations for $\psi_n(p)$:

$$-\frac{\partial^2 \bar{\psi}}{\partial \tau^2} + \bar{\mu}_0 \sin \bar{\psi} - j(1 - \cos \bar{\psi}) = \quad (286)$$
$$-\frac{D}{2\lambda_{ab}} \sum_{n=1}^{\infty} \int_0^{\infty} dp \, e^{-p\tau} \cos(k_n x) \frac{k_n^2 \chi_n(p)}{\nu_n(p)} \psi_n(p),$$

where

$$\bar{\psi}(\tau, x) = \sum_{n=0}^{\infty} \int_0^{\infty} dp \, e^{-p\tau} \cos(k_n x) \, \psi_n(p). \quad (287)$$

If the current through the junction is comparable with $J_c$, then the plasma frequency is renormalized [68] as $\omega_J(j) = \omega_J(0)(1 - j^2)^{1/4}$. The characteristic time of MQT is certainly lower than $1/\omega_J(j)$. On the other hand, the only time scale in Eqs. (274) and (275) is $1/\omega_J(j)$, and we can consider the MQT as a quasi-static process assuming $p = 0$ in $\nu_n(p)$ and $\chi_n(p)$ in Eqs. (286). In the case under consideration, $d \ll 1$, we have $\nu_n = k_n/(\bar{\mu}_0)^{1/2}$ for $n > 0$. So, we reduce Eqs. (286) and (287) to

$$-\frac{\partial^2 \bar{\psi}}{\partial \tau^2} + \bar{\mu}_0 \sin \bar{\psi} - j(1 - \cos \bar{\psi})$$
$$= \int_0^d dx' K(x; x') \frac{\partial^2 \bar{\psi}(\tau, x')}{\partial x'^2}, \quad (288)$$

where the kernel $K(x; x')$ is

$$K(x; x') = \frac{\gamma D \sqrt{\bar{\mu}_0}}{L_x} \sum_{n=1}^{\infty} \cos k_n x \cos k_n x' \frac{\chi_N^l(an)}{k_n}, \quad (289)$$

and

$$\chi_N^l(a) = \frac{2 \cosh(al) \cosh(a(N-l))}{\sinh(aN)}, \quad a = \frac{\pi \gamma D}{L_x \sqrt{\bar{\mu}_0}}. \quad (290)$$

Equation (288) is a generalization of the non-local expression (192) for the Josephson vortex to the case of a finite sample. If $l, N \gg 1$ and $a \sim 1$, then $\chi_N^l(a) \cong 1$, and the kernel can be calculated explicitly,

$$K(x; x') = -\frac{\gamma D (\bar{\mu}_0)^{1/2}}{2\pi \lambda_c}$$
$$\times \ln \left| 4 \sin \left( \frac{\pi(x - x')}{2d} \right) \sin \left( \frac{\pi(x + x')}{2d} \right) \right|. \quad (291)$$

Equation (288), in contrast to Eq. (274), has a Lagrangian, which can be written in imaginary time $t = i\tau$ as

$$\mathcal{L}_{\text{eff}}(\tau) = \varepsilon_0 \int_0^d dx \left[ -\frac{1}{2}\left(\frac{\partial \bar{\psi}}{\partial \tau}\right)^2 - \bar{\mu}_0(1 - \cos\bar{\psi}) + \right.$$
$$\left. j(\bar{\psi} - \sin\bar{\psi}) + \frac{1}{2}\bar{\psi} \int_0^d dx' K(x;x') \frac{\partial^2 \bar{\psi}}{\partial x'^2} \right], \tag{292}$$

where $\varepsilon_0 = J_c L_z \lambda_c/(2e\omega_J)$. Indeed, the variation of Eq. (292) with respect to $\bar{\psi}$ gives Eq. (288). The effective Lagrangian depends on $l$ and $N$ via the functions $\chi_N^l$.

3. *Field tunnelling: Numerical approach*

The tunnelling escape rate $\Gamma$, that is the tunnelling probability per unit time, can be calculated in the semi-classical approach for a system with a Lagrangian given in the general form [159]. For the case of tunnelling of a fluxon, $\Gamma$ can be presented as [149, 155, 156, 160]

$$\Gamma = \omega_p(j) \sum_{l=0}^{N} \sqrt{\frac{30 B_N^l}{\pi}} \exp\left(-B_N^l\right), \tag{293}$$

where we take into account that the fluxon can tunnel through any junction $0 < l < N$ of the stack. The tunnelling exponent $B_N^l$ can be expressed via the Lagrangian as

$$B_N^l = -2 \int_0^\infty d\tau \, \mathcal{L}_{\text{eff}}(\tau). \tag{294}$$

The current $I$ in the stack is close to the critical value $I_c(d)$. Thus, $|\bar{\psi}| \ll 1$, and we can expand the Lagrangian (292) and the equation of motion (288) in series over $\bar{\psi}$. We seek the function $\bar{\psi}$ in the form [157, 158, 161]

$$\bar{\psi}(\tau, x) = \sum_{n=0}^{\infty} c_n(\tau) \psi_n(x) \tag{295}$$

where $\psi_n$ are orthogonal eigenfunctions of the operator

$$\hat{L} = \bar{\mu}_0 - \int_0^d dx' \, K(x;x') \frac{\partial^2}{\partial x'^2} \cdot \,. \tag{296}$$

The tunnelling exponent can be expanded as [158]

$$B_N^l = \frac{\varepsilon_0}{6} \int_0^\infty d\tau \left[ \sum_{nmk} U_{nmk}^{(3)} c_n c_m c_k \right.$$
$$\left. + \frac{1}{2} \sum_{nmkl} U_{nmkl}^{(4)} c_n c_m c_k c_l + \ldots \right] \tag{297}$$

where
$$U^{(i)}_{n\ldots k} = \int_0^d dx\, \frac{\partial^i(\cos\varphi_0)}{\partial\varphi_0^i}\, \psi_n\ldots\psi_k \bigg|_{\varphi_0=\arcsin(j)}. \tag{298}$$

The functions $c_n(\tau)$ satisfy the set of equations,
$$\ddot{c}_n - \mu_n c_n = -\frac{1}{2}\sum_{mk} U^{(3)}_{nmk} c_m c_k - \frac{1}{6}\sum_{mkl} U^{(4)}_{nmkl} c_m c_k c_l - \ldots, \tag{299}$$

with the initial conditions
$$\dot{c}_n(0) = 0, \quad \lim_{\tau\to\infty} c_n(\tau) = 0. \tag{300}$$

Here dot denotes the imaginary time derivative, and $\mu_n$ are eigenvalues of $\hat{L}$.

From the expansion (289) for $K(x;x')$, we derive the orthogonal eigenfunctions $\psi_n(x)$ of the operator $\hat{L}$,
$$\psi_0(x) = \sqrt{\frac{1}{d}}, \quad \psi_n(x) = \sqrt{\frac{2}{d}}\cos k_n x, \quad n > 0, \tag{301}$$

and the corresponding eigenvalues,
$$\mu_n = \bar{\mu}_0\left[1 + \frac{an}{2}\chi^l_N(an)\right], \quad a = \frac{\pi\gamma D}{L_x\sqrt{\bar{\mu}_0}}. \tag{302}$$

Equation (298) gives
$$\begin{aligned} U^{(3)}_{0mk} &= j\delta_{mk}/\sqrt{d}, \\ U^{(3)}_{nmk} &= j\frac{\delta_{n,m+k} + \delta_{m,n+k} + \delta_{k,n+m}}{\sqrt{2d}}, \quad n,\,m,\,k > 0. \end{aligned} \tag{303}$$

Note that Eqs. (301) and (302) for $\psi_n$ and $\mu_n$ are derived in the limit $L_x/\lambda_c \ll 1$ when the stationary solution is almost uniform, $\varphi_0(x) \approx$ const.

The lowest eigenvalue, $\mu_0 = \sqrt{1-j^2}$, is small when $j$ is close to 1. Therefore, the functions $c_n(\tau)$ should be small, and we can neglect all terms in the right-hand-side of Eq. (299), except the first one. Introducing new variables,
$$\alpha_n(\eta) = \frac{jc_n(\tau)}{3\bar{\mu}_0\sqrt{d}}, \quad \eta = \sqrt{\bar{\mu}_0}\tau, \tag{304}$$

and substituting Eqs. (303) and (304) into the set (299), we derive
$$\frac{d^2\alpha_0}{d\eta^2} - \alpha_0 = -\frac{3}{2}\sum_{m=0}^\infty \alpha_n^2, \tag{305}$$
$$\begin{aligned} \frac{d^2\alpha_n}{d\eta^2} - \lambda_n\alpha_n &= -3\bigg(\alpha_n\alpha_0 + \frac{1}{\sqrt{2}}\sum_{m=1}^\infty \alpha_m\alpha_{n+m} + \\ &\quad \frac{1}{2\sqrt{2}}\sum_{m=1}^{n-1}\alpha_m\alpha_{n-m}\bigg), \quad n > 0, \end{aligned}$$

where $\lambda_n = \mu_n/\bar{\mu}_0$. Equation (297) for the tunnelling exponent $B_N^l$ can be rewritten as

$$B_N^l = \frac{J_c}{2e\omega_J} \frac{24(1-j^2)^{5/4}}{5j^2} F(\{\lambda_n\}), \qquad (306)$$

where

$$F(\{\lambda_n\}) = \frac{15}{16} \int_0^\infty d\eta \left( \alpha_0^3 + 3\alpha_0 \sum_{n=1}^\infty \alpha_n^2 + \frac{3}{\sqrt{2}} \sum_{n,m=1}^\infty \alpha_n \alpha_m \alpha_{n+m} \right). \qquad (307)$$

Substituting Eq. (306) into Eq. (293), we can calculate the escape rate of the fluxon through a set of the junctions.

Below we assume that $N \gg 1$ and $a \sim 1$. In this case, the functions $\chi_N^l(an) \cong 1$ and $B_N^l \cong B$, for all $l$ with the exception of $l = 0$ and $l = N$. Neglecting these two contributions to tunnelling, we derive

$$\Gamma \cong N\omega_J \sqrt{\frac{30\bar{\mu}_0 B}{\pi}} \exp(-B). \qquad (308)$$

Under the conditions considered here, we obtain from Eq. (302)

$$\lambda_n = 1 + an/2.$$

Therefore, the function $F(\{\lambda_n\})$ depends only on the single parameter $a$, that is, $F(\{\lambda_n\}) = F(a)$. Eq. (308) for $\Gamma$ is derived here in the limit $N \gg 1$. To find the dependence $\Gamma(N)$, we should solve numerically the problem for each contact $0 < l < N$ and perform the summation according to Eq. (293). However, the results of both procedures are qualitatively the same as shown in Fig. 1 in Ref. 155.

Our analysis of Eq. (305) shows that, for any $\eta$, we have $\alpha_0(\eta) > \alpha_1(\eta) > \alpha_2(\eta) > \ldots$. Thus, for a given accuracy, we can consider only the first $n_0$ equations of the set (305), taking $\alpha_n = 0$ for $n \geq n_0$. This closed set of equations was solved numerically. The number of equations that we should take into account depends on the value of $a$: the smaller $a$, the larger $n_0$. The analysis also shows that there exists a critical value $a_c = 2.5$, or a critical value of the junction's width

$$L_c = \frac{2\pi\gamma D}{5(1-j^2)^{1/4}}. \qquad (309)$$

If $L_x < L_c$, all the solutions of Eq. (305), except $\alpha_0(\eta)$, are equal to zero. In this case, $F \equiv 1$ and the tunnelling exponent in Eq. (306) coincides with the value calculated under the approximation of the fluxon tunnelling by the tunnelling of a single quantum particle in the effective potential well [149, 160]. Using for an estimate $\gamma = 300$–500, $D = 1.5$ nm, we find that $L_c \approx 1$ $\mu$m.

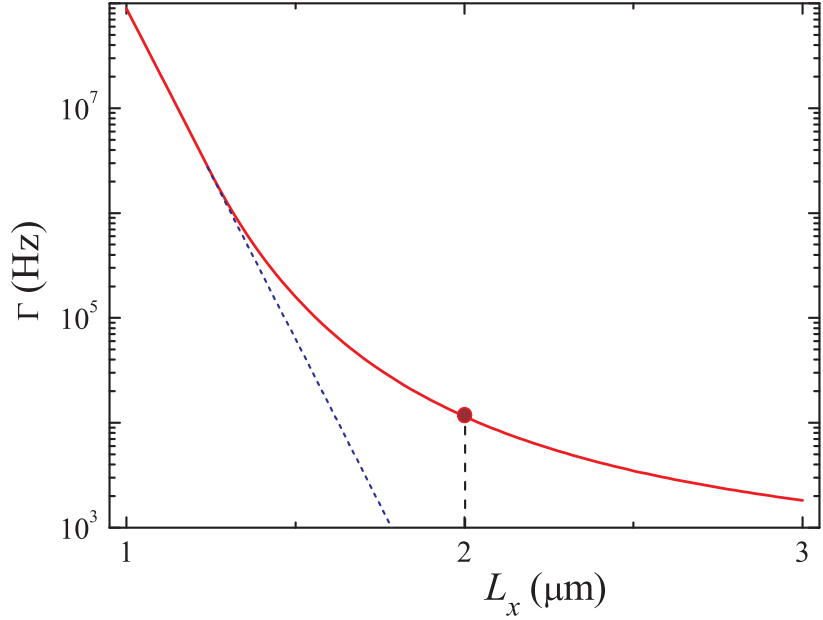

FIG. 38: (Color online) The escape rate $\Gamma$ versus sample width $L_x$ (From Ref. 157). The red solid curve is obtained numerically using the equation (308) for the sample US1 from Ref. 149. The dashed blue line corresponds to the particle tunnelling approximation. The parameters $\omega_J$, $J_c$, and $L_x$ are taken from Table I in Ref. 149. The anisotropy coefficient $\gamma$ and interlayer distance $D$ are chosen as $\gamma = 350$ and $D = 1.5$ nm. The red point $L_x \approx 2\,\mu$m indicates the experimental result [149].

The function $\Gamma(L_x)$ is shown in Fig. 38 by the red solid curve. This dependence is calculated by means of the numerical procedure described above. In our calculations, we used the parameters of the $Bi_2Sr_2CaCu_2O_{8+\delta}$ sample US1 from Ref. 149. The same figure shows the curve $\Gamma(L_x)$ calculated using the quantum particle approach (Eq. (308) with $F = 1$ in Eq. (306) for the tunnelling exponent $B_N^l$). It follows from Fig. 38 that the difference between the particle and field approaches becomes significant if $L_x$ exceeds the value of about 1.3–1.4 $\mu$m.

The dependence of the escape rate $\Gamma$ on the dimensionless current $j$, calculated using Eq. (308) and the numerical procedure described above, is shown in Fig. 39. The calculations were performed for two $Bi_2Sr_2CaCu_2O_{8+\delta}$ samples described in Ref. 149. The only adjustable parameter is the product $\gamma D$, which is about 400–800 nm for $Bi_2Sr_2CaCu_2O_{8+\delta}$. It is seen from Fig. 39 that the agreement between the calculated and measured value of $\Gamma$ is quite good. Small discrepancies can be attributed to the violation of the semiclassical approximation.

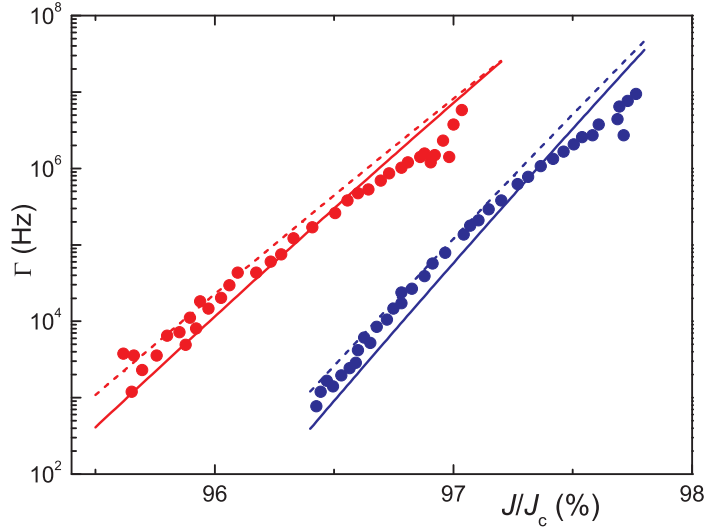

FIG. 39: (Color online) The escape rate $\Gamma$ versus dimensionless current $j = J/J_c$ (From Ref. 157). The red points (on the left) are for the experiment on the $Bi_2Sr_2CaCu_2O_{8+\delta}$ sample US1 and the blue points (on the right) are for the sample US4 from Ref. 149; the red and blue solid curves are obtained numerically using formula (308) for the samples US1 and US4, respectively. The parameters $\omega_J$, $J_c$, and $L_x$ are taken from Table 1 in Ref. 149; the anisotropy coefficient and interlayer distance were chosen as $\gamma = 350$ and $D = 1.5$ nm. Dashed red and dashed blue lines are obtained using Eqs. (308) and (316) with $C = 0.45$ to calculate $B$ for the same samples. The value of gamma for the blue curve is the same as used for numerical calculations, while $\gamma = 455$ for the red curve.

If $L_x > L_c$, $\alpha_n(\eta) \neq 0$ for $n > 0$ and we should consider $n_0$ equations in set (305). The tunnelling of the fluxon in this case is similar to the tunnelling of a quantum particle in $n_0$ dimensions, where the $\alpha_n$ play the role of the particle coordinates in $n_0$ dimensional space. The field $\bar{\psi}$ is strongly inhomogeneous. In Fig. 40 we show the spatial profile $\bar{\psi}(\tau, x)$ of the tunnelling fluxon, calculated numerically for different values of the imaginary time $\tau$, which changes from $-\infty$ to zero. The maximum value of $\bar{\psi}$ increases monotonically with $\tau$, while its characteristic size remains practically constant. This analysis shows that the characteristic size of the fluxon is about $aL_x \sim \gamma D$, as it was mentioned above.

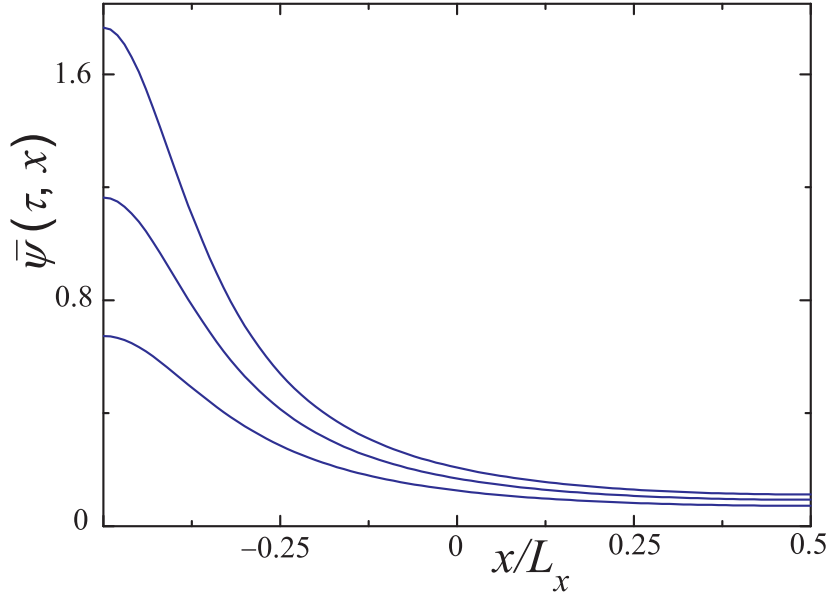

FIG. 40: Spatial profile of tunnelling fluxon for different values of the imaginary time $\tau$: for the curves from the bottom to top $\eta = \sqrt{\tilde{\mu}_0}\tau = -1.6$, $\eta = -1$, and $\eta = 0$; $j = 0.96$, $\gamma D/L_x = 0.22$ (From Ref. 157).

### 4. Analytical approach

In this subsection, we obtain the analytical formula for calculating the tunnelling exponent $B_N^l$ in the case $L_x \ll \lambda_c$. For this purpose, we reduce the problem of field tunnelling to the tunnelling of a quantum particle. However, in contrast to the usual approach, we take into account the spatial variation of the gauge-invariant phase difference $\varphi$ when deriving the effective potential $U$ [155–157].

We now use the real time $t$ in Eq. (288) instead of imaginary $\tau$. According to the numerical result shown in Fig. 40, the function $\bar\psi(t,x)$ can be presented as $\bar\psi(t,x) \approx f(x)\,r(t)$, where $df/dx = 0$ at $x = 0, d$. The function $f(x)$ is normalized by the condition,

$$\int_0^d dx\, f^2(x) = 1. \tag{310}$$

We substitute $\bar\psi = f(x)\,r(t)$ into Eq. (288), then multiply both sides of this equation by $f$, and integrate along the junction. As a result, we obtain the equation of motion for a hypothetical particle with coordinate $r(t)$,

$$\frac{d^2 r}{dt^2} + \bar\mu_0 r - \frac{jr^2}{2}\int_0^d dx\, f^3(x)$$
$$= -r \int_0^d dx \int_0^d dx' \frac{df(x)}{dx} P(x;x') \frac{df(x')}{dx'}, \tag{311}$$

where the new kernel $P(x;x')$ is expressed through the kernel $K(x;x')$, Eq. (289). Under conditions $L_x \ll \lambda_c$ and $N \gg 1$, $a \sim 1$, we derive the explicit analytical formula for $P(x;x')$

$$P(x;x') = \frac{\gamma D \sqrt{\bar\mu_0}}{2\pi \lambda_c} \ln\left|\frac{\sin\left(\frac{\pi(x+x')}{2d}\right)}{\sin\left(\frac{\pi(x-x')}{2d}\right)}\right|. \tag{312}$$

We approximate $f(x)$ by a step function $f(x) = \theta(x_0 - x)/\sqrt{x_0}$. After substitution of this function into Eq. (311) and integration, we see that the term in the right hand side of this equation has a logarithmic singularity since $\partial f/\partial x = -\delta(x - x_0)/\sqrt{x_0}$. To cut off this singularity, we take into account that the characteristic scale of change of the phase $\bar\psi$ in the stack of Josephson junctions is $\gamma D$. Thus, performing integration, we put $|x - x'| = C\gamma D$ at $x' \to x$, where $C$ is a constant of the order of unity. Thus, we obtain the equation of motion for the effective particle in the form $d^2 r/dt^2 = -\partial U(r)/\partial r$, where the effective potential $U(r)$ can be written as

$$U(r) = \frac{jr^2}{6\sqrt{x_0}}(r_0 - r),$$
$$r_0 = \frac{3\sqrt{x_0}}{j}\left[\bar\mu_0 + \frac{\gamma D \sqrt{\bar\mu_0}}{2\pi \lambda_c x_0} \ln\left(\frac{2L_x}{C\pi\gamma D}\right)\right]. \tag{313}$$

Here we neglect the small term $\propto \ln|\sin(\pi(x+x')/2d)|$ in the kernel.

The tunnelling exponent for the effective particle in the potential $U(r)$ in a semiclassical approximation reads [159]

$$B = 2\varepsilon_0 \int_0^{r_0} \sqrt{2U(r)}\, dr. \tag{314}$$

Performing an integration we obtain

$$B_N^l = \frac{24\varepsilon_0 x_0}{5j^2}\left[\bar{\mu}_0 + \frac{\gamma D\sqrt{\bar{\mu}_0}}{2\pi\lambda_c x_0}\ln\left(\frac{2L_x}{C\pi\gamma D}\right)\right]^{5/2}. \tag{315}$$

The optimal value for the tunnelling of the fluxon is found from the condition of minimum of $B$: $dB/dx_0 = 0$. Thus, finally, we derive

$$B_N^l = \sqrt{\frac{5}{3}}\frac{5J_c}{e\omega_J}\left(\frac{1-j^2}{j^2}\right)\frac{\gamma D}{\pi L_x}\ln\left(\frac{2L_x}{C\pi\gamma D}\right). \tag{316}$$

The dependence of $\Gamma$ on $j$, calculated by means of Eqs. (308) and (316), is shown by dashed lines in Fig. 39. Calculating these curves, we use $\gamma$ and $C$ as adjustable parameters. The analytical approach is appropriate to estimate the value of $\ln\Gamma$. However, the numerical results are more accurate.

The theory presented here, of MQT in stacks of Josephson junctions, is based on a quantum-field-theory approach, in contrast to the phenomenological treatment [162, 163] of capacitively-coupled Josephson junctions. It allows us to describe quantitatively this effect in agreement with the experimental observations. The obtained value of $\Gamma$ is strongly nonlinear with respect to the number of superconducting layers $N$, and changes to $\Gamma \propto N$ when $N$ exceeds a certain critical value $N_c$. The proposed numerical approach can also be used to describe quantum tunnelling in arrays of Josephson junctions [164, 165], as well as in electromechanical [166] and magnetic [167] systems, where the "particle approximation" can be invalid.

## VII. CONCLUSION

As this review shows, Josephson plasma waves in layered superconductors have become a subject of intense study of considerable physical interest. They are interesting for the physics of superconductors, nonlinear physics, and can be used for designing future devices including emitters, filters, detectors, and waveguides working in the sub-THz and THz frequency ranges, which could be very important for various applications.

Electromagnetic waves in layered superconductors display many remarkable features that are very unusual for conducting media. For example, as we show in Sec. II of this review, the electric and magnetic components of the Josephson plasma waves can be of the same order. This property of JPWs can be useful to *avoid the impedance-mismatch problem* when generating THz radiation from moving Josephson vortices (see Sec. IV).

It is also of interest that the interface between the vacuum and a layered superconductor can support surface Josephson plasma waves. As we discuss in Sec. II, the resonant excitation of these waves provide Wood-like anomalies in the reflection of sub-THz and THz electromagnetic waves. This phenomenon could be used for designing future THz detectors.

Layered superconductors represent nonlinear media with a very specific kind of nonlinearity. In Sec. III, we review a number of nonlinear phenomena regarding the propagation of JPWs, which originate from the nonlinear Josephson relation, $J = J_c \sin\varphi$, for the interlayer current. Along with phenomena known from nonlinear optics (e.g., generation of higher harmonics, self-focusing, pumping of weak waves by a stronger one, nonlinear resonance), nontrivial effects can be observed in layered superconductors. A very interesting example of such a phenomenon is the stop-light effect that occurs due to both nonlinearity and dissipation.

Several animations demonstrating the unusual properties of linear and nonlinear Josephson plasma waves are available in Refs. 168, 169. The study of the linear and nonlinear electrodynamics of layered superconductors should also provide important information for future studies in THz science.

**Acknowledgement.**

We gratefully acknowledge partial support from the National Security Agency (NSA), Laboratory of Physical Sciences (LPS), Army Research Office (ARO), National Science Foundation (NSF) grant No. EIA-0130383, JSPS-RFBR 06-02-91200, and Core-to-Core (CTC) program supported by Japan Society for Promotion of Science (JSPS). S.S. acknowledges support from the Ministry of Science, Culture and Sport of Japan via the Grant-in Aid for Young Scientists No 18740224, the EPSRC via No. EP/D072581/1, EP/F005482/1, and ESF network-programme "Arrays of Quantum Dots and Josephson Junctions".

[1] R. Kleiner, F. Steinmeyer, G. Kunkel, and P. Muller, Phys. Rev. Lett. **68**, 2394 (1992).

[2] R. Kleiner and P. Muller, Phys. Rev. B **49**, 1327 (1994).

[3] G. Blatter, M.V. Feigel'man, V.B. Geshkenbein, A.I. Larkin, and V.M. Vinokur, Rev. Mod. Phys. **66**, 1125 (1994).

[4] E.H. Brandt, Rep. Prog. Phys. **58**, 1465 (1995).

[5] V.L. Pokrovsky, Phys. Rep. **288**, 325 (1997).

[6] M.M. Mishonov, Phys. Rev. B **44**, 12033 (1991).

[7] K. Tamasaku, Y. Nakamura, and S. Uchida, Phys. Rev. Lett. **69**, 1455 (1992).

[8] M. Tachiki, T. Koyama, and S. Takahashi, Phys. Rev. B **50**, 7065 (1994).

[9] L.N. Bulaevskii, M.P. Maley, and M. Tachiki, Phys. Rev. Lett. **74**, 801 (1995).

[10] C.C. Homes, T. Timusk, R. Liang, D.A. Bonn, and W.N. Hardy, Phys. Rev. Lett. **71**, 1645 (1993).

[11] Y. Matsuda, M.B. Gaifullin, K. Kumagai, K. Kadowaki, and T. Mochiku, Phys. Rev. Lett. **75**, 4512 (1995).

[12] Y. Matsuda, M.B. Gaifullin, K. Kumagai, K. Kadowaki, T. Mochiku, and K. Hirata, Phys. Rev. B **55**, R8685 (1997).

[13] K. Kadowaki, I. Kakea, M.B. Gaifullin, T. Mochiku, S. Takahashi, T. Koyama, and M. Tachiki, Phys. Rev. B **56**, 5617 (1997).

[14] H.B. Wang, P.H. Wu, and T. Yamashita, Phys. Rev. Lett. **87**, 17002 (2001).

[15] N. Kameda, M. Tokunaga, T. Tamegai, M. Konczykowski, and S. Okayasu, Phys. Rev. B **69**, 180502(R) (2004).

[16] Y. Tominari, T. Kiwa, H. Murakami, M. Tonouchi, H. Wald, P. Seidel, and H. Schneidewind, Appl. Phys. Lett. **80**, 3147 (2002).

[17] J. Zitzmann, A.V. Ustinov, M. Levitchev, and S. Sakai, Phys. Rev. B **66**, 064527 (2002).

[18] Special Issue of Philosophical Transactions: Mathematical, Physical & Engineering Science **362**, Number 1815 (2004), e.g., X.C. Zhang, *Three-dimensional terahertz wave imaging* on page 283; N.C.J. van der Valk and P.C.M. Planken *Towards terahertz near-field microscopy* on page 315; P.H. Bolívar, M. Nagel, F. Richter, M. Brucherseifer, H. Kurz, A. Bosserhoff, and R. Büttner, *Label-free THz sensing of genetic sequences: towards "THz biochips"* on page 323.

[19] M. Chamberlain and M. Smith Eds., *Proceeding of the first international conference on Biomedical imaging and sensing applications of THz technology*, Physics in Medicine and Biology, **47** number 21 (2002).

[20] M. Tonouchi, Nature Photonics **1**, 97 (2007).

[21] A. Dobroiu, C. Otani, and K. Kawase, Measurement Science and Technology **17**, R161 (2006)

[22] A. Dobroiu, Y. Sasaki, T. Shibuya, C. Otani, and K. Kawase, Proc. IEEE **95**, 1566 (2007).

[23] F. Miyamaru, M.W. Takeda, T. Suzuki, and C. Otani, Optic Express **15**, 14804 (2007).

[24] M. Tonouchi, Nat. Phot. **1**, 97 (2007).

[25] F. Capasso, C. Gmachl, D.L. Sivco, and A.Y. Cho, Phys. Today **55**, 34 (2002).

[26] C. Gmachl, F. Capasso, D.L. Sivco, and A.Y. Cho, Rep. Progr. Phys **64**, 1533 (2001).

[27] V.P. Koshelets and S.V. Shitov, Supercond. Sci. Technol. **13**, R53 (2000).

[28] R. Kleiner, Science **318**, 1254 (2007).

[29] K. Tamasaku, Y. Nakamura, and S. Uchida, Phys. Rev. Lett. **69**, 1455 (1992).

[30] K.C. Tsui Ophelia, N.P. Ong, Y. Matsuda, Y.F. Yan, and J.B. Peterson, Phys. Rev. Lett. **73**, 724 (1994).

[31] Y. Matsuda, M.B. Gaifullin, K. Kumagai, K. Kadowaki, and T. Mochiku, Phys. Rev. Lett. **75**, 4512 (1995).

[32] K.C. Tsui Ophelia, N.P. Ong, and J.B. Peterson, Phys. Rev. Lett. **76**, 819 (1996).

[33] T. Hanaguri, Y. Tsuchiya, S. Sakamoto, A. Maeda, and D.G. Steel, Phys. Rev. Lett. **78**, 3177 (1997).

[34] S.E. Shafranjuk, M. Tachiki, and T. Yamashita, Phys. Rev. B **55**,8425 (1997).

[35] N.F. Pedersen and S. Sakai, Phys. Rev. B **58**, 2820 (1998).

[36] S. Sakai and N.F. Pedersen, Phys. Rev. B **60**, 9810 (1999).

[37] Ch. Helm and L.N. Bulaevskii, Phys. Rev. B **66**, 094514 (2002).

[38] E.J. Singley, M. Abo-Bakr, D.N. Basov, J. Feikes, P. Guptasarma, K. Holldack, H.W. Hübers, P. Kuske, M.C. Martin, W.B. Peatman, U. Schade, and G. Wüstefeld, Phys. Rev. B **69**, 092512 (2004).

[39] M.M. Mola, J.T. King, C.P. McRaven, S. Hill, J.S. Qualls, and J.S. Brooks, Phys. Rev. B **62**, 5965 (2000).

[40] S. Sakamoto, A. Maeda, T. Hanaguri, Y. Kotaka, J. Shimoyama, K. Kishio, Y. Matsushita, M. Hasegava, H. Takei, H. Ikeda, and R. Yoshizaki, Phys. Rev. B **53**, R14749 (1996).

[41] Y. Matsuda, M.B. Gaifullin, K. Kumagai, M. Kosugi, and K. Hirata, Phys. Rev. Lett. **78**, 1972 (1997).

[42] T. Shibauchi, A. Sato, A. Mashio, T. Tamegai, H. Mori, S. Tajima, and S. Tanaka, Phys.

Rev. B **55**, R11977 (1997).

[43] A. Maeda, Y. Tsuchiya, S. Sakamoto, H. Ikeda, and R. Yoshizaki, Physica C **293**, 143 (1997).

[44] A.E. Koshelev, L.N. Bulaevskii, and M.P. Maley, Phys. Rev. Lett. **81**, 1292 (1998).

[45] N.F. Pedersen and S. Sakai, Physica C **332**, 297 (2000).

[46] Yu.I. Latishev, A.E. Koshelev, and L.N. Bulaevskii, Phys. Rev. B **68**, 134504 (2003).

[47] S. Sakai, P. Bodin, and N.F. Pedersen, J. Appl. Phys **73**, 2411 (1993).

[48] L.N. Bulaevskii, M. Zamora, D. Baeriswyl, H. Beck, and J.R. Clem, Phys. Rev. B **50**, 12831 (1994).

[49] T. Koyama and M. Tachiki, Phys. Rev. B **54**, 16183 (1996).

[50] S.N. Artemenko and S.V. Remizov, JETP Lett. **66**, 853 (1997).

[51] M. Tachiki and M. Machida, Physica C **341-348**, 1493 (2000).

[52] M. Machida, T. Koyama, A. Tanaka, and M. Tachiki, Physica C **331**, 85 (2000).

[53] S.N. Artemenko and S.V. Remizov, Physica C **362**, 200 (2001).

[54] Yu.H. Kim and J. Pokharel, Physica C **384**, 425 (2003).

[55] L.D. Landau, E.M. Lifshitz, and L.P. Pitaevskii, *Electrodynamics of Continuous Media*, (Butterworth-Heinemann, Oxford, 1995).

[56] R.G. Mints and I.B. Snapiro, Phys. Rev. B **52**, 9691 (1995).

[57] G. Hechtfischer, R. Kleiner, A.V. Ustinov, and P. Muller, Phys. Rev. Lett. **79**, 1365 (1997).

[58] T. Koyama and M. Tachiki, Solid State Commun. **96**, 367 (1995).

[59] G. Hechtfischer, R. Kleiner, A.V. Ustinov, and P. Muller, Appl. Supercond. **5**, 303 (1997).

[60] G. Hechtfischer, W. Walkenhorst, G. Kunkel, K. Schlenga, R. Kleiner, P. Muller, and L. Johnson, IEEE Trans. Appl. Supercond. **7**, 2723 (1997).

[61] Yu.I. Latyshev, M.B. Gaifullin, T. Yamashita, M. Machida, and Y. Matsuda, Phys. Rev. Lett. **87**, 247007 (2001).

[62] M. Tachiki, M. Iizuka, K. Minami, S. Tejima, and H. Nakamura, Phys. Rev. B **71**, 134515 (2005).

[63] V.V. Kurin and A.V. Yulin, Phys. Rev. B **55**, 11659 (1997).

[64] E. Goldobin, A. Wallraff, N. Thyssen, and A.V. Ustinov, Phys. Rev. B **57**, 130 (1998).

[65] V.M. Krasnov and D. Winkler, Phys. Rev. B **60**, 13179 (1999).

[66] E. Goldobin, B.A. Malomed, and A.V. Ustinov, Phys. Rev. B **62**, 1414 (2000).

[67] V.M. Krasnov, Phys. Rev. B **63**, 064519 (2001).

[68] A. Barone and G. Paterno, *Physics and Applications of the Josephson Effect* (Wiley, New York, 1982).

[69] A. Gurevich, Phys. Rev. B **46**, 3187 (1992).

[70] S. Savel'ev, V. Yampol'skii, A. Rakhmanov, and F. Nori, Phys. Rev. B **72**, 144515 (2005).

[71] S. Savel'ev, V. Yampol'skii, A. Rakhmanov, and F. Nori, Physica C **437-438**, 281 (2006).

[72] S. Savel'ev, V. Yampol'skii, A.L. Rakhmanov, and F. Nori, Physica C **445-448**, 175 (2006).

[73] J.H. Choy, S.J. Kwon, and G.S. Park, Science **280**, 1589 (1998).

[74] D. Dulić, A. Pimenov, D. van der Marel, D.M. Broun, S. Kamal, W.N. Hardy, A.A. Tsvetkov, I.M. Sutjaha, R.X. Liang, A.A. Menovsky, A. Loidl, and S.S. Saxena, Phys. Rev. Lett. **86**, 4144 (2001).

[75] S.J. Kwon, J.H. Choy, D. Jung, and P.V. Huong, Phys. Rev. B **66**, 224510 (2002).

[76] K. Endo, H. Sato, K. Yamamoto, T. Mizukoshi, T. Yoshizawa, K. Abe, P. Badica, J. Itoh, K. Kajimura, and H. Akoh, Physica C **372-376**, 1075 (2002).

[77] W.Y. Chen, Q. Li, L. Qiu, W.M. Chen, and S.S. Jiang, J. Supercond. **17**, 525 (2004).

[78] R. Rajaraman, *Solitons and instantons, An introduction to solitons and instantons in quantum field theory*, (North-Holland Publishing, Amsterdam, 1982).

[79] O.H. Olsen and M.R. Samuelsen, Phys. Rev. B **34**, 3510 (1986).

[80] S. Savel'ev, A.L. Rakhmanov, V.A. Yampol'skii, and F. Nori, Nat. Phys. **2**, 521 (2006).

[81] S. Savel'ev, V.A. Yampol'skii, A.L. Rakhmanov, and F. Nori, Phys. Rev. B **75**, 184503 (2007).

[82] V.A. Yampol'skii, S. Savel'ev, T.M. Slipchenko, A.L. Rakhmanov, and F. Nori, Physica C **468**, 499 (2008).

[83] V.A. Yampol'skii, S. Savel'ev, A.L. Rakhmanov, and F. Nori, Phys. Rev. B. **78**, 024511 (2008).

[84] A.E. Koshelev, Phys. Rev. B **62**, R3616 (2000).

[85] D.A. Ryndyk, Phys. Rev. Lett. **80**, 3376 (1998).

[86] A. Gurevich and M. Tachiki, Phys. Rev. Lett. **83**, 183 (1999).

[87] Ch. Helm, J. Keller, Ch. Peris, and A. Sergeev, Physica C **362**, 43 (2001).

[88] Y.M. Shukrinov and F. Mahfouzi, Phys. Rev. Lett. **98**, 157001 (2007).

[89] Y.M. Shukrinov, F. Mahfouzi, and N.F. Pedersen, Phys. Rev. B **98**, 104508 (2007).

[90] B.D. Josephson, Adv. Phys. **14**, 419 (1965).

[91] P.W. Anderson and A.H. Dayem, Phys. Rev. Lett. **13**, 195 (1964).

[92] H. Suhl, Phys. Rev. Lett. **14**, 226 (1965).

[93] S. Rother, Y. Koval, P. Muller, R. Kleiner, D.A. Ryndyk, J. Keller, and C. Helm, Phys. Rev. B **67**, 024510 (2003).

[94] P.M. Platzman and P.A. Wolff, *Waves and Interactions in Solid State Plasmas* (Academic Press, London, 1973).

[95] V.M. Agranovich and D.L. Mills, *Surface Polaritons* (Nauka, Moscow, 1985).

[96] H. Raether, *Surface Plasmons* (Springer-Verlag, New York, 1988).

[97] R. Petit, *Electromagnetic Theory of Gratings* (Springer, Berlin, 1980).

[98] A.V. Zayats, I.I. Smolyaninov, A.A. Maradudin, Phys. Rep. **408**, 131 (2005).

[99] A.V. Kats and I.S. Spevak, Phys. Rev. B **65**, 195406 (2002).

[100] W.L. Barnes, A. Dereux, and T.W. Ebbesen, Nature (London) **424**, 824 (2003).

[101] T.W. Ebbesen, H.J. Lezec, H.F. Ghaemi, T. Tio, and P.A. Wolff, Nature (London) **391**, 667 (1998).

[102] H.F. Ghaemi, T. Thio, D.E. Grupp, T.W. Ebbesen, and H.J. Lezec, Phys. Rev. B **58**, 6779 (1998).

[103] S.C. Hohng, Y.C. Yoon, D.S. Kim, V. Malyarchuk, R.Müller, Ch. Lienau, J.W. Park, K.H. Yoo, J. Kim, H.Y. Ryu, and Q.H. Park, Appl. Phys. Lett. **81**, 3239 (2002).

[104] L. Martin-Moreno, F.J. Garcia-Vidal, H.J. Lezec, K.M. Pellerin, T. Thio, J.B. Pendry, and T.W. Ebbesen, Phys. Rev. Lett. **86**, 1114 (2001).

[105] A.V. Kats and A.Yu. Nikitin, Phys. Rev. B **7**0, 235412 (2004).

[106] A.V. Kats, M.L. Nesterov, and A.Yu. Nikitin, Phys. Rev. B **72**, 193405 (2005).

[107] S. Savel'ev, V. Yampol'skii, and F. Nori, Phys. Rev. Lett. **95**, 187002 (2005).

[108] S. Savel'ev, V. Yampol'skii, A.L. Rakhmanov, and F. Nori, Physica C **445-448**, 183 (2006).

[109] V.A. Yampol'skii, A.V. Kats, M.L. Nesterov, A.Yu. Nikitin, T.M. Slipchenko, S. Savel'ev, and F. Nori, Phys. Rev. B **76**, (2007).

[110] A. Otto, Zeitschr. Phys. **216**, 398 (1968).

[111] H.G. Winful, Phys. Rep. **436**, 1 (2006).

[112] H. Shibata and T. Yamada, Phys. Rev. B , **54**, 7500 (1996).

[113] A.E. Koshelev, Phys. Rev. Lett. **83**, 187 (1999).

[114] S. Savel'ev and F. Nori, Nature Materials **1**, 179 (2002).

[115] S. Savel'ev, A.L. Rakhmanov, and F. Nori, Phys. Rev. Lett. **94**, 157004 (2005).

[116] E. Yablonovitch, Phys. Rev. Lett. **58**, 2059 (1987).

[117] J.D. Joannopoulos, R.D. Meade, and J.N. Winn, *Photonic Crystals* (Princeton, Princeton, 1995).

[118] R.E. Slusher and B.J. Eggleton, *Nonlinear Photonic Crystals* (Springer, NY, 2003).

[119] K. Sakoda, *Optical Properties of Photonic Crystals* (Springer, NY, 2001).

[120] H. Susanto, E. Goldobin, D. Koelle, R. Kleiner, and S.A. van Gils, Phys. Rev. B **71**, 174510 (2005).

[121] H. Takeda, K. Yoshino, A.A. Zakhidov, Phys. Rev. B **70**, 085109 (2004).

[122] S. Savel'ev, A.L. Rakhmanov, and F. Nori, Phys. Rev. B **74**, 184512 (2006).

[123] I.O. Kulik, JETP **24**, 1307 (1967).

[124] A.L. Fetter and M.J. Stephen, Phys. Rev **168**, 475 (1968).

[125] D.R. Gulevich, S. Savel'ev, F. Kusmartsev, V.A. Yampol'skii, and F. Nori, J. Appl. Phys. (submitted).

[126] R.W. McGowan, G. Gallot, and D. Grischkowsky, Opt. Lett. **24**, 1431 (1999).

[127] G. Gallot, S.P. Jamison, R.W. McGowan, and D. Grischkowsky, J. Opt. Soc. Am. B **17**, 851 (2000).

[128] K. Wang and D.M. Mittleman, Nature **432**, 376 (2004).

[129] R. Mendis and D. Grischkowsky, J. Appl. Phys. **88**, 4449 (2000).

[130] S.P. Jamison, R.W. McGown, and D. Grischkowsky, Appl. Phys. Lett. **76**, 1987 (2000).

[131] D.L. Mills, *Nonlinear Optics*, (Springer, Berlin, 1998).

[132] N. Bloembergen, *Nonlinear Optics*, (World Scientific, Singapore, 1996).

[133] L.D. Landau and E.M. Lifshitz, *Mechanics*, (Butterworth-Heinemann, Oxford, 1995).

[134] A.V. Ustinov, T. Doderer, R.P. Huebener, N.F. Pedersen, B. Mayer, and V.A. Oboznov, Phys. Rev. Lett. **69**, 1815 (1992).

[135] A.V. Ustinov, T. Doderer, R.P. Huebener, B. Mayer, and V.A. Oboznov, Europhys. Lett. **19**, 63 (1992).

[136] J.C. Swihart, J. Appl. Phys. **32**, 461 (1961).

[137] I.E. Batov, X.Y. Jin, S.V. Shitov, Y. Koval, P. Müller, and A.V. Ustinov, Appl. Phys. Lett. **88**, 262504 (2006).

[138] A.E. Koshelev and I. Aranson, Phys. Rev. B **64**, 174508 (2001).

[139] S.N. Artemenko and S.V. Remizov, Phys. Rev. B **67**, 144516 (2003).

[140] L. Ozyuzer, A.E. Koshelev, C. Kurter, N. Gopalsami, Q. Li, M. Tachiki, K. Kadowaki, T. Yamamoto, H. Minami, H. Yamaguchi, T. Tachiki, K.E. Gray, W.-K. Kwok, and U. Welp, Science **318**, 1291 (2007).

[141] L.N. Bulaevskii and A.E. Koshelev, Phys. Rev. Lett. **99**, 057002 (2007).

[142] Yu.I. Latyshev, T. Yamashita, L.N. Bulaevskii, M.J. Graf, A.V. Balatsky, and M.P. Maley, Phys. Rev. Lett. **82**, 5345 (1999).

[143] L.N. Bulaevskii and A.E. Koshelev, Phys. Rev. B **77**, 014530 (2008).

[144] A.L. Rakhmanov, S. Savel'ev, M. Gaifullin, and F. Nori (unpublished).

[145] S.Z. Lin and X. Hu, Phys. Rev. Lett. **100**, 247006 (2008).

[146] K. Inomata, S. Sato, K. Nakajima, A. Tanaka, Y. Takano, H.B. Wang, M. Nagao, H. Hatano, and S. Kawabata, Phys. Rev. Lett. **95**, 107005 (2005).

[147] T. Bauch, F. Lombardi, F. Tafuri, A. Barone, G. Rotoli, P. Delsing, and T. Claeson, Phys. Rev. Lett. **94**, 087003 (2005).

[148] T. Bauch, T. Lindstrom, F. Tafuri, G. Rotoli, Per Delsing, T. Claeson, and F. Lombardi, Science **311**, 57 (2006).

[149] X.Y. Jin, J. Lisenfeld, Y. Koval, A. Lukashenko, A.V. Ustinov, and P. Müller, Phys. Rev. Lett. **96**, 177003 (2006).

[150] J.Q. You and F. Nori, Phys. Today **58** (11), 42 (2005).

[151] A.J. Berkley, H. Xu, R.C. Ramos, M.A. Gubrud, F.W. Strauch, P.R. Johnson, J.R. Anderson, A.J. Dragt, C.J. Lobb, and F.C. Wellstood, Science **300**, 1548 (2003).

[152] R. McDermott, R.W. Simmonds, M. Steffen, K.B. Cooper, K. Cicak, K.D. Osborn, S. Oh, D.P. Pappas, and J.M. Martinis, Science **307**, 1299 (2005).

[153] S. Kawabata, S. Kashiwaya, Y. Asano, and Y. Tanaka, Phys. Rev. B **70**, 132505 (2004).

[154] S. Kawabata, S. Kashiwaya, Y. Asano, and Y. Tanaka, Phys. Rev. B **72**, 052506 (2005).

[155] S. Savel'ev, A.L. Rakhmanov, and F. Nori, Phys. Rev. Lett. **98**, 077002 (2007).

[156] S. Savel'ev, A.L. Rakhmanov, and F. Nori, Phys. Rev. Lett. **98**, 269901 (2007).

[157] S. Savel'ev, A.O. Sboychakov, A.L. Rakhmanov, and F. Nori, Phys. Rev. B **77**, 014509 (2008).

[158] A.O. Sboychakov, S. Savel'ev, A.L. Rakhmanov, and F. Nori, Europhys. Lett., **80** 17009 (2007).

[159] S. Coleman, Phys. Rev. D **15**, 2929 (1977).

[160] A.O. Caldeira and A.J. Leggett, Phys. Rev. Lett. **46**, 211 (1981).

[161] T. Kato and M. Imada, J. Phys. Soc. Jpn. **66**, 1445 (1997).

[162] M. Machida, T. Koyama, Supercond. Sci. Technol. **20**, S23 (2007).

[163] M.V. Fistul, Phys. Rev. B **75**, 014502 (2007).

[164] E. Goldobin, D. Koelle, and R. Kleiner, Phys. Rev. B **67** 224515 (2003).

[165] E. Goldobin, K. Vogel, O. Crasser, R. Walser, W.P. Schleich, D. Koelle, and R. Kleiner, Phys. Rev. B **72**, 054527 (2005).

[166] S. Savel'ev, A.L. Rakhmanov, X. Hu, A. Kasumov, and F. Nori, Phys. Rev. B **75**, 165417 (2007).

[167] B.A. Ivanov, D.D. Sheka, V.V. Kryvonos, and F.G. Mertens, Phys. Rev. B **75**, 132401 (2007).

[168] http://dml.riken.jp/waveguides

[169] http://dml.riken.go.jp

|  | Nonlinear Josephson plasma waves | Traditional nonlinear optics |
|---|---|---|
| **Nonlinearity** | Due to nonlinear current-phase relation, $J = J_c \sin(\varphi)$ | Due to, e.g., nonlinear dependence of the refraction coefficient on the electric field, $n(E) = n_0 + n_2 E^2$ |
| **Higher harmonic frequencies generated from the basic $\omega$** | Only $3\omega$, $5\omega$, ... | $2\omega$, $3\omega$, $4\omega$, ... |
| **Wave propagation below gap** | Propagation of plane wave with $\omega < \omega_J$ | — |
| **Slowing down EMW** | THz wave can slow down significantly if $\omega < \omega_J$ | Light can be slowed down |
| **Transparency due to nonlinearity** | Weak waves with $\omega < \omega_J$, which cannot originally propagate, do propagate assisted by nonlinear JPWs | Self-induced transparency |
| **Nonlinear pumping** | Weak wave grows while the nonlinear JPW is attenuated | Pumping of nonlinear waves in plasma |
| **Focusing** | Below $\omega_J$, focused THz beam propagates | Self-focusing due to nonlinearity of $n(E)$ |
| **Wave packet spreading** | Open problem | Can propagate without spreading |
| **Loading-unloading cycles due to nonlinearity** | Frequency hysteresis of nonlinear geometric resonance converts continuous radiation to amplified pulses (analogy with nonlinear mechanical resonance) | Nonlinear optical bistable devises |

TABLE II: Comparison between nonlinear electromagnetic waves in optics and nonlinear Josephson plasma waves.